\documentclass[11pt,canadian,preprint,preprintnumbers,amsmath,amssymb,nofootinbib,superscriptaddress]{revtex4}
\usepackage[latin9]{inputenc}
\usepackage[a4paper]{geometry}
\usepackage{color}
\usepackage{babel}
\usepackage{rotating}
\usepackage{amsmath}
\usepackage{amssymb}
\usepackage{graphicx,subfigure}
\usepackage{esint}
\usepackage{epstopdf}
\usepackage[unicode=true,pdfusetitle,
 bookmarks=true,bookmarksnumbered=false,bookmarksopen=false,
 breaklinks=false,pdfborder={0 0 1},backref=false,colorlinks=false]
 {hyperref}

\makeatletter
\@ifundefined{textcolor}{}
{%
 \definecolor{BLACK}{gray}{0}
 \definecolor{WHITE}{gray}{1}
 \definecolor{RED}{rgb}{1,0,0}
 \definecolor{GREEN}{rgb}{0,1,0}
 \definecolor{BLUE}{rgb}{0,0,1}
 \definecolor{CYAN}{cmyk}{1,0,0,0}
 \definecolor{MAGENTA}{cmyk}{0,1,0,0}
 \definecolor{YELLOW}{cmyk}{0,0,1,0}
}
\newcommand{\tcr}{\textcolor{red}}

\usepackage{color}\usepackage{graphics}\usepackage{dcolumn}\usepackage{bm}\usepackage{afterpage}

\numberwithin{equation}{section}

\usepackage{hhline}
\usepackage[normalem]{ulem}
\usepackage{placeins}

\makeatother

\begin{document}
\allowdisplaybreaks

\title{Gauss-Bonnet Boson Stars with a Single Killing Vector}

\author{Laura J. Henderson}
\email{l7hender@uwaterloo.ca}

\author{Robert B. Mann}
\email{rbmann@sciborg.uwaterloo.ca}

\affiliation{Department of Physics and Astronomy, University of Waterloo,\\
 Waterloo, Ontario, Canada, N2L 3G1}

\author{Sean Stotyn}
\email{sstotyn@phas.ubc.ca}

\affiliation{Department of Physics and Astronomy, University of British Columbia,\\
 Vancouver, British Columbia, Canada, V6T 1Z1}

\date{\today}

\begin{abstract}
We  construct asymptotically anti-de Sitter boson stars in Einstein-Gauss-Bonnet gravity coupled to a $\frac{D-1}{2}$-tuplet of complex massless scalar fields both perturbatively and numerically
in $D=5,7,9,11$ dimensions. These solutions possess just a single helical Killing symmetry due to the choice of scalar fields.  The energy density at the centre of the star characterizes the solutions, and for
each choice of the Gauss-Bonnet coupling $\alpha$ we obtain a one parameter family of solutions. All solutions respect the first
law of thermodynamics; in the numerical case to within 1 part in $10^6$.
We describe the dependence of the angular velocity, mass, and angular momentum of the boson stars on $\alpha$ and on the dimensionality.  For $D>5$,  these quantities  exhibit damped oscillations 
  about finite central values as the central energy density tends to infinity, where the amplitude of oscillation increases nonlinearly with $\alpha$.  In the limit of diverging central energy density, the Kretschmann invariant at the centre of the boson star also diverges.  This is in contrast to the $D=5$ case, where the Kretschmann invariant diverges at a finite value of the central energy density.
\end{abstract}

\maketitle

\newpage{}

\section{Introduction}

The discovery of the Higgs boson has given a clear indication of the existence of at least one scalar field in nature.  In addition to the fact that  scalar fields are pervasive in many versions of quantum gravity, this has generated renewed interest in obtaining solutions to quantum-inspired effective theories of gravitation coupled to scalar fields.  

Prominent amongst such solutions are boson stars: these are 
 smooth, horizonless geometries composed of self-gravitating and (possibly) self-interacting bosonic matter\cite{Kaup:1968zz,Ruffini:1969qy}.  
 Unlike ordinary stars and planets, they typically do not have a sharp edge (though solutions with such edges have recently been found \cite{Kichakova:2013sza}) but instead are  bundles of field energy that decay at large distances from their cores.  Their astrophysical relevance is not clear at this point in time but they may be possible dark matter halo candidates \cite{Jetzer:1991jr}.  In an excited state they typically produce a more physically realistic, flatter rotation curve than  in the ground state. However, such excited states decay to the ground state unless they  are in rather specific mixed states \cite{Bernal:2009zy}.   Furthermore, boson stars can provide dark alternatives to astrophysical black hole candidates, which could potentially be discerned by gravitational wave astronomy \cite{Berti:2006qt,Kesden:2004qx}.     
 
  From a theoretical perspective, asymptotically anti-de Sitter (AdS)  boson stars may play an important role in holographic gauge theories through the AdS/CFT correspondence \cite{Liebling:2012fv}.    All such solutions are zero-temperature objects without horizons and, as such, they describe   finite energy excitations above the vacuum state.   If AdS boundary conditions are present, their (non-linear) stability is of key import, as it determines whether or not the corresponding state in the holographic dual CFT thermalizes and on what time-scale.   While recent work indicates that asymptotically AdS space-times suffer from a gravitational turbulent instability \cite{Dias:2011ss,Bizon:2011gg,Jalmuzna:2011qw,Maliborski:2013jca}, which might lead one to expect that AdS boson stars are non-linearly unstable to black hole formation, it has also been shown that there exists a wide range of initial data that are immune to this instability \cite{Buchel:2013uba}.  This suggests that  on the gauge theory side there is a family of strongly coupled CFT states that do not thermalize in finite time.
  
 A natural question to ask is how generic this feature of the initial data considered in Ref. \cite{Buchel:2013uba} is.  In that respect, one of the immediate ongoing tasks is to map out the  territory of boson star solutions. Quite a wide  range of   solutions have been obtained for various forms of scalar matter.  These include  a complex doublet of massive \cite{Astefanesei:2003qy,Brihaye:2013hx} and massless  \cite{Stotyn:2011ns,Dias:2011at} scalar fields,  self-interacting scalar fields \cite{Colpi:1986ye,Ho:2002vz}, scalars with gauge charges \cite{Brihaye:2004nd,Dias:2011tj}, scalars in space times with de Sitter \cite{Fodor:2010hg,Cai:1997ij}, flat \cite{Colpi:1986ye,Hartmann:2012gw}, or anti-de Sitter \cite{Astefanesei:2003qy,Bjoraker:1999yd} boundary conditions, solutions in $(2+1)$ dimensions \cite{Astefanesei:2003qy,Stotyn:2012ap,Stotyn:2013spa}, and rotating doublets \cite{Dias:2011at} and multiplets \cite{Stotyn:2011ns,Stotyn:2013yka} of scalars.
  
  These latter solutions are intriguing insofar as they posses only a single Killing vector, in contrast to most other solutions that impose a relatively high level of symmetry to yield equations that can feasibly be solved, either analytically or numerically.   Solutions possessing less symmetry have gained interest lately, in part because of recent work suggesting that the turbulent instability of global AdS is due a high level of symmetry \cite{Dias:2012tq}: the normal mode frequencies are all integer multiples of the AdS frequency, leading to a large number of resonances responsible for the nonlinear instability.  Though a clear relationship  between boson star properties and physically observed states in CFTs is not currently known,  these lower-symmetry solutions are thus expected to be nonlinearly stable and their holographic dual CFT states not to thermalize.

Construction of analytic solutions to gravitational field equations have commonly exploited space-time symmetries in the form of Killing vector fields, which generate the space-time's isometry group.  This is the case for both Einstein gravity and its   quantum-inspired generalizations, such as Lovelock theories.  A rigidity theorem states that if a space-time is stationary, 
it must also possess an axis of symmetry, implying the horizon of the (multiply rotating) black hole must be a Killing horizon with respect to the Killing field
$$
\zeta = \frac{\partial}{\partial t} + \sum_{I}\Omega_I \frac{\partial}{\partial \phi_I}
$$
where $\frac{\partial}{\partial t}$ and $\frac{\partial}{\partial \phi_I}$ generate the respective stationary and rotational symmetries, with $ \Omega_I$ the associated angular velocities. This
yields a minimum of 2 Killing symmetries \cite{Hawking:1971vc,Hollands:2006rj,Moncrief:2008mr}.
Under physically reasonable assumptions 
a variety of theorems indicate that Killing symmetries are ubiquitous. 

If matter fields are introduced, it is possible to reduce the overall degree of symmetry a solution might have whilst retaining a higher degree of symmetry in the metric itself. An interesting case that has been explored recently in a few settings is that of rotating boson star solutions possessing a single Killing vector (SKV).  In 5 dimensions there is a clever choice of ansatz  \cite{Hartmann:2010pm} for a doublet of scalar fields that admits SKV rotating boson star solutions to the Einstein equations with AdS boundary conditions \cite{Dias:2011at}. The scalar doublet has an harmonic time dependence that breaks the continuous rotational symmetry.  This ansatz is straightforwardly generalizable to any odd dimension  \cite{Stotyn:2011ns,Stotyn:2012ap}, allowing for perturbative solutions (where energy and angular momentum are small) and for full numerical solutions \cite{Stotyn:2013yka,Stotyn:2013spa}.

In this paper we obtain both perturbative and numerical solutions for SKV asymptotically AdS boson stars in $D=5,7,9,11$ dimensions in Einstein-Gauss-Bonnet (EGB) gravity.  EGB gravity is a special case of Lovelock gravity \cite{Lovelock:1971yv}, a class of higher curvature theories of gravity that
are distinguished by having their field equations not containing terms with more than two derivatives of the metric.  They are of interest in quantum gravity insofar as one expects the Einstein action to be an effective gravitational action, valid for small curvature (or low energies), that will be modified by higher-curvature terms. A given contribution to the action in $D$-dimensions consists of a product of the Euler density $\mathcal{L}^{\left(k\right)}$ of a $2k$ dimensional manifold and an arbitrary constant $\hat{\alpha}_{\left(k\right)}$ and so contributes to the equations of motion only if $D>2k$.   The simplest case beyond Einstein gravity is EGB gravity for which $k=2$, implying $D\geq 5$.

   As with the Einstein case, we obtain for any given value of the Gauss-Bonnet coupling constant $\alpha$ (except at a certain critical value $\alpha_{\text{cr}}$) a one parameter family of solutions parameterized by the central energy density of the boson star.   For $D>5$, in the limit that this central energy density tends to infinity, the mass, angular momentum, and angular velocity all display damped oscillations that limit to finite values, while the Kretschmann scalar at the centre of the boson star diverges.  We find that as $\alpha$ increases, the amplitude of the damped oscillations increases, while it decreases with increasing space-time dimension.  For $D=5$ we find markedly different behaviour: the Kretschmann scalar at the centre diverges for a finite value of the central energy density, with this critical value decreasing as $\alpha$ increases.     All of the solutions constructed in this paper, both perturbative and numerical,  limit to  the  Einstein case \cite{Stotyn:2011ns,Stotyn:2012ap} as $\alpha \to 0$; we do not consider alternate possible branches of solutions that do not have this limit, though such branches may indeed exist with the addition of the Gauss-Bonnet term in the gravitational action.
  
The critical value of $\alpha$ exists in any dimension and corresponds to a class of quadratic curvature theories that have different properties than those of the standard EGB-type \cite{Crisostomo:2000bb}.
 Locally maximally symmetric solutions admit only one fixed radius of curvature, and while
they admit  spherically symmetric  black hole  solutions, these solutions do not obey the falloff boundary conditions that solutions in $\alpha\neq \alpha_{\text{cr}}$ theories require.  Indeed, this class of theories admits BTZ-like solutions, in which there is a mass gap between the zero mass black hole and AdS spacetime. While we shall not consider these theories here, analysis of boson star solutions in this case would  be interesting, and require an approach similar to that recently employed for   $D=3$ rotating boson stars  \cite{Stotyn:2013yka}.

Numerically we are  unable to explore values of $\alpha$ in the range $0.90\alpha_{\text{cr}} <\alpha < \alpha_{\text{cr}}$.  However it is possible to examine boson-star solutions for $\alpha > \alpha_{\text{cr}}$. Even though these solutions have
ghostlike excitations about an AdS-vacuum, their  physical mass and angular momentum are both multiplied by an $\alpha$-dependent factor that ensures the mass remains positive.  We shall briefly comment on these solutions near the end of our paper, leaving a more detailed analysis for future study.

The remainder of this paper is structured as follows: in section \ref{Setup} we present the metric and scalar field ans\"atze and obtain the constraint equations and ordinary differential equations (ODEs) that must be solved. We describe the boundary conditions and the basic physical properties of the boson stars  in  section \ref{BosonConditions}.  We then analytically construct perturbative solutions in section \ref{pertsolns}. In section \ref{numconst} we describe the numerical methods used to construct the full non-perturbative solutions, which are presented in section \ref{Results} along with a discussion about their salient features.  In section \ref{beyond} we discuss a preliminary investigation of the region of parameter space for which $\alpha > \alpha_{\text{cr}}$.  Finally, in section \ref{Conclusion} we provide some concluding remarks.

\section{Setup}

\label{Setup}

We begin with $D=n+2$ dimensional Einstein-Gauss-Bonnet gravity with negative
cosmological constant minimally coupled to an $\frac{n+1}{2}$-tuplet of
complex massless scalar fields
\begin{equation}
S=\frac{1}{16\pi}\int{d^{D}x\sqrt{-g}\left(R - 2\Lambda +\alpha L_{GB} -2\big|\nabla\vec{\Pi}\big|^{2}\right)}\label{eq:action}
\end{equation}
 where
\begin{equation}
 L_{GB} =  R_{a b c d}R^{abcd} - 4 R_{ab}R^{ab} +R^2
 \label{eq:GB}
\end{equation}
and where $\Lambda$ is the cosmological constant.  The equations of motion resulting from this action are 
\begin{eqnarray}
G_{ab} +\Lambda g_{ab}  - 2\alpha \Big(-R_{a c d e}R_b^{\phantom{b}cde} + 2R_{a c b d}R^{cd} + 2R_{a c}R^c_{\phantom{c}b} - R R_{ab} +\frac{1}{4} g_{ab}L_{GB}   \Big)
&=&T_{ab} \label{eq:GFE}\\
\nabla^{2}\vec{\Pi} &=& 0
 \label{eq:PiFE}
\end{eqnarray}
where 
\begin{equation}
T_{ab}=\left(\partial_{a}\vec{\Pi}^{*}\partial_{b}\vec{\Pi}+\partial_{a}\vec{\Pi}\partial_{b}\vec{\Pi}^{*}\right)-g_{ab}\left(\partial_{c}\vec{\Pi}\partial^{c}\vec{\Pi}^{*}\right).\label{eq:Tab}
\end{equation}
is the stress-energy tensor of the scalar fields.  We will only be interested in solutions that asymptote to AdS and we thus require global AdS to be a solution to eq. (\ref{eq:GFE}) with $\vec\Pi=0$.  This requirement fixes the bare cosmological constant $\Lambda$ to
\begin{equation}
\label{Lamalpha}
\Lambda=\frac{n(n + 1)\big((n - 1)(n- 2)\alpha - \ell^2\big)}{2\ell^4},
\end{equation}
where $\ell$ is the effective AdS length.  Note that global AdS is still a solution for $\Lambda$ positive, negative, or zero, provided $\alpha$ is appropriately chosen; this is because the Gauss-Bonnet term, in the absence of matter sources, acts like a negative cosmological constant.

We shall employ the following ans\"atze for the metric and scalar fields \cite{Stotyn:2011ns}
\begin{equation}
ds^{2}=-f(r)g(r)dt^{2}+\frac{dr^{2}}{f(r)}+r^{2}\bigg(h(r)\big(d\chi+A_{i}dx^{i}-\Omega(r)dt\big)^{2}+g_{ij}dx^{i}dx^{j}\bigg)\label{eq:metric}
\end{equation}
\begin{equation}
\Pi_{i}=\Pi(r)e^{-i\omega t}z_{i},\quad\quad\quad\quad i=1...\frac{n+1}{2}\label{eq:ScalarField}
\end{equation}
where $z_{i}$ are complex coordinates such that ${\displaystyle \sum_{i}dz_{i}d\bar{z}_{i}}$
is the metric of a unit $n-$sphere.  It is straightforward to show using the choice
 \begin{equation}
z_{i}=\left\{ {\genfrac{.}{.}{0pt}{0}{e^{i(\chi+\phi_{i})}\cos\theta_{i}{\displaystyle \prod_{j<i}\sin\theta_{j},\quad\quad\quad i=1...\frac{n-1}{2}}}{{e^{i\chi}{\displaystyle \prod_{j=1}^{\frac{n-1}{2}}\sin\theta_{j}}},\quad\quad\quad\quad\quad i=\frac{n+1}{2}}}\right.\label{eq:zi}
\end{equation}
that   ${\displaystyle \sum_{i}dz_{i}d\bar{z}_{i}=(d\chi+A_{i}dx^{i})^{2}+g_{ij}dx^{i}dx^{j}}$
is the Hopf fibration of the unit $n-$sphere, where 
\begin{equation}
A_{i}dx^{i}=\sum_{i=1}^{\frac{n-1}{2}}{\cos^{2}\theta_{i}\left[\prod_{j<i}\sin^{2}\theta_{j}\right]d\phi_{i}}
\end{equation}
and $g_{ij}$ is the metric on a unit complex projective space $\mathbb{CP}^{\frac{n-1}{2}}$.  In these coordinates, $\chi$ and the $\phi_{i}$ all have period $2\pi$ while the $\theta_{i}$ take value in the range $[0,\frac{\pi}{2}]$.  This construction only works in odd dimensions, so we shall restrict our attention  to $n=3,5,7,9$ (though  our results can be extended to any odd value of $n\ge3$).

The $n=3$ form of this construction was first used in Ref. \cite{Hartmann:2010pm} to construct boson star solutions in 5-dimensional Einstein gravity coupled to a doublet of self-interacting scalars.     The construction above is crucial in obtaining rotating boson star solutions in any odd dimension.  The scalar fields can be regarded as
coordinates on ${\mathbb{C}}^{\frac{n+1}{2}}$, with   $\vec{\Pi}$ tracing
out a round $n$-sphere (for any value of $r$) with a time-varying but otherwise constant
phase. Constant $r$ surfaces in the metric (\ref{eq:metric}) correspond to squashed rotating $n$-spheres. The first
term in the stress-energy tensor $T_{ab}$ is the pull-back of the round metric on the $n$-sphere and the
second term is proportional to $g_{ab}$. Hence $T_{ab}$   has
the same symmetries as the metric (\ref{eq:metric}). 

Note that while  the matter stress tensor has the same symmetries as the metric,
the scalar fields themselves do not. The scalar field
(\ref{eq:ScalarField}) is only invariant under the combination 
\begin{equation}
\kappa =\partial_{t}+\omega\partial_{\chi}.\label{eq:KV}
\end{equation}
whereas  the metric (\ref{eq:metric}) is indeed
  invariant under $\partial_{t}$, $\partial_{\chi}$ as well as the rotations of $\mathbb{CP}^{\frac{n-1}{2}}$.  Any solution with non-trivial scalar field content will only be invariant under the single Killing vector field 
  $\kappa$  given by (\ref{eq:KV}).

The equations of motion yield a  system of five coupled second
order ODEs.  These are rather cumbersome to write down and so we have relegated them to   appendix  \ref{FEs}.

\section{Boundary Conditions and Physical Charges}

\label{BosonConditions}

In this section, we write down the boundary conditions that boson stars in EGB gravity must satisfy; we shall employ the same boundary conditions as in the Einstein case \cite{Stotyn:2013yka}, both 
at the origin and asymptotically. 
 
\subsection{Boundary Conditions at the Origin}

Since a boson star geometry must be smooth and horizonless, all
metric functions must be regular at the origin. Furthermore, 
surfaces of constant $t$ in the vicinity of the origin are described
by round $n$-spheres (with $r$  the proper radial distance).
Multiplying (\ref{eq:PiEq})
by $r^{2}$, we note that $\Pi$ must vanish at the origin in order
to yield consistent equations of motion. 
 Thus, the boundary conditions at the centre of the boson star take the form 
\begin{eqnarray}
\left.f\right|_{r\rightarrow0}=1+\mathcal{O}(r^{2}),\quad\left.g\right|_{r\rightarrow0}=g(0)+\mathcal{O}(r),\quad\left.h\right|_{r\rightarrow0}=1+\mathcal{O}(r^{2}),\label{eq:OriginBC}\\
\left.\Omega\right|_{r\rightarrow0}=\Omega(0)+\mathcal{O}(r),\quad\left.\Pi\right|_{r\rightarrow0}=q_{0}\frac{r}{\ell}+{\cal O}(r^{2}),\quad\quad\quad\quad\nonumber 
\end{eqnarray}
for all $n$, where $q_{0}$ is a dimensionless parameter
such that the energy density of the scalar field, $T^{00}$, at the  
origin (hereafter called the central energy density) is proportional to $q_{0}^{2}$. In fact for a given value of $\alpha$, $q_{0}$ uniquely parameterizes the one-parameter family of boson star solutions in
each dimension.  Formally, it is defined by $q_0\equiv\ell\Pi'(0)$.

\subsection{Asymptotic Boundary Conditions}

In order to simplify the asymptotic boundary conditions, we first
make note of a residual gauge freedom. It is straightforward to show
that the transformation 
\begin{eqnarray}
\chi=\tilde{\chi}+\lambda t,\quad\quad\Omega(r)=\tilde{\Omega}(r)+\lambda,\quad\quad\omega=\tilde{\omega}+\lambda\label{eq:PsiGaugeFreedom}
\end{eqnarray}
for some arbitrary constant $\lambda$, leaves both the metric (\ref{eq:metric})
and scalar field (\ref{eq:ScalarField}) unchanged. We find it convenient
in our numerical analysis to set $\lambda=\omega$ so that we can
set $\tilde{\omega}=0$: in this frame, the coordinates are rigidly
rotating asymptotically so that $\tilde{\Omega}(r)\rightarrow-\omega$
as $r\rightarrow\infty$. In what follows, we use $\tilde{\chi}$, $\tilde{\Omega}(r )$, and ${\omega}$ but we drop the tildes for notational convenience.

As $r\rightarrow\infty$, the boson stars asymptote to AdS with corrections for mass and angular
momentum. This determines the metric functions up to constants $C_{f},~C_{h},~\mathrm{and}~C_{\Omega}$.
The boundary condition for the scalar field is set by requiring $\Pi$ to be normalizable.
Explicitly we obtain
\begin{align}
\left.f\right|_{r\rightarrow\infty}={} & \frac{r^{2}}{\ell^{2}}+1+\frac{C_{f}\ell^{n-1}}{r^{n-1}}+\mathcal{O}(r^{-n}),\quad\>\>\left.g\right|_{r\rightarrow\infty}=1-\frac{C_{h}\ell^{n+1}}{r^{n+1}}+\mathcal{O}(r^{-(n+2)}),\nonumber \\
\left.h\right|_{r\rightarrow\infty}={} & 1+\frac{C_{h}\ell^{n+1}}{r^{n+1}}+\mathcal{O}(r^{-(n+2)}),\quad\quad\left.\Omega\right|_{r\rightarrow\infty}=-\omega+\frac{C_{\Omega}\ell^{n}}{r^{n+1}}+\mathcal{O}(r^{-(n+2)}),\label{eq:AsymptoticBC}\\
\left.\Pi\right|_{r\rightarrow\infty}={} & \frac{\epsilon\ell^{n+1}}{r^{n+1}}+\mathcal{O}(r^{-(n+2)}),\nonumber 
\end{align}
where $\epsilon$ is a dimensionless measure
of the amplitude of the scalar field at infinity.   The constants $C_f, C_h,$ and $C_\Omega$ are  determined from solving the equations; note that the leading order corrections to $g(r)$ and $h(r)$ are not independent.  In the next section we shall solve the equations perturbatively, and will find (as in the Einstein case   \cite{Stotyn:2011ns})  that $\epsilon$ uniquely parameterizes the boson star
solutions.  Non-perturbatively the situation is different, as we shall explore in section \ref{Results}.

At this point, it is important to note that we are using $\ell$ to define the asymptotic behaviour of $f(r)$, and as long as $\ell^2>0$, the space-time will be asymptotically AdS.  By defining the AdS length in this way, the relation (\ref{Lamalpha}) for the cosmological constant  follows from the field equations.  This convention differs from another common convention \cite{Crisostomo:2000bb,Brihaye:2011}, which sets
\begin{equation}
\Lambda=\frac{n\left(n+1\right)}{2L^2}
\end{equation}
(where $L$ is the would-be AdS length in the absence of the Gauss-Bonnet term) and leads to an upper bound on $\alpha$
\begin{equation}
\alpha<\frac{L^2}{4\left(n-1\right)\left(n-2\right)}
\end{equation}
 in order to satisfy AdS boundary conditions \cite{Brihaye:2011}.
In terms of the convention used here, this inequality can be expressed as
\begin{equation}
\big(2\left(n-1\right)\left(n-2\right)\alpha - \ell^2\big)^2>0,
\end{equation}
which immediately implies
\begin{equation}
\alpha \neq \frac{\ell^2}{2\left(n-1\right)\left(n-2\right)} \equiv \alpha_{\text{cr}}.
\end{equation}
Note that this gives the  critical value of $\alpha$ for the Gauss-Bonnet coupling  in any dimension \cite{Crisostomo:2000bb}, which also appears in the perturbative boson star solutions as will be seen shortly.

\subsection{Physical Charges}

The SKV boson stars are invariant under the single
Killing field (\ref{eq:KV}). However since the scalar fields vanish at infinity with sufficient fall-off, both
 $\partial_{t}$ and $\partial_{\chi}$ are each asymptotic Killing fields (the metric alone being invariant
under them for the full solution).  Consequently conserved charges can be defined where $\partial_{t}$ and $\partial_{\chi}$ are readily associated with a conserved energy and angular momentum respectively. We use the definitions proposed in reference \cite{Brihaye:2008mj}, given by
\begin{align}\label{eq:EJ}
M & =\frac{(n+1)\pi^{\frac{n-1}{2}}\ell^{n-1}}{16\left(\frac{n+1}{2}\right)!}\left(1-\frac{\alpha}{\alpha_{\text{cr}}}\right)\big((n+1)C_{h}-nC_{f}\big),\\
J & =\frac{(n+1)^{2}\pi^{\frac{n-1}{2}}\ell^{n}}{16\left(\frac{n+1}{2}\right)!}\left(1-\frac{\alpha}{\alpha_{\text{cr}}}\right)C_{\Omega}
\end{align}
where $C_{f},~C_{h},~\mathrm{and}~C_{\Omega}$ are the constants appearing  in the asymptotic boundary conditions (\ref{eq:AsymptoticBC}).  This prescription emerges from a semiclassical treatment whereby the actions are calculated using the boundary counterterms found in \cite{Brihaye:2008mj}.  The above definitions ensure that the first law of thermodynamics is satisfied for black holes in EGB theory, at least in the spherically symmetric case where an explicit solution is known.

The existence of these asymptotic Killing symmetries also guarantees that the boson stars satisfy the first law of thermodynamics.  They have vanishing temperature and entropy, so the first law takes the form
\begin{equation}\label{eq:firstlaw}
dM=\omega dJ.
\end{equation}
This relation provides a useful and  important numerical tool: by requiring the solutions to respect the first law (\ref{eq:firstlaw}) to at least one part in $10^{6}$, we obtain a primary cross-check on the validity of the numerical methods used.

\section{Perturbative Solutions}
\label{pertsolns}

In this section, we consider perturbative solutions to the Einstein-Gauss-Bonnet equations that satisfy the  boundary conditions of section \ref{BosonConditions}.  These solutions  are horizonless and are  constructed as perturbations around AdS; they generalize those found previously for Einstein gravity in \cite{Stotyn:2011ns}.  We take the scalar field condensate parameter, $\epsilon$, as the perturbative expansion parameter and we give results up to order $\epsilon^6$ for the space-time dimensions of interest in string theory, namely $D=5,7,9,11$.  As a perturbative construction, these results will only be valid for small energies and angular momenta. They will provide another useful check on the numeric solutions we present later in the paper.

\subsection{Perturbative Boson Star}\label{PertBS}

Since we require global AdS to be a solution when the scalar field amplitude vanishes, we employ the expansion
\begin{equation}
F(r,\epsilon)=\sum_{i=0}^m{\tilde{F}_{2i}(r)\epsilon^{2i}}\quad\quad\quad \Pi(r,\epsilon)=\sum_{i=0}^m{\tilde{\Pi}_{2i+1}(r)\epsilon^{2i+1}}\quad\quad\quad \omega(\epsilon)=\sum_{i=0}^m{\tilde{\omega}_{2i}\epsilon^{2i}} \label{eq:FieldExpansion}
\end{equation}
where $F\in\{f,g,h,\Omega\}$ is shorthand for each of the metric functions in (\ref{eq:metric}).  Note that the scalar fields are expanded in odd powers of $\epsilon$ whereas the metric functions are expanded in even powers.  The perturbative solution to the equations is thus obtained 
by starting with global AdS -- the 0th terms in the metric function expansions above.

The scalar equation (\ref{eq:PiEq}) is then solved in this background, yielding  $\tilde{\Pi}_1(r)$.  The full set of equations of motion (\ref{CC}--\ref{RR}) are then satisfied up to order $\epsilon$.  The stress-energy for these scalars is then of order $\epsilon^2$, which induces source corrections to the gravitational fields $\tilde{F}_2(r)$ (the $i=1$ terms in the metric functions).  These in turn back-react on the scalar fields, and the next-order solution to (\ref{eq:PiEq}) yields
 $\tilde{\Pi}_3(r)$, thereby satisfying the equations of motion  up to order $\epsilon^3$.  
 
This bootstrapping procedure can be carried out to arbitrary order in $\epsilon$.   The angular frequency must also be expanded in even powers of $\epsilon$  because the scalar fields back react on the metric, inducing nontrivial frame-dragging effects, which in turn affect the rotation of the scalar fields.  These corrections to $\omega$ are found by imposing the boundary conditions.

Global AdS is given by
\begin{equation}
f_0=1+\frac{r^2}{\ell^2},\quad\quad g_0=1,\quad\quad h_0=1,\quad\quad \Omega_0=0
\end{equation}
The most general massless scalar field solution to (\ref{eq:PiEq}) in this background that  is consistent with the asymptotic boundary conditions (\ref{eq:AsymptoticBC}) is  
\begin{equation}
\Pi_1(r)=\frac{r\ell^{n+1} }{(r^2+\ell^2)^{\frac{n+2}{2}}}{_2F_1}\left[\frac{n+2-\omega \ell}{2},\frac{n+2+\omega\ell}{2};\frac{n+3}{2};\frac{\ell^2}{r^2+\ell^2}\right]\label{eq:Pi1}
\end{equation}
where $_2F_1$ is the hypergeometric function. Note that the Gauss-Bonnet parameter $\alpha$ does not enter at this order. In order to satisfy the boundary conditions at the origin (\ref{eq:OriginBC}) we must require
\begin{equation}
\omega\ell=n+2+2k,\quad\quad\quad\quad k=0,1,2,...
\end{equation}
where the non-negative integer, $k$, describes the various possible radial modes of the scalar field.  While any radial profile for the scalars is possible, it must be built up from a linear combination of the radial modes that necessitate the inclusion of 
 multiple frequency parameters, $\omega_k$.    This is inconsistent with the existence of the Killing vector field (\ref{eq:KV}).  These higher
 modes represent excited states, whereas the $k=0$ mode is the ground state.  In what follows, we only consider the ground state, in which case (\ref{eq:Pi1}) simplifies to
\begin{equation}
\Pi_1(r)=\frac{r\ell^{n+1} }{(r^2+\ell^2)^{\frac{n+2}{2}}}.   \label{eq:Piepsilon}
\end{equation}

 Next, we insert (\ref{eq:Piepsilon}) and the expansion (\ref{eq:FieldExpansion}) into the equations of motion (\ref{CC}--\ref{RR}), expand in $\epsilon$ and solve at order $\epsilon^2$.  Using the boundary conditions to fix the  two constants of integration that emerge, the order $\epsilon^2$ corrections to the metric functions, $\tilde{F}_2(r)$, are then inserted into the equation of motion for $\Pi(r)$ to find its order $\epsilon^3$ correction. This procedure, carried out up to but not including order $\epsilon^6$, yields
\begin{equation}
f(r)=1+\frac{r^2}{\ell^2}-\frac{r^2 \ell^{n+1}f_{n;2}}{(r^2+ \ell^2)^{n+1}\big(\ell^2-2(n-1)(n-2)\alpha\big)}\epsilon^2-\frac{r^2 \ell^{n+3}f_{n;4}}{(r^2+ \ell^2)^{2n+3}\big(\ell^2-2(n-1)(n-2)\alpha\big)^3}\epsilon^4+{\cal O}(\epsilon^6)\label{eq:fExpansion}
\end{equation}
\begin{equation}
g(r)=1-\frac{2\ell^{2n+4}\big((n+1)r^2+\ell^2\big)}{n(r^2+\ell^2)^{n+2}\big(\ell^2-2(n-1)(n-2)\alpha\big)}\epsilon^2-\frac{\ell^{n+5}g_{n;4}}{(r^2+\ell^2)^{2n+4}\big(\ell^2-2(n-1)(n-2)\alpha\big)^3}\epsilon^4+{\cal O}(\epsilon^6)
\end{equation}
\begin{equation}
h(r)=1+\frac{r^2\ell^{n+5}h_{n;4}}{(r^2+\ell^2)^{2n+3}\big(\ell^2-2(n-1)(n-2)\alpha\big)^3}\epsilon^4+{\cal O}(\epsilon^6)
\end{equation}
\begin{equation}
\Omega(r)=\frac{\ell^{n+2}\Omega_{n;2}}{(r^2+\ell^2)^{n+1}\big(\ell^2-2(n-1)(n-2)\alpha\big)}\epsilon^2+\frac{\ell^{n+4}\Omega_{n;4}}{(r^2+\ell^2)^{2n+3}\big(\ell^2-2(n-1)(n-2)\alpha\big)^3}\epsilon^4+{\cal O}(\epsilon^6)
\end{equation}
\begin{equation}
\begin{split}
\Pi(r)=\frac{r\ell^{n+1} }{(r^2+\ell^2)^{\frac{n+2}{2}}}\epsilon+\frac{r\ell^{n+5}\Pi_{n;3}}{(r^2+\ell^2)^{\frac{3n+4}{2}}\big(\ell^2-2(n-1)(n-2)\alpha\big)}\epsilon^3&\\
+\frac{r\ell^{n+7}\Pi_{n;5}}{(r^2+\ell^2)^{\frac{5(n+2)}{2}}\big(\ell^2-2(n-1)(n-2)\alpha\big)^3}\epsilon^5+{\cal O}(\epsilon^7)\label{eq:PiExpansion}
\end{split}
\end{equation}
where the fields $\{f_{n;s},g_{n;s},h_{n;s},\Omega_{n;s},\Pi_{n;s}\}$ are simple polynomials in $r$; in this notation $s$ labels the order in $\epsilon$ and $n=D-2$ labels the space-time dimension\footnote{The notation is the same as that of ref \cite{Stotyn:2011ns}, except that
an additional ``$0$'' index (relevant for perturbative black hole solutions) is dropped.}.   These fields are catalogued in Appendix 
\ref{pertanswers} for $n=3,5,7,9$ up to order $\epsilon^6$, with the associated energies and angular momenta in Appendix \ref{conscharges}.  It is straightforward to check that these latter quantities obey the first law (\ref{eq:firstlaw}) to order $\epsilon^6$.

The perturbative solutions all reduce to those of the Einstein case \cite{Stotyn:2011ns} in the limit $\alpha\to 0$.  Perturbative boson star solutions obeying the boundary conditions (\ref{eq:AsymptoticBC}) do not exist for $\alpha=\alpha_{\text{cr}}$ in any dimension.

\section{Numerical Construction}

\label{numconst}

We numerically construct boson star solutions by approximating the metric and scalar field functions $\{f(r),g(r),h(r),\Omega(r),\Pi(r)\}$ as Chebyshev polynomials and using a relaxation method on a Chebyshev grid. A detailed review of this procedure can be found in \cite{Pfeiffer:2005zm}.

 Before the Chebyshev grid can be constructed, we must first compactify the infinite domain $r\in[0,\infty)$ to the finite domain $y\in[0,1]$, which is done via the simple coordinate transformation
\begin{equation}
y=\frac{r^{2}}{r^{2}+\ell^{2}}.
\end{equation}
The grid then consists of the set of points $y_k= (\cos(k\pi/N) + 1)/2, k=0,...,N$, which are the $N+1$ extrema of the $N^{th}$ order Chebyshev polynomial, $T_N(2y-1)$.

Next, we define auxiliary functions $\{q_{f},q_{g},q_{h},q_{\Omega},q_{\Pi}\}$, which are analytic over the entire domain $y\in[0,1]$, by taking into account the boundary conditions (\ref{eq:OriginBC}) and (\ref{eq:AsymptoticBC}).
\begin{align}
 & f(y)=\frac{y}{1-y}+1+(1-y)^{\frac{n-1}{2}}q_{f}(y), \label{fq}\\
 & g(y)=1+(1-y)^{\frac{n+1}{2}}q_{g}(y),\label{gq}\\
 & h(y)=1+(1-y)^{\frac{n+1}{2}}q_{h}(y),\label{hq}\\
 & \Omega(y)=\frac{q_{\Omega}(y)}{\ell},\label{Oq}\\
 & \Pi(y)=\sqrt{y}(1-y)^{\frac{n+1}{2}}q_{\Pi}(y) . \label{Pq}
\end{align}
The remaining discussion will use the coordinate $y$, with a prime $'$ denoting a derivative with respect to $y$.

The boundary conditions for the $q$-functions are found by first converting the field equations (\ref{RR})--(\ref{TT}) into functions of $y$, then Taylor expanding them around the two end points $y_0=0,1$ and requiring that they vanish order by order.  The lowest order terms give non-trivial relationships between the $q$-functions and their first derivatives.
At $y_0=1$ we obtain
\begin{align} \label{eq:BC1}
&q_{f}'(1)+\frac{n-1}{2}q_{f}(1)+q_{h}(1)=0,\qquad\qquad q_{g}(1)+q_{h}(1)=0, \qquad\qquad q_{\Omega}'(1)=0,\\
&q_{h}'(1)+\frac{n+1}{2}q_{h}(1)=0, \qquad\qquad q_{\Pi}'(1)+\frac{1}{2(n+3)}\big((n+2)^{2}-q_{\Omega}(1)^{2}\big)q_{\Pi}(1)=0. \label{eq:BC2}
\end{align}
At $y_0=0$ we find
\begin{align} \label{eq:BC0}
& q_{f}(0)=0,\qquad q_{h}(0)=0,\qquad    q_{\Pi}(0)-q_{0}=0, \\
&\frac{2 \ell^2 q_0^2 q_{\Omega }(0)}{(n+3) \left(2 \alpha  (n-1) \left((n-2)
   \left(q_f'(0)+1\right)+3 q_h'(0)\right)-\ell^2\right)}+q_{\Omega }'(0)=0,\\
&q_g'(0)+\frac{1}{2 \alpha  (n-2) (n-1) \left(n \left(q_f'(0)+1\right)+3 q_h'(0)\right)-\ell^2 n}& \nonumber\\
&\times\Bigg(q_f' (0)\left(\alpha  (n-2) (n-1) \left(3 (n-1) \left(q_g(0)+1\right) q_h'(0)+n
   (n+1)\right)-\frac{1}{2} \ell^2 n (n+1) \left(q_g(0)+1\right)\right) \nonumber\\
&+\frac{1}{2} \alpha 
   (n-2) (n-1) n (n+1) \left(q_g(0)+1\right) \left(q_f'(0)\right)^2 - \ell^2 (n-1) q_0^2 \left(q_g(0)+1\right) \nonumber\\
&+(n-1) q_h'(0)
   \left(\alpha  (n-2) \left(3 (n-1)-6 q_g(0)\right)-\frac{3}{2} \ell^2
   \left(q_g(0)+1\right)\right)  \nonumber\\
&+\frac{15}{2} \alpha  (n-3) (n-1) \left(q_g(0)+1\right)
   \left(q_h'(0)\right)^2+n (n+1) q_g(0) \left(\frac{\ell^2}{2}-\alpha  (n-2)
   (n-1)\right) \Bigg)=0, 	
 \end{align}
where   $ q_f'(0)$ is given by
\begin{align}\label{eq:qfprime}
q_f'(0)={}&-\frac{1}{2 \alpha  (n-2) (n-1) n}\bigg[\Big(-24 \alpha ^2 (n-2) (n-1)^2 (n+3) q_h'(0)^2+\ell^4 n^2-4 \alpha  \ell^2 (n-2) (n-1) n^2\nonumber\\&
+8 \alpha  \ell^2 (n-2) (n-1) n q_0^2+4 \alpha ^2 (n-2)^2 (n-1)^2 n^2\Big)^{1/2}\nonumber\\&
+6 \alpha  (n-2) (n-1) q_h'(0)+\ell^2 (-n)+2
   \alpha  (n-2) (n-1) n\bigg].
 \end{align}

We generate solutions to the equations of motion by approximating each of the $q$-functions as an order $N$ Chebyshev expansion, using the method described in \cite{Pfeiffer:2005zm}.  These expansions are substituted into the equations of motion (\ref{CC})--(\ref{RR}) at the $N-1$ interior grid points and into the boundary conditions  (\ref{eq:BC0}) and (\ref{eq:BC1}) at the 2 boundary points.  In so doing, the numerical integration is reduced to a set of $5(N+1)$ algebraic equations in the spectral coefficients of the Chebyshev expansions.  These equations are then linearized with respect to each spectral coefficient and a standard Newton-Raphson method is used to solve the resulting system of linear equations, with convergence defined as a change in the spectral coefficients less than $10^{-30}$ between iterations.

A one parameter family of solutions, parameterized by $q_0$, is built up using a step size of $\Delta q_0=0.01$; along this chain of discrete solutions, the previous numeric solution is used as the seed for the next one. The initial seed we use to start the procedure has $q_0=0.01$ and is given by the perturbative solution (\ref{eq:fExpansion})--(\ref{eq:PiExpansion}), where the numeric parameter $q_0$ is related to the pertubative parameter $\epsilon$ through the equality
\begin{equation}
q_0=\lim_{r \to 0}\frac{\left(r^2+\ell^2\right)^{\frac{n+2}{2}}}{r\ell^{n+1}}{\Pi}(r)\nonumber,
\end{equation}
and $\Pi(r)$ is taken to be the perturbative solution of equation (\ref{eq:PiExpansion}).
 
 To ensure the numeric solutions are physical, we demand that they obey the first law of thermodynamics (\ref{eq:firstlaw}) to one part in $10^{6}$.  When this limit is no longer satisfied, this is taken as an indication that the Chebyshev grid is too coarse; at this juncture we increase $N$, refining the grid, and  continue the procedure.  This provides a useful cross-check on the validity of the solutions, but not the only one.
  Due to the exponential convergence of the Chebyshev approximation (for example, see Ref. \cite{Pfeiffer:2005zm}),  the truncation error in the approximations of the $q$-functions can be easily estimated via  
    \begin{equation}
|\text{error}|\le\sum_{j=N+1}^{\infty}C_j\sim \int_{N}^{\infty}{C_0 e^{-kj}dj} = \frac{C_0 e^{-kN}}{k}
\end{equation}
for some real number $k$, characteristic of the Chebyshev expansion.  Here we have used the property that  $-1\le T_j(x) \le 1$ and that $C_j \approx C_0e^{-kj}$.  Using the $\mathrm{\%~error}\approx \frac{C_N}{C_0}$ to quantify the truncation error, we impose that it must be less than $10^{-7}$ for all solutions.

The numerical integration method requires that the equations of motion are satisfied only on the grid points.  This does not demand that they are solved everywhere.  We find that, in general, the largest interpolation error occurrs at the midpoint between  grid points, \emph{i.e.} the zeroes of the $N^{\mathrm {th}}$ Chebyshev polynomial.  As long as the grid is dense enough to satisfy the constraint on the first law, this error is less than $10^{-5}$.  

Finally, we note that the critical value of $\alpha$ imposes constraints on the efficiency of our numerical work: in seeking to obtain solutions for $\alpha$ less than but close to $\alpha_{\text{cr}}$, we find that significantly larger resolution is required, necessitating a corresponding increase in the number of grid points and thus the computational processing time;  above a certain value of $\alpha$ this becomes intractable.  We have not been able to obtain solutions for $ 0.9 \alpha_{\text{cr}} < \alpha < \alpha_{\text{cr}}$ in any dimension.

\section{Results}
\label{Results}

We find it convenient to write the expressions for the physical quantities of the boson star in terms of the boundary values of the $q$-functions and their derivatives.  For example, the thermodynamic quantities are given by \cite{Brihaye:2008mj}
\begin{align}
M={}&\frac{\pi^{\frac{n-1}{2}}\ell^{n-1}}{8\left(\frac{n-1}{2}\right)!}\left(1-\frac{\alpha}{\alpha_{\text{cr}}}\right)\big((n+1)q_h(1)-nq_f(1)\big),\\
J={}&\frac{(-1)^{\frac{n+1}{2}}\pi^{\frac{n-1}{2}}\ell^n}{4\left(\left(\frac{n-1}{2}\right)!\right)^2}\left(1-\frac{\alpha}{\alpha_{\text{cr}}}\right)q_\Omega^{(\frac{n+1}{2})}(1),\\
\omega={}&-q_\Omega(1)
\end{align}
which can be easily calculated using (\ref{eq:EJ}) and (\ref{fq} -- \ref{Pq}).  
 The Kretschmann scalar $K=R_{abcd}R^{abcd}$ at the center of the boson star is given by
\begin{equation}
\begin{split}
K_{n}={}&\frac{2 (n+1)}{\ell^4} \Bigg(\frac{2 q_f'(0) \left(2 q_g'(0)+3 \left(q_h(0)+1\right) q_h'(0)+q_h(0)+n+2\right)}{q_h(0)+1}+(n+2) q_f'(0)^2
\\
&-\frac{2 q_g'(0) \left((n-1) q_h(0)-2\right)}{\left(q_h(0)+1\right)^2}+\frac{2
   q_g'(0)^2}{\left(q_h(0)+1\right)^2}+3 (n+2) q_h'(0)^2
\\
&+\frac{\left(\frac{1}{2} (n-1) (n+1)+1\right) q_h(0)^2}{\left(q_h(0)+1\right)^2}+\frac{n+2}{\left(q_h(0)+1\right)^2}+6 q_h'(0)+\frac{2 q_h(0)}{\left(q_h(0)+1\right)^2}\Bigg),
\end{split}
\end{equation}
which further simplifies using the boundary conditions (\ref{eq:BC0}), yielding an expression in terms of just $q_h'(0)$. 
 The resulting expressions are rather cumbersome, and so we have placed them appendix \ref{Kscalar}.

\begin{figure}[ht!]
	\begin{center}
		\subfigure[~$M$ vs $\epsilon$ for $D=5$ ($n=3$) for various $\tilde{\alpha}$;  except for $\tilde\alpha=0$, the curves all terminate at their endpoints, at which the Kretschmann scalar diverges.  The dashed lines correspond to the perturbative solutions.]{%
			\label{fig:5DEve}
			\includegraphics[width=0.48\textwidth]{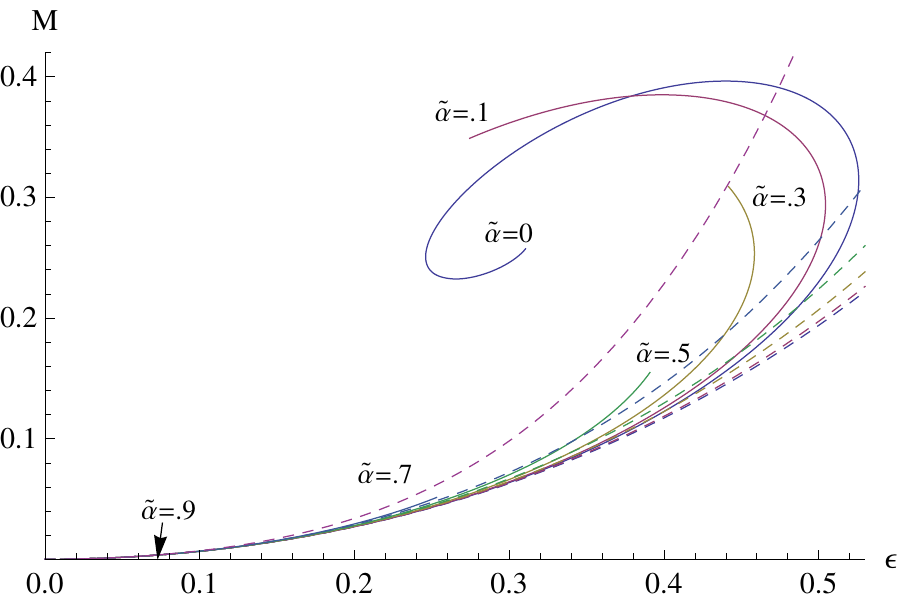}
		}%
		\subfigure[~$M$ vs $\epsilon$ for $D=7$ ($n=5$) for various $\tilde{\alpha}$. Dashed lines correspond to perturbative solutions. ]{%
			\label{fig:7DEve}
			\includegraphics[width=0.48\textwidth]{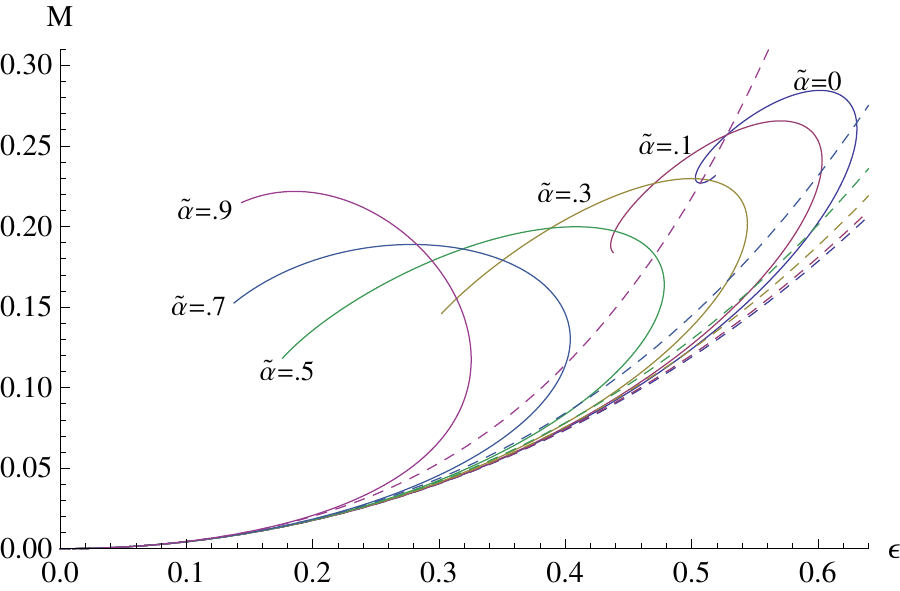}
		}\\ 
		\subfigure[~$M$ vs $\epsilon$ for $D=9$ ($n=7$) for various $\tilde{\alpha}$. Dashed lines correspond to perturbative solutions.]{%
			\label{fig:9DEve}
			\includegraphics[width=0.48\textwidth]{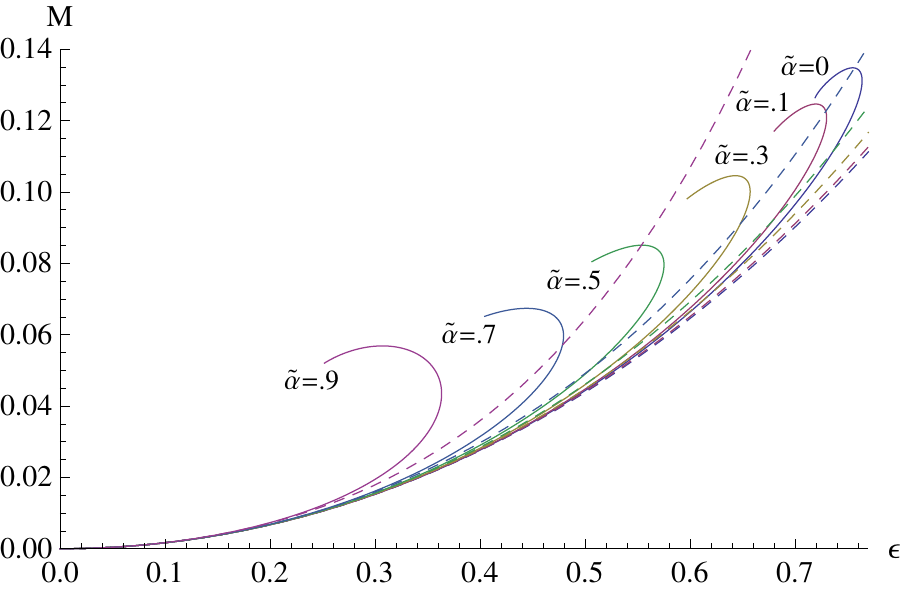}
		}%
        \subfigure[~$M$ vs $\epsilon$ for $D=11$ ($n=9$) for various $\tilde{\alpha}$. Dashed lines correspond to perturbative solutions.]{%
		\label{fig:11DEve}
		\includegraphics[width=0.48\textwidth]{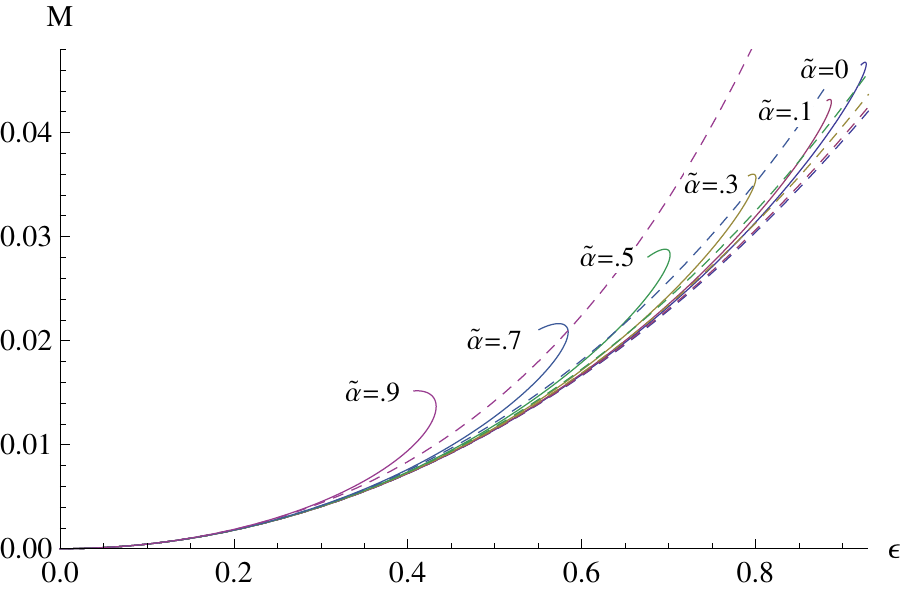}
	}%
	\end{center}
	\caption{%
		The boson star mass plotted against the perturbative parameter $\epsilon$ in (a) 5, (b) 7, (c) 9, and (d) 11 dimensions for various
values of $\tilde\alpha$.  In the non-perturbative regime, $\epsilon$ no longer uniquely parameterizes the boson star family.   The spirals tighten with increasing dimension and the truncated behaviour for $D=5$ is associated with diverging curvature at finite $q_0$.
	}%
	\label{fig:Eve}
\end{figure}

\begin{figure}[ht!]
	\begin{center}
		\subfigure[~$J$ vs $\epsilon$ for $D=5$ ($n=3$) for various $\tilde{\alpha}$;  except for $\tilde\alpha=0$, the curves all terminate at their endpoints, at which the Kretschmann scalar diverges.  The dashed lines correspond to the perturbative solutions.]{%
			\label{fig:5DJve}
			\includegraphics[width=0.48\textwidth]{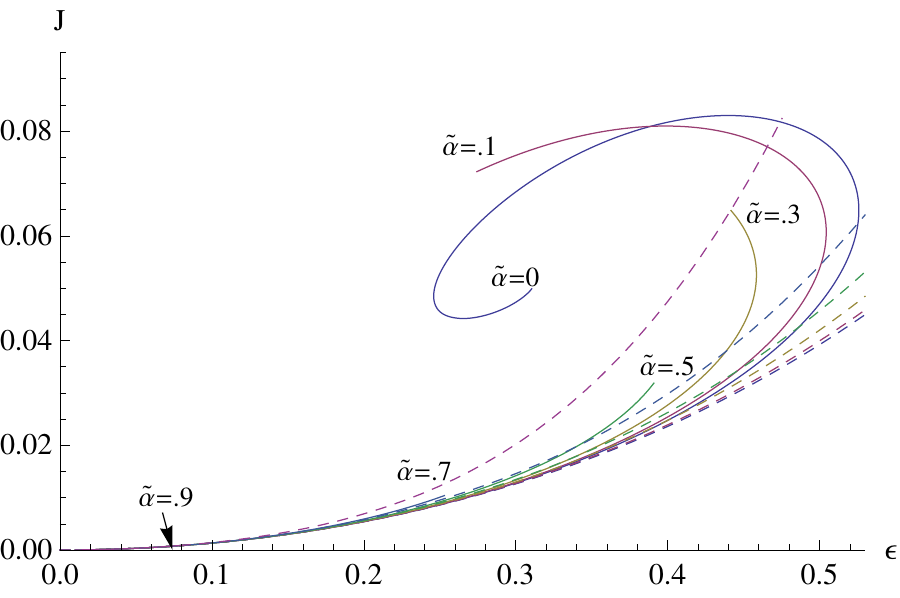}
		}%
		\subfigure[~$J$ vs $\epsilon$ for $D=7$ ($n=5$) for various $\tilde{\alpha}$. Dashed lines correspond to perturbative solutions. ]{%
			\label{fig:7DJve}
			\includegraphics[width=0.48\textwidth]{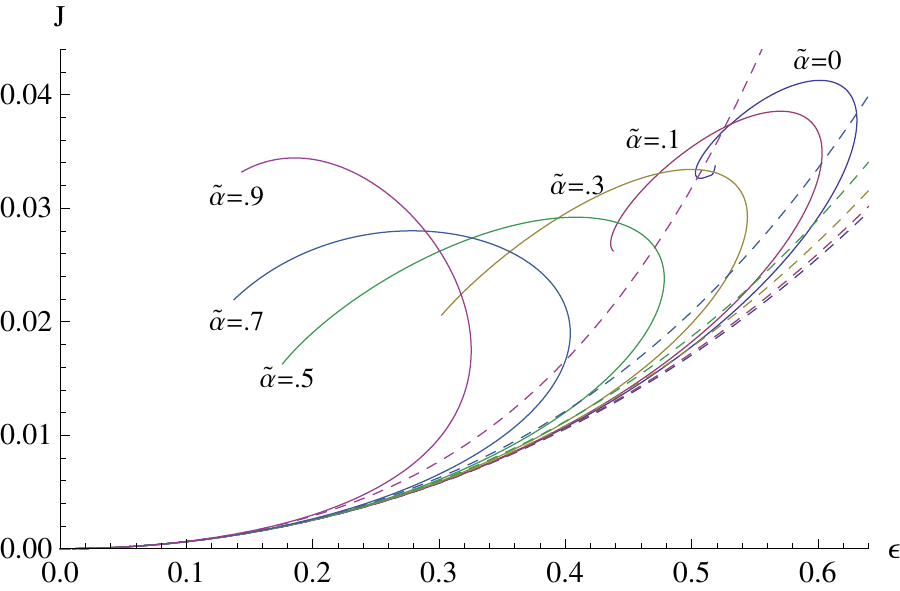}
		}\\ 
		\subfigure[~$J$ vs $\epsilon$ for $D=9$ ($n=7$)for various $\tilde{\alpha}$. Dashed lines correspond to perturbative solutions. ]{%
			\label{fig:9DJve}
			\includegraphics[width=0.48\textwidth]{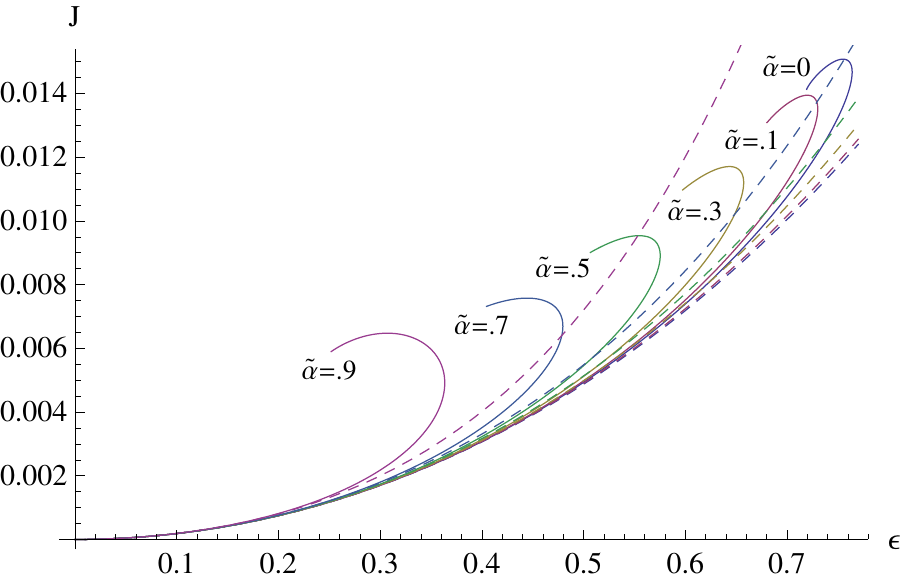}
		}%
		\subfigure[~$J$ vs $\epsilon$ for $D=11$ ($n=9$) for various $\tilde{\alpha}$. Dashed lines correspond to perturbative solutions.]{%
			\label{fig:11DJve}
			\includegraphics[width=0.48\textwidth]{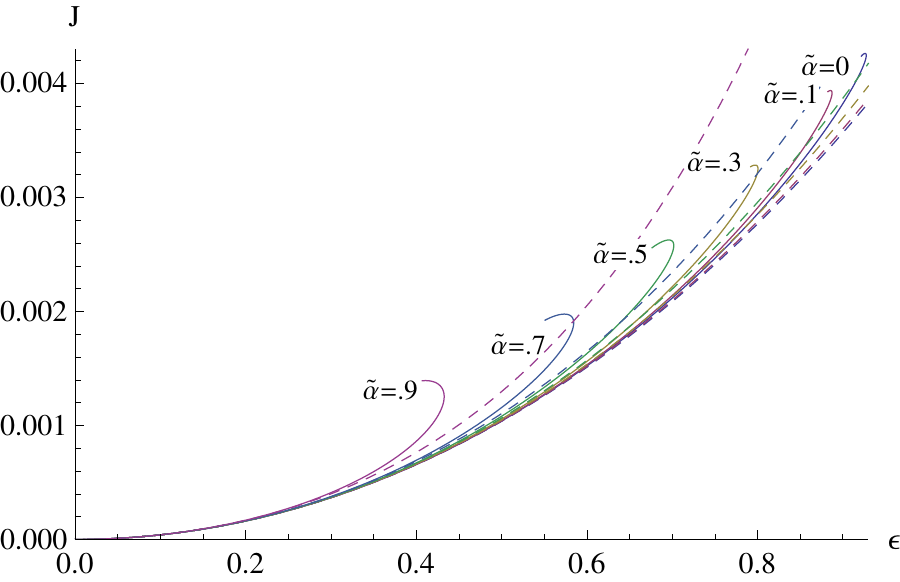}
		}%
	\end{center}
	\caption{%
		The boson star angular momentum plotted against the perturbative parameter $\epsilon$ in (a) 5, (b) 7, (c) 9, and (d) 11 dimensions for various  values of $\tilde\alpha$.  In the non-perturbative regime, $\epsilon$ no longer uniquely parameterizes the boson star family. As with the mass, the spirals tighten with increasing dimension and the truncated behaviour for $D=5$ is associated with diverging curvature at finite $q_0$.
	}%
	\label{fig:Jve}
\end{figure}

\begin{figure}[ht!]
	\begin{center}
		\subfigure[~Detail of $M$ vs $\epsilon$ for $D=5$ ($n=3$) for $\tilde{\alpha}=0.7,\ 0.9$.]{%
			\label{fig:5DEveZ}
			\includegraphics[width=0.48\textwidth]{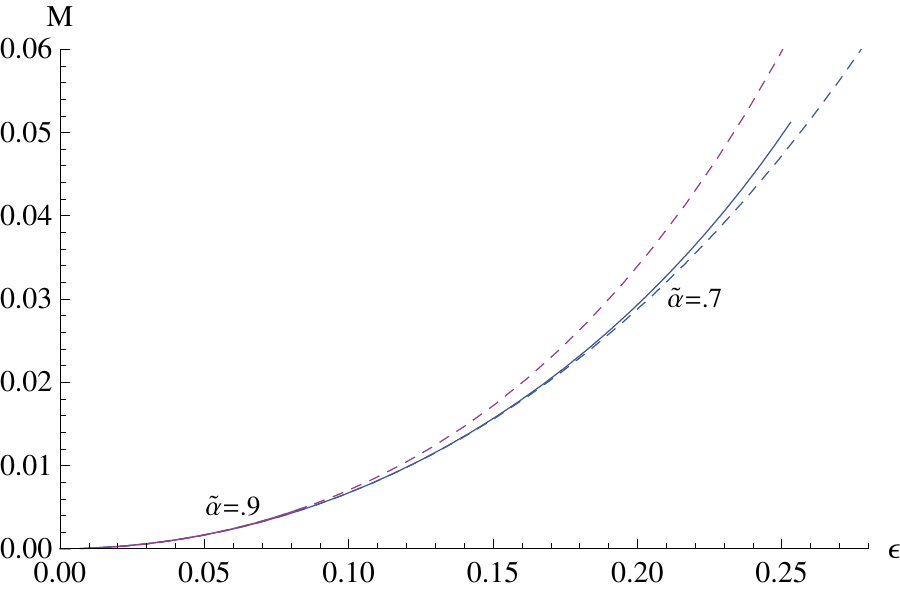}
		}%
		\subfigure[~Detail of $J$ vs $\epsilon$ for $D=5$ ($n=3$) for $\tilde{\alpha}=0.7,\ 0.9$. ]{%
			\label{fig:7DJveZ}
			\includegraphics[width=0.48\textwidth]{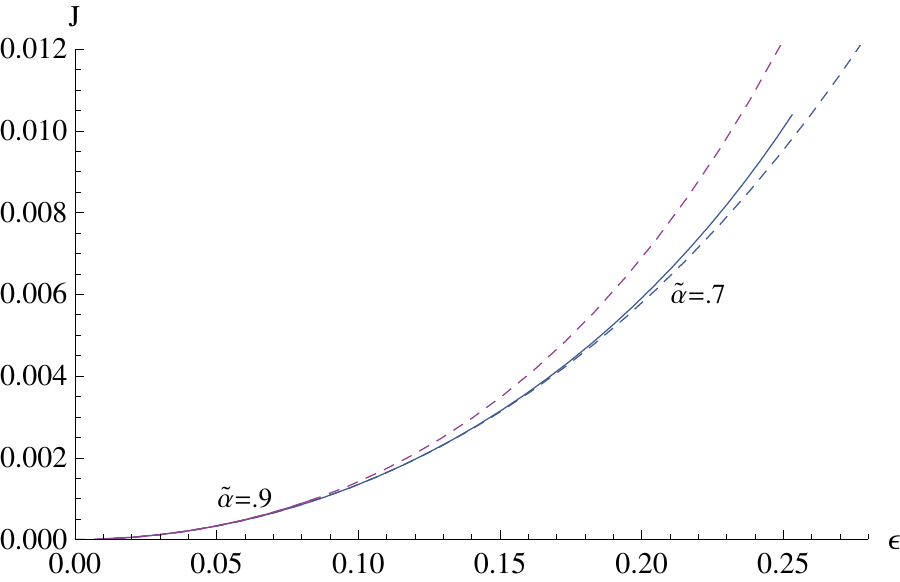}
		}%
	\end{center}
	\caption{%
		Detail plots of mass (a) and angular momentum (b) against the perturbative parameter $\epsilon$ in 5 dimensions for $\tilde{\alpha}=\frac{7}{10}$ and $\tilde{\alpha}=\frac{9}{10}$.  Perturbative results are plotted as dashed lines. The truncation, which is associated with diverging curvature, occurs at smaller values of $\epsilon$ as $\alpha$ increases.
	}%
	\label{fig:5DEJveZoom}
\end{figure}
  
\begin{figure}[ht!]
	\begin{center}
		\subfigure[~$\omega$ vs $\epsilon$ for $D=5$ ($n=3$) for various $\tilde{\alpha}$.]{%
			\label{fig:5Domve}
			\includegraphics[width=0.48\textwidth]{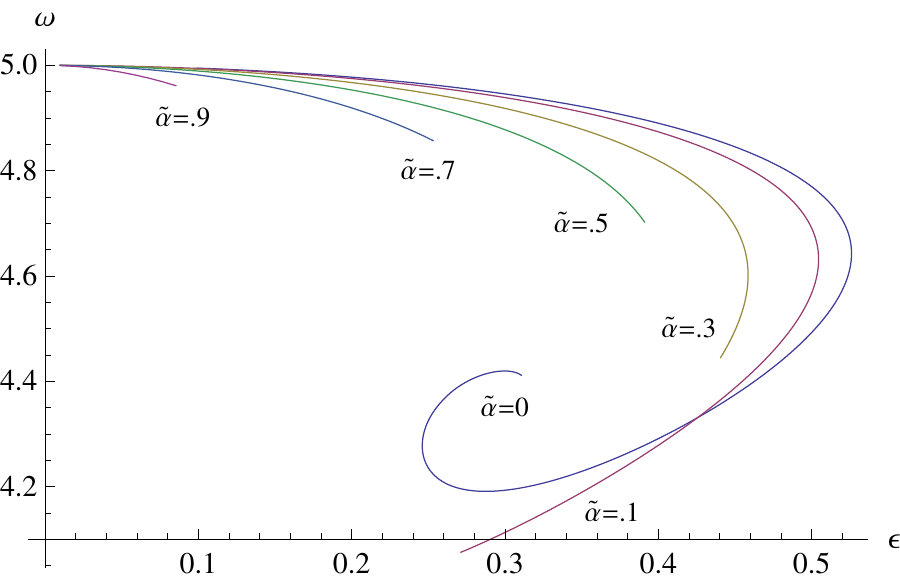}
		}%
		\subfigure[~$\omega$ vs $\epsilon$ for $D=7$ ($n=5$) for various $\tilde{\alpha}$. ]{%
			\label{fig:7Domve}
			\includegraphics[width=0.48\textwidth]{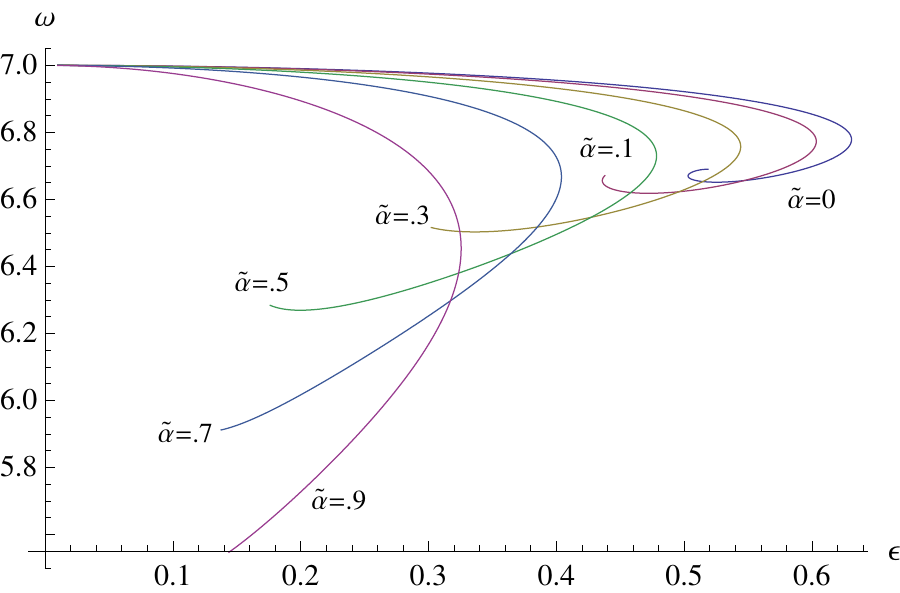}
		}\\
		\subfigure[~$\omega$ vs $\epsilon$ for $D=9$ ($n=7$) for various $\tilde{\alpha}$.]{%
			\label{fig:9Domve}
			\includegraphics[width=0.48\textwidth]{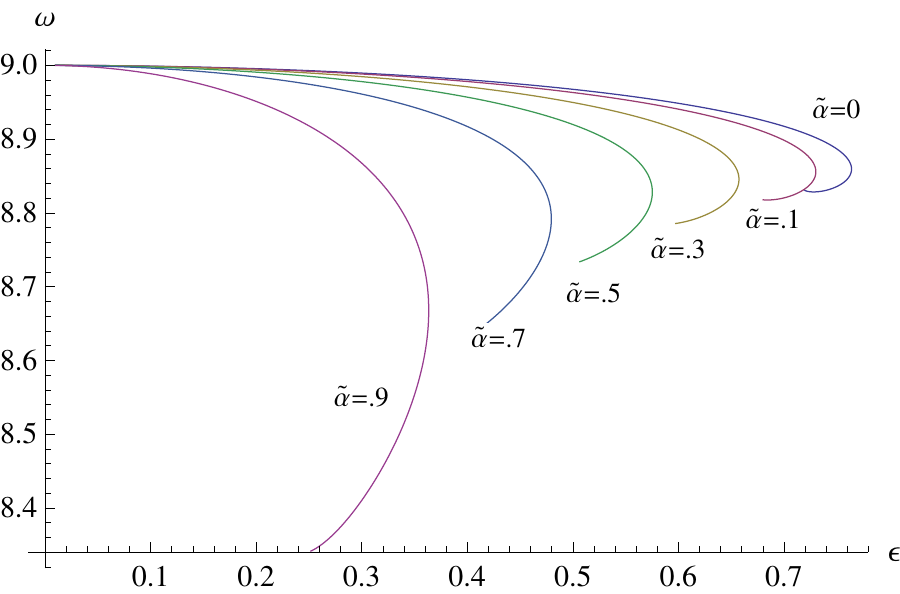}
		}%
		\subfigure[~$\omega$ vs $\epsilon$ for $D=11$ ($n=9$) for various $\tilde{\alpha}$.]{%
			\label{fig:11Domve}
			\includegraphics[width=0.48\textwidth]{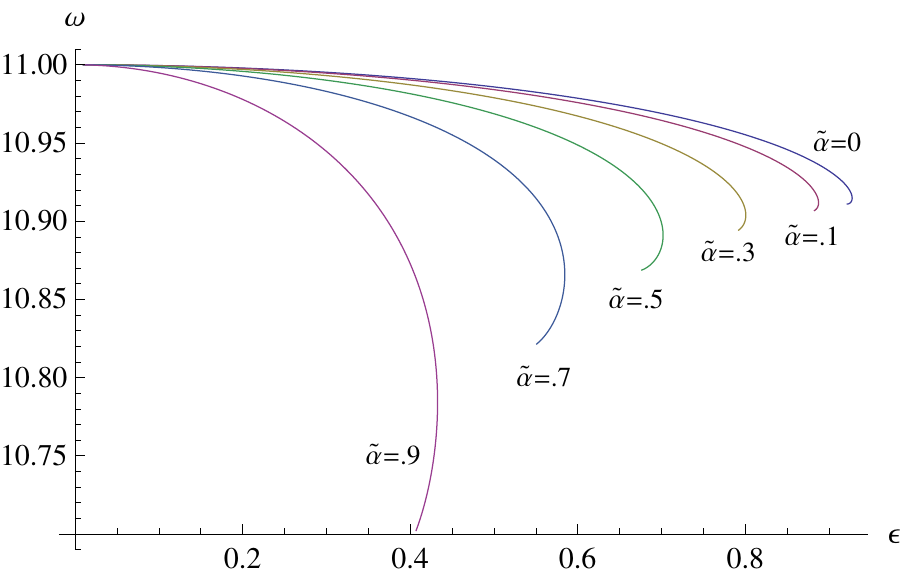}
		}%
	\end{center}
	\caption{%
		The angular frequency $\omega$ plotted against the perturbative parameter $\epsilon$ in (a) 5, (b) 7, (c) 9, and (d) 11 dimensions for various  values of $\tilde\alpha$.
	}%
	\label{fig:omve}
\end{figure}
 
\begin{figure}[ht!]
	\begin{center}
		\subfigure[~$M$ vs $q_0$ for $D=5$ ($n=3$), with $\tilde\alpha$ increasing from bottom to top; except for $\tilde\alpha=0$, the curves all terminate at their endpoints, at which the Kretschmann scalar diverges.]{%
			\label{fig:5DEvq}
			\includegraphics[width=0.48\textwidth]{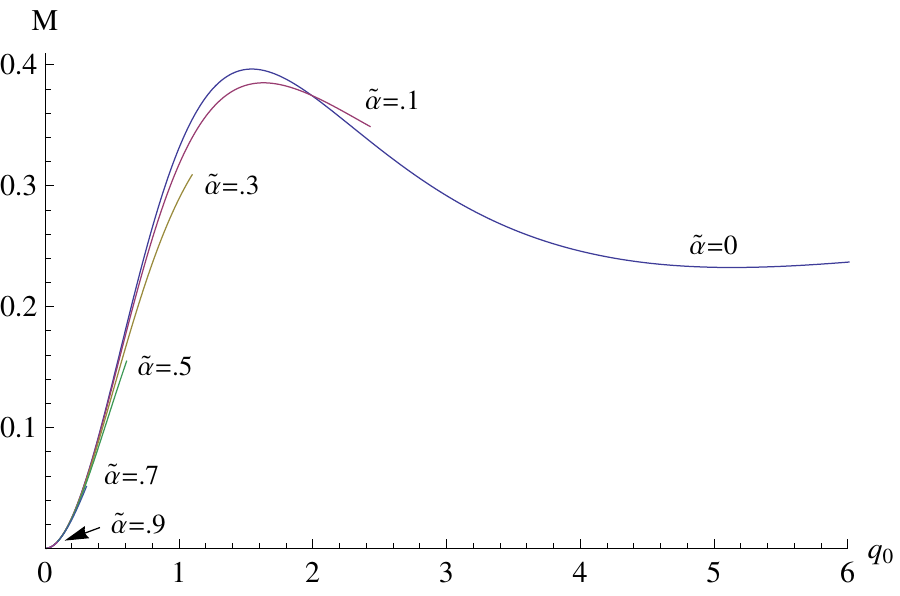}
		}%
		\subfigure[~$M$ vs $q_0$ for $D=7$ ($n=5$), with $\tilde\alpha$ increasing from bottom to top. ]{%
			\label{fig:7DEvq}
			\includegraphics[width=0.48\textwidth]{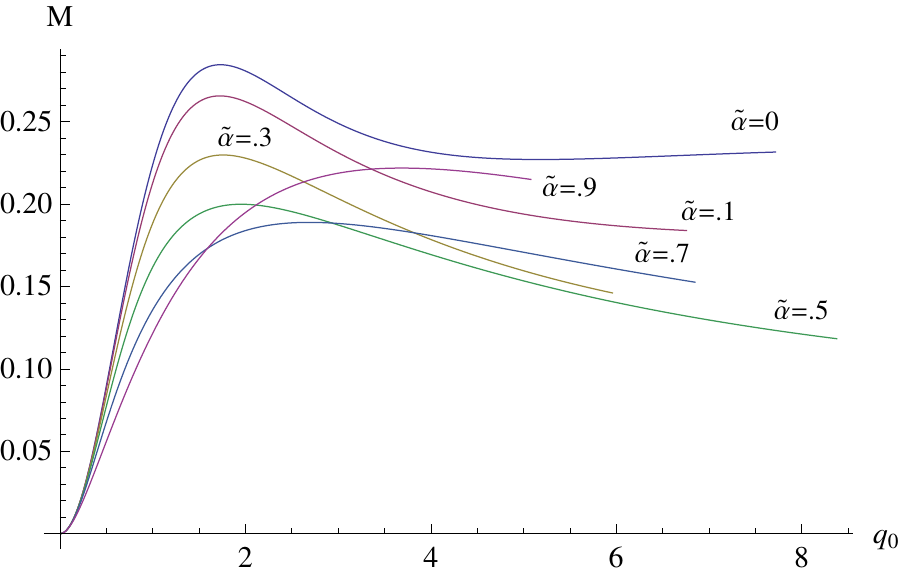}
		}\\ 
		\subfigure[~$M$ vs $\epsilon$ for $D=9$ ($n=7$), with $\tilde\alpha$ increasing from bottom to top.]{%
			\label{fig:9DEvq}
			\includegraphics[width=0.48\textwidth]{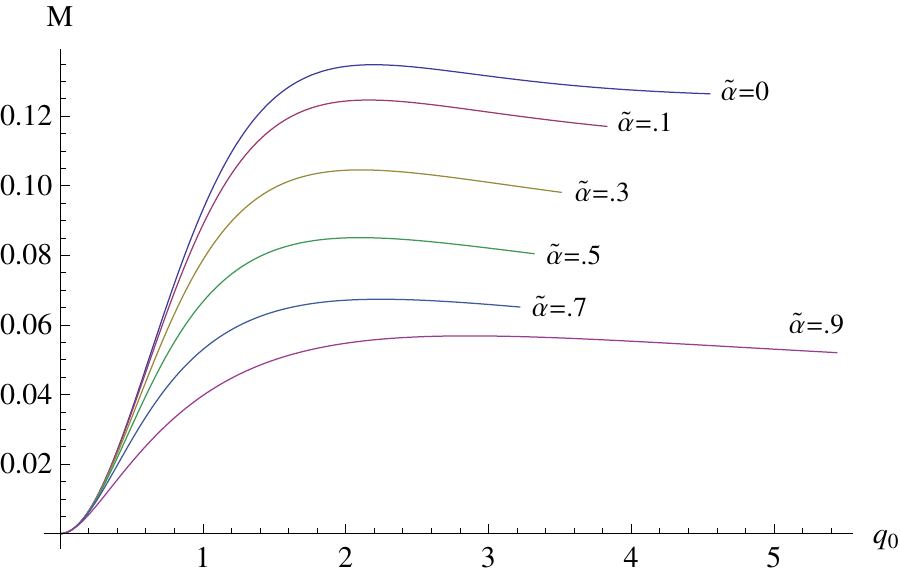}
		}%
		\subfigure[~$M$ vs $\epsilon$ for $D=11$ ($n=9$), with $\tilde\alpha$ increasing from bottom to top.]{%
			\label{fig:11DEvq}
			\includegraphics[width=0.48\textwidth]{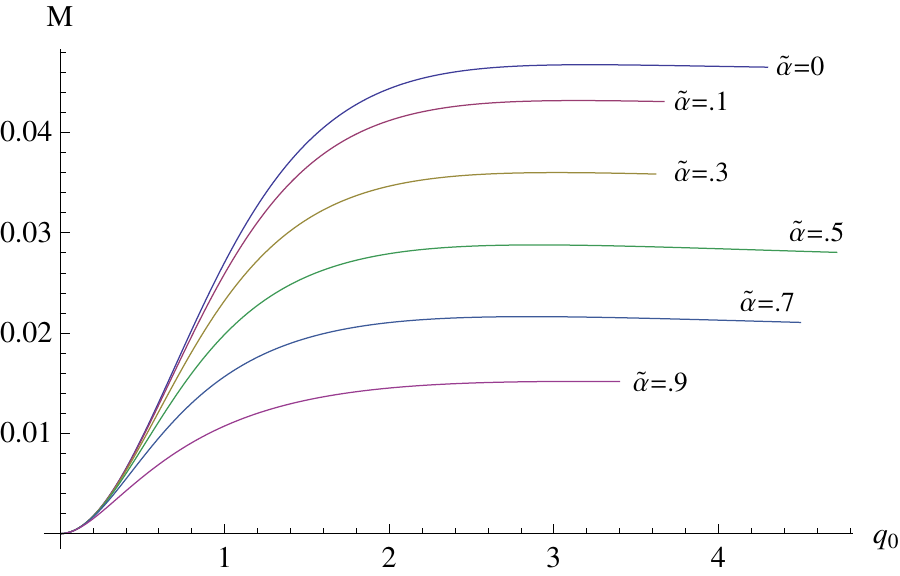}
		}%
	\end{center}
	\caption{%
		The boson star mass plotted against the  parameter $q_0$ in (a) 5, (b) 7, (c) 9, and (d) 11 dimensions for various
values of $\tilde\alpha$.  The maximum mass decreases nonlinearly with the space-time dimension.
	}%
	\label{fig:Evq}
\end{figure}

\begin{figure}[ht!]
	\begin{center}
		\subfigure[~$J$ vs $q_0$ for $D=5$ ($n=3$), with $\tilde\alpha$ increasing from bottom to top; except for $\tilde\alpha=0$, the curves all terminate at their endpoints, at which the Kretschmann scalar diverges.]{%
			\label{fig:5DJvq}
			\includegraphics[width=0.48\textwidth]{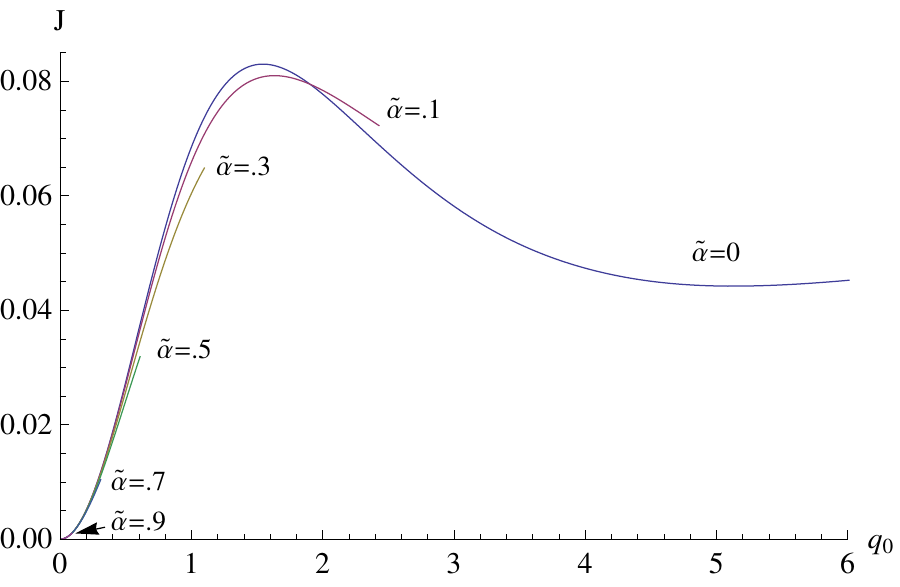}
		}%
		\subfigure[~$J$ vs $q_0$ for $D=7$ ($n=5$), with $\tilde\alpha$ increasing from bottom to top. ]{%
			\label{fig:7DJvq}
			\includegraphics[width=0.48\textwidth]{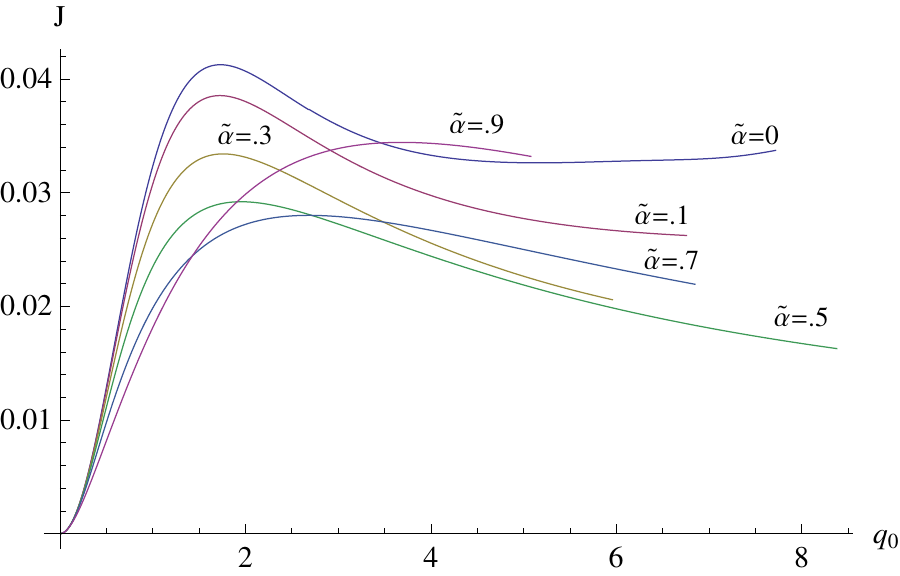}
		}\\ 
		\subfigure[~$J$ vs $\epsilon$ for $D=9$ ($n=7$), with $\tilde\alpha$ increasing from bottom to top.]{%
			\label{fig:9DJvq}
			\includegraphics[width=0.48\textwidth]{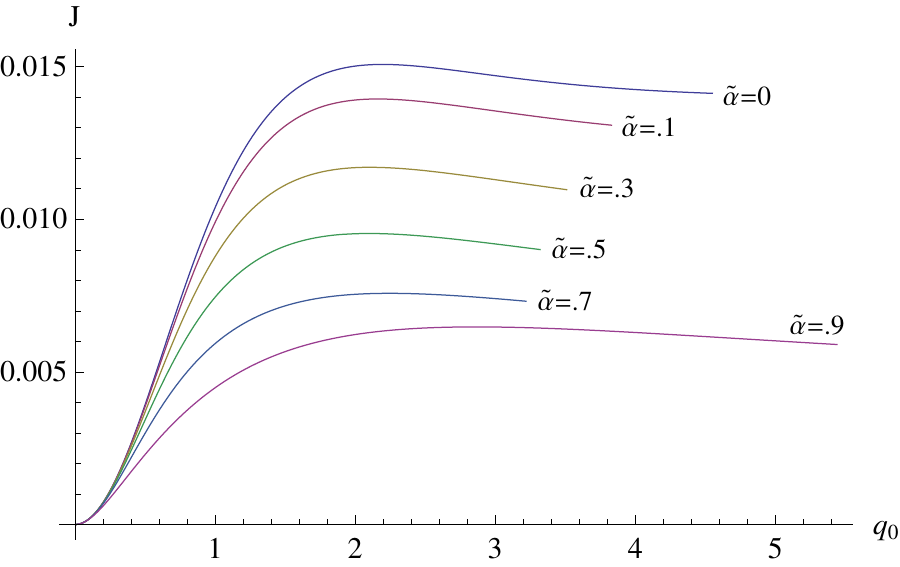}
		}%
		\subfigure[~$J$ vs $\epsilon$ for $D=11$ ($n=9$), with $\tilde\alpha$ increasing from bottom to top.]{%
			\label{fig:11DJvq}
			\includegraphics[width=0.48\textwidth]{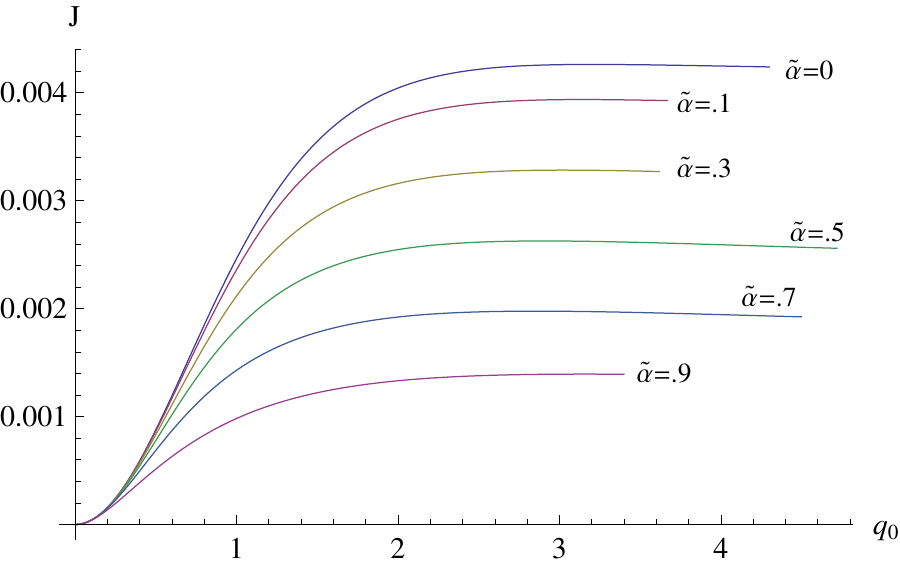}
		}%
	\end{center}
	\caption{%
		The boson star angular momentum plotted against the  parameter $q_0$ in (a) 5, (b) 7, (c) 9, and (d) 11 dimensions for various
values of $\tilde\alpha$. As with the mass, the maximum angular momentum decreases nonlinearly with the space-time dimension.
	}%
	\label{fig:Jvq}
\end{figure}	

\begin{figure}[ht!]
	\begin{center}
		\subfigure[~Detail of $M$ vs $q_0$ for $D=5$ ($n=3$).]{%
			\label{fig:7DEvqZ}
			\includegraphics[width=0.48\textwidth]{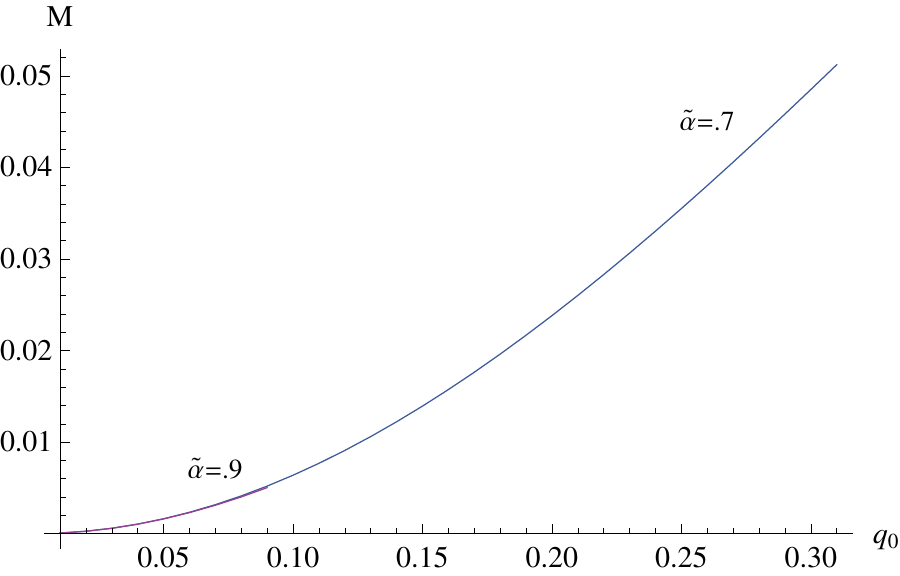}
		}%
		\subfigure[~Detail of $J$ vs $q_0$ for $D=5$ ($n=3$). ]{%
			\label{fig:7DJvqZ}
			\includegraphics[width=0.48\textwidth]{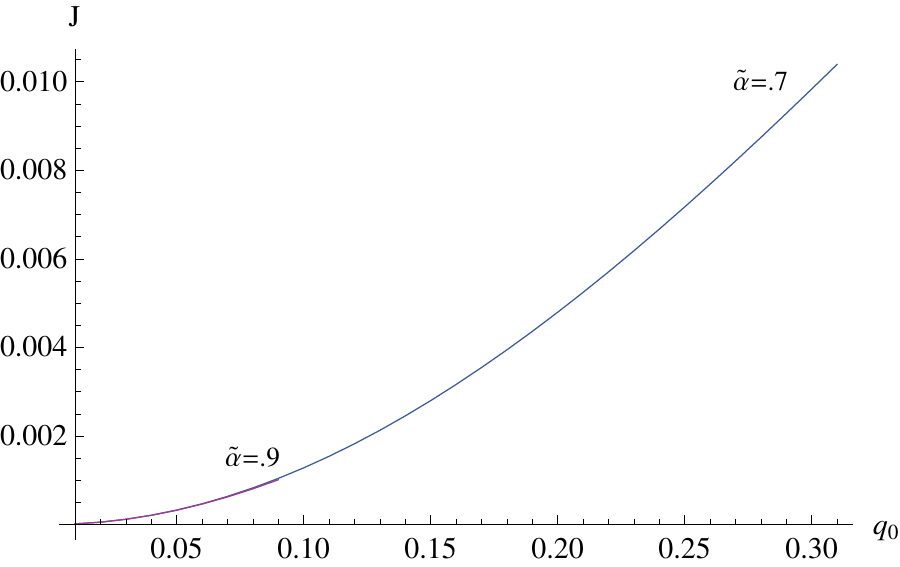}
		}%
	\end{center}
	\caption{%
		Detail plots of (a) mass and (b) angular momentum against the perturbative parameter $q_0$ in 5 dimensions for various values of $\tilde\alpha$.  The truncation, which is a result of the divergent curvature, occurs at smaller values of $q_0$  for as $\alpha$ increases.
	}%
	\label{fig:5DEJvqZoom}
\end{figure}

We use these quantities to study the behaviour of boson stars as a function of the Gauss-Bonnet parameter $\alpha$ for $n=3,5,7,9$.  In order to easily make comparisons between the different dimensions, we have chosen to look at values of $\alpha$ in fractions of the critical value: $\tilde{\alpha}\equiv \frac{\alpha}{\alpha_{\textrm{cr}}} = (0,\frac{1}{10}, \frac{3}{10}, \frac{1}{2}, \frac{7}{10}, \frac{9}{10})$. 
The $\alpha=0$ case  yields results that are commensurate with those obtained for the Einstein case in Ref. \cite{Stotyn:2013yka}.  This provides a useful cross-check on our analysis, since the form of the field equations used here (\ref{CC})--(\ref{RR}) differs significantly from the form in Ref. \cite{Stotyn:2013yka}.

 The most noteworthy feature to emerge from our study is the distinction between the $D=5$ ($n=3$) case and that of higher dimensions. We shall first discuss the $D>5$ cases.
 
 The $M$ and $J$ curves as functions of $\epsilon$ have the familiar spiral-like behaviour that was seen in the Einstein case \cite{Stotyn:2013yka}, evident in figures \ref{fig:Eve} and \ref{fig:Jve},  with better resolution given for the $D=5$ values in figure \ref{fig:5DEJveZoom}.    As $\tilde\alpha$ increases, the spirals become larger.  The spirals also tighten with increasing dimension -- they are barely visible for $D=11$ ($n=9$).  Similar spiral-type behaviour is also present for $\omega$ plotted as a function of $\epsilon$, as seen in figure \ref{fig:omve}.

Plotting both $M$ and $J$ as functions of $q_0$, we see from figures \ref{fig:Evq} and \ref{fig:Jvq} (with detail for $D=5$ given in figure \ref{fig:5DEJvqZoom})  that  for the $n=5,7,9$ cases  the damped oscillations of the Einstein case \cite{Stotyn:2013yka}
are still present.  Except for $D=7$, the maximum of these oscillations  decrease as $\alpha$ increases.

We find notably different behaviour for the $n=3$ ($D=5$) case.  A perusal of figures \ref{fig:Eve} and \ref{fig:Jve} indicates that even
for $\tilde{\alpha}=0.1$ the spiral behaviour is eliminated.  The curves terminate at finite values of $(M,J)$, with our code breaking down at critical values of $q_0$.  This behaviour is seen in more striking terms in figures \ref{fig:Evq}, \ref{fig:Jvq} and \ref{fig:5DEJvqZoom}:  as
$\tilde \alpha$ increases, the curves terminate at such small values of $q_0$ that no oscillatory behaviour is manifest.

\begin{figure}[ht!]
	\begin{center}
		\subfigure[~$K$ vs $q_0$ for $D=5$ ($n=3$) for various $\tilde{\alpha}$.]{%
			\label{fig:5DKvq}
			\includegraphics[width=0.48\textwidth]{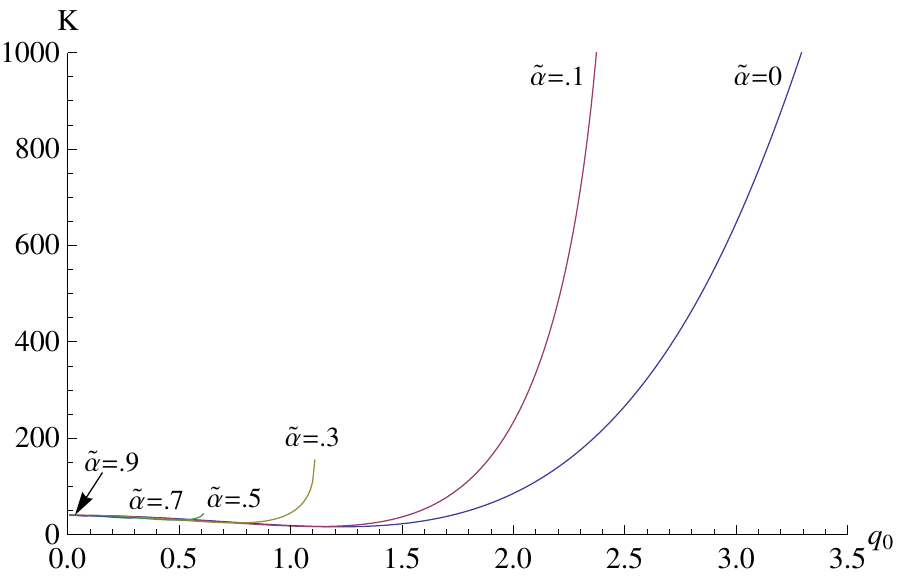}
		}%
		\subfigure[~$K$ vs $q_0$ for $D=7$ ($n=5$) for various $\tilde{\alpha}$. ]{%
			\label{fig:7DKvq}
			\includegraphics[width=0.48\textwidth]{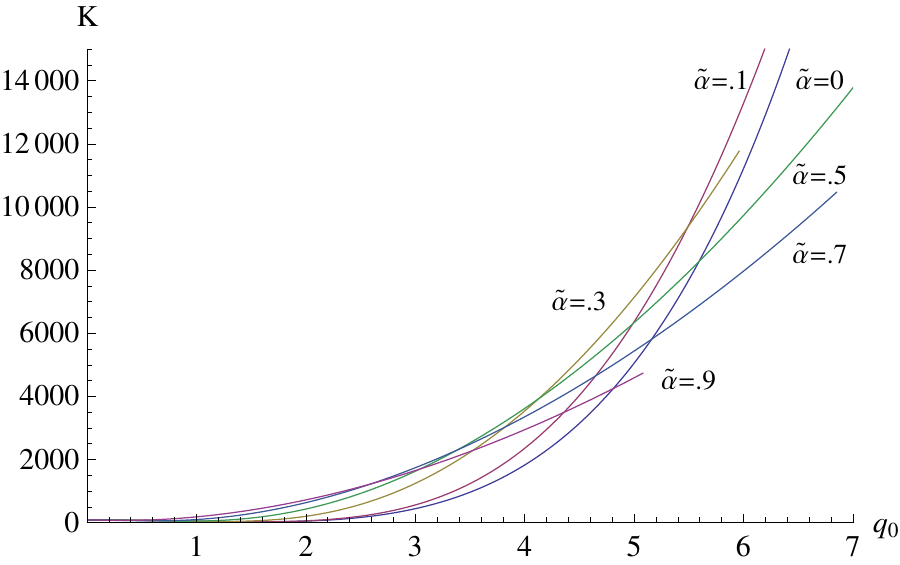}
		}\\ 
		\subfigure[~$K$ vs $q_0$ for $D=9$ ($n=7$) for various $\tilde{\alpha}$.]{%
			\label{fig:9DKvq}
			\includegraphics[width=0.48\textwidth]{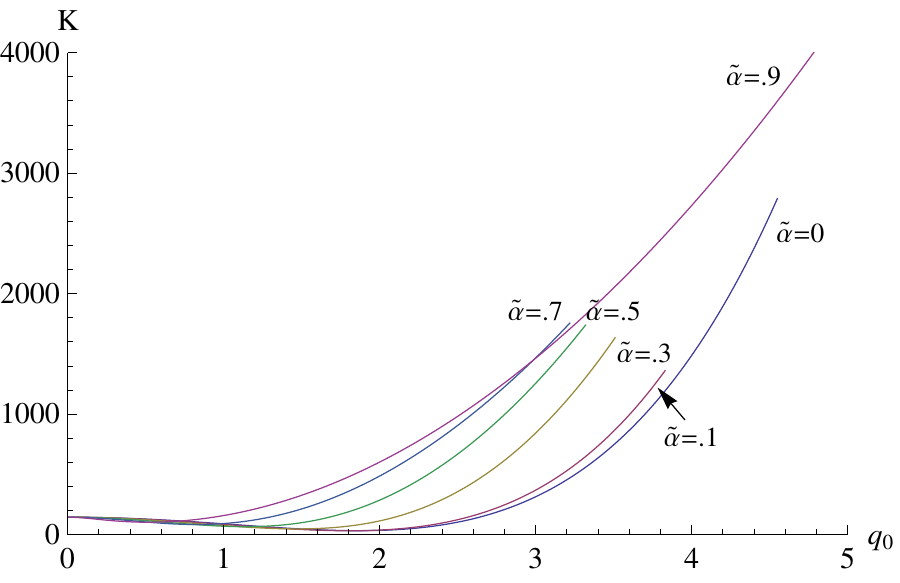}
		}%
		\subfigure[~$K$ vs $q_0$ for $D=11$ ($n=9$) for various $\tilde{\alpha}$.]{%
			\label{fig:11DKvq}
			\includegraphics[width=0.48\textwidth]{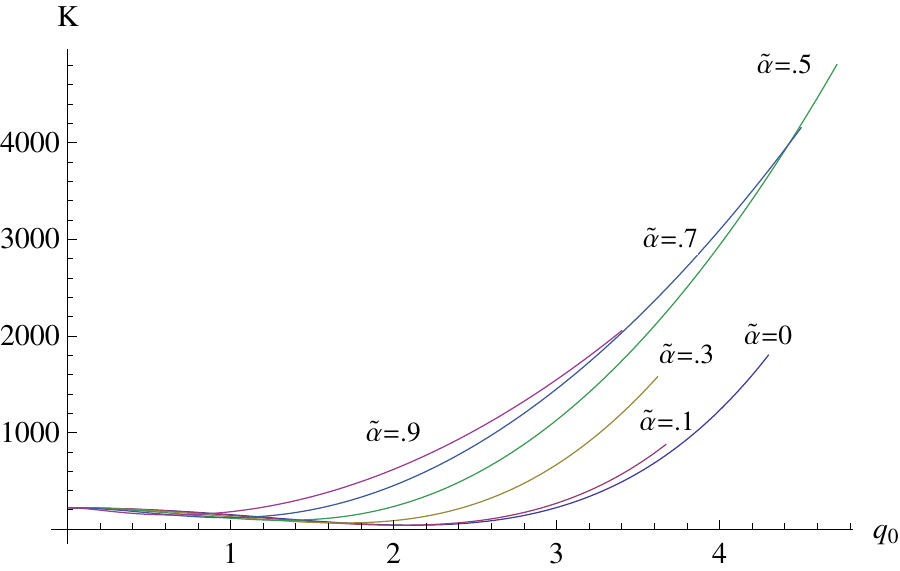}
		}%
	\end{center}
	\caption{%
		The Kretschmann scalar $K=R^{abcd}R_{abcd}$ plotted against the perturbative parameter $q_0$ in (a) 5, (b) 7, (c) 9, and (d) 11 dimensions for various  values of $\tilde\alpha$.  The divergent behaviour of $K$ in $D=5$ is drastically different than in higher dimensions for reasons explained in the text.
	}%
	\label{fig:Kvq}
\end{figure}

\begin{figure}[ht!]
	\begin{center}
		\subfigure[~$q_{h}'^*-q_h'(0)$ vs $q_0$ for $\tilde{\alpha}=0.1$ in $D=5$ ($n=3$).]{%
			\label{fig:5Dqh10}
			\includegraphics[width=0.48\textwidth]{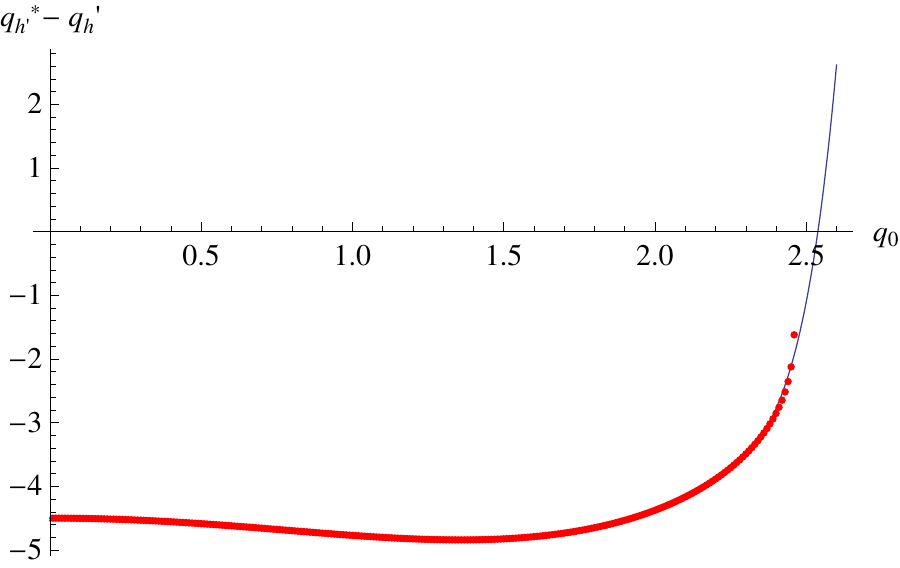}
		}%
		\subfigure[~$q_{h}'^*-q_h'(0)$ vs $q_0$ for $\tilde{\alpha}=0.9$ in $D=5$ ($n=3$).]{%
			\label{fig:5Dqh90}
			\includegraphics[width=0.48\textwidth]{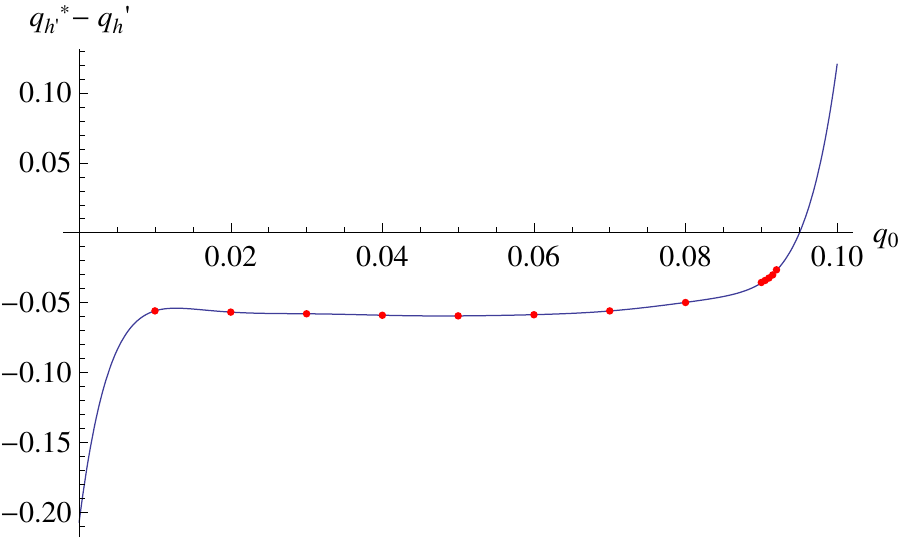}
		}\\ 
		\subfigure[~$q_{h}'^*-q_h'(0)$ vs $q_0$ for $\tilde{\alpha}=0.1$ \hspace{\textwidth} in $D=7, 9, 11$ ($n=5, 7, 9$).]{%
			\label{fig:7911Dqh10}
			\includegraphics[width=0.48\textwidth]{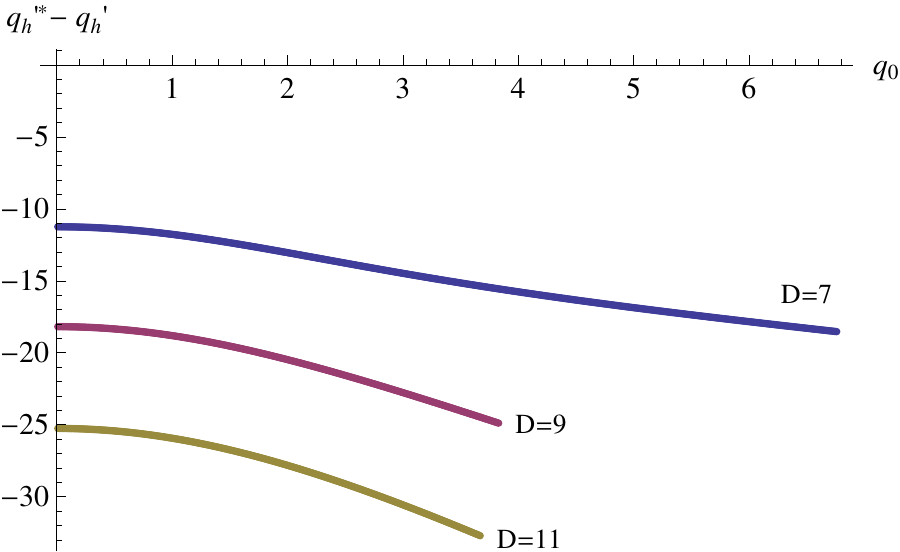}
		}%
		\subfigure[~$q_{h}'^*-q_h'(0)$ vs $q_0$ for $\tilde{\alpha}=0.9$ \hspace{\textwidth} in $D=7, 9, 11$ ($n=5, 7 ,9$).]{%
			\label{fig:7911Dqh90}
			\includegraphics[width=0.48\textwidth]{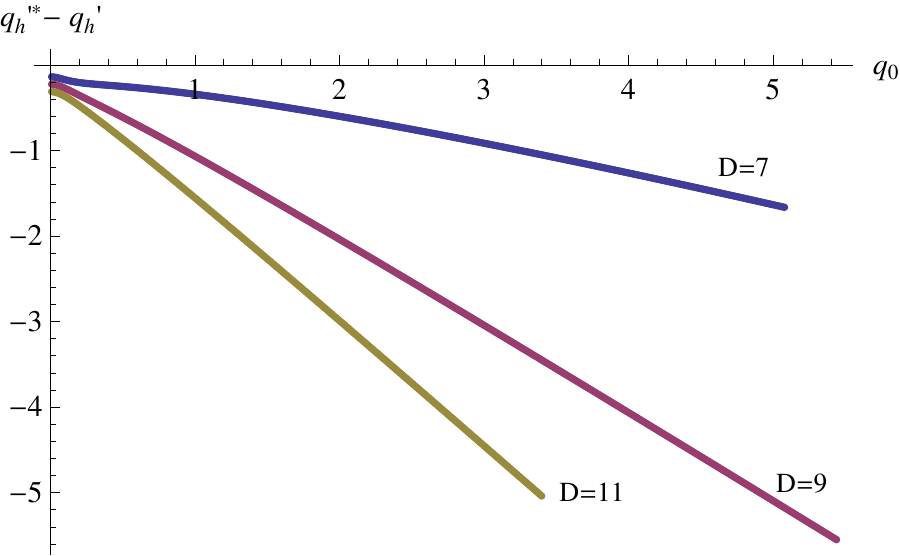}
		}%
	\end{center}
	\caption{%
		Plots of $q_{h}'^*-q_h'(0)$ against the perturbative parameter $q_0$  in 5, 7, 9, and 11 dimensions for various  values of $\tilde\alpha$.  The solid curves in sub-figures (a), (b) are the best-fit 9th-order polynomials to the points used to obtain $q_{0 \textrm{critical}}$.
	}%
	\label{fig:qh}
\end{figure}

The underlying reason for this behaviour can be traced to the Kretschmann scalar, as delineated in figure \ref{fig:Kvq}.
For all values of $n$ we
find numerically that the Kretchmann scalar initially decreases to a minimum and then becomes a strictly increasing function of $q_0$.  However 
for  $n=5,7,9$ we find that it remains finite for finite values of $q_0$, 
whereas for $n=3$  it  diverges at a critical value $q_{h}'^*$ of $q_h'(0)$, where 
\begin{equation}
q_{h}'^*={}-\frac{\sqrt{144 \alpha ^2+9 \ell^4-72 \alpha  \ell^2+48 \alpha  \ell^2 q_0^2}}{24 \alpha }.
\end{equation}
Although there is a critical value $q_{h}'^*$ in any dimension (see appendix \ref{Kscalar}), we find numerically that as $q_0$ increases $q_h'(0)$ departs further from this critical value, implying that the Kretschmann scalar remains finite for all finite $q_0$.  However,
for $n=3$ we find that $q_h'(0)\to q_{h}'^*$ at finite $q_0$, signaling a divergence in the Kretschmann scalar.  Furthermore, this critical value of $q_0$ decreases as $\tilde\alpha$ increases; to estimate the critical value of $q_0$, in figure \ref{fig:qh} we plot $q_h'(0)-q_{h}'^*$ and employ a 9th order polynomial fit to interpolate the curve.  The specific values are shown in table \ref{tab:critq0}.

\begin{table}
\caption{~Estimate of the critical value of $q_0$ at which the Kretschmann scalar diverges in $D=5$ for various values of $\tilde\alpha$ using the best fit polynomials in figure \ref{fig:qh}.}
\begin{tabular}{|c|c|}
\hline
$\tilde{\alpha}$ & $q_{0\textrm{critical}}$\\
\hhline{|=|=|}
0.1 & 2.54\\
\hline
0.3 & 1.15\\
\hline
0.5 & 0.648\\
\hline
0.7 & 0.329\\
\hline
0.9 & 0.0950\\
\hline
\end{tabular}
\label{tab:critq0}
\end{table}

This behaviour is neither present in the Einstein case nor  in the perturbative solutions.   It suggests the existence of a critical central energy density in $D=5$ at or before which the behaviour of the boson star must radically change.  This situation is reminiscent of the numerical boson star solution obtained in $D=3$ \cite{Stotyn:2013spa} insofar as the mass, angular momentum, and angular velocity all approach finite terminal values at a critical value of the central energy density, whereas in higher dimensions the central energy density is unbounded and the corresponding physical quantities all exhibit damped harmonic oscillations about finite limiting values. However numerical evidence for the $D=3$ case indicates the formation of an extremal BTZ black hole at the critical value, with vanishing scalar field in the exterior,  whereas in the present case we do not find the scalar field to vanish and we find no indication of the formation of a horizon.  Nevertheless, we expect that these solutions will be dynamically unstable to hairy black hole formation for some $q_0 < q_{0 \textrm{critical}}$.

In $D=3$ the Ricci scalar is the highest-curvature term that
can appear in that dimension;  in $D=5$ it is the Gauss-Bonnet term that fulfills this criterion.  It is tempting to conjecture that 
such critical energies appear in all odd dimensions in which the gravitational theory includes its highest-possible curvature term.  
While this would be challenging to numerically check, it is conceivably feasible for the $D=7$ 3rd-order Lovelock and $D=9$ 4th order
Lovelock cases.

\begin{figure}[ht!]
	\begin{center}
		\subfigure[~$\omega$ vs $q_0$ for $D=5$ ($n=3$) for various $\tilde{\alpha}$.]{%
			\label{fig:5Domvq}
			\includegraphics[width=0.48\textwidth]{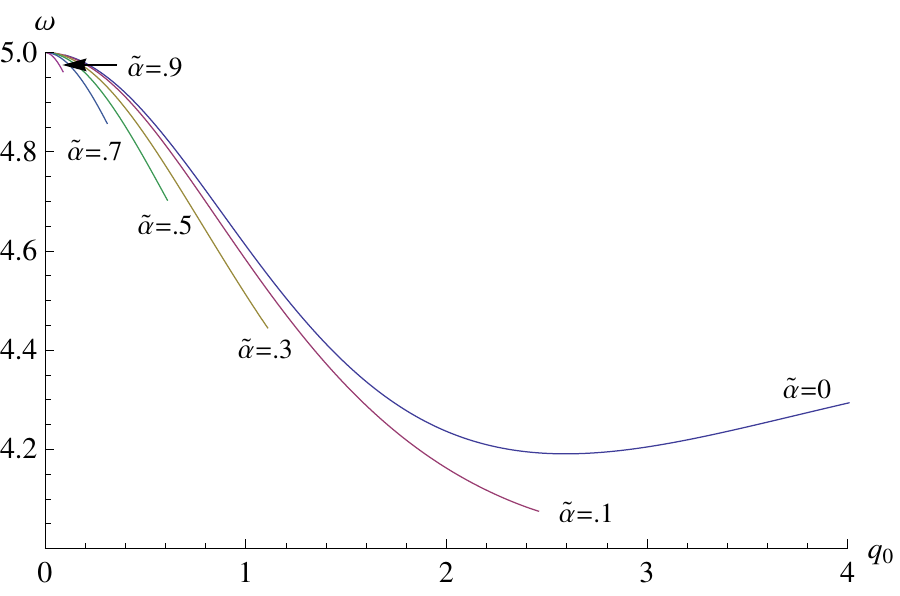}
		}%
		\subfigure[~$\omega$ vs $q_0$ for $D=7$ ($n=5$) for various $\tilde{\alpha}$. ]{%
			\label{fig:7Domvq}
			\includegraphics[width=0.48\textwidth]{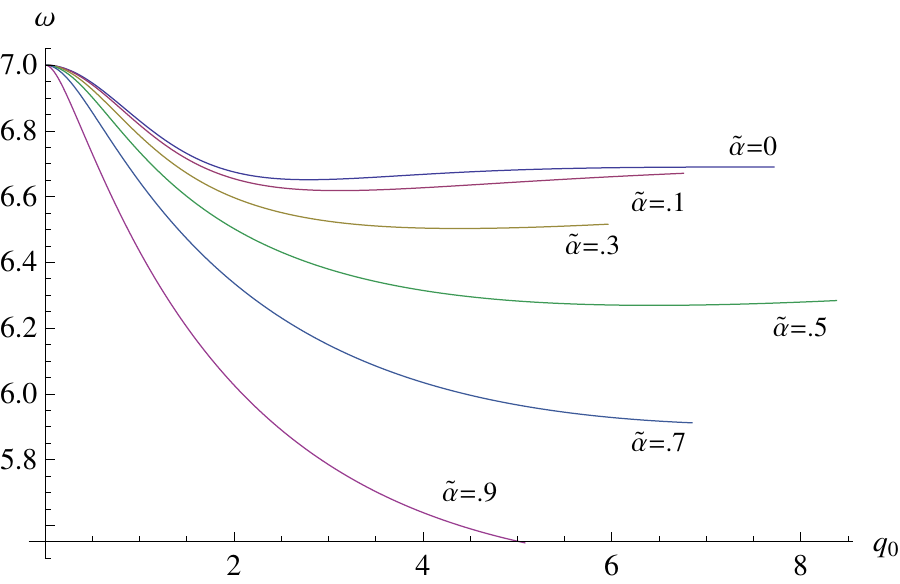}
		}\\
		\subfigure[~$\omega$ vs $q_0$ for $D=9$ ($n=7$) for various $\tilde{\alpha}$.]{%
			\label{fig:9Domvq}
			\includegraphics[width=0.48\textwidth]{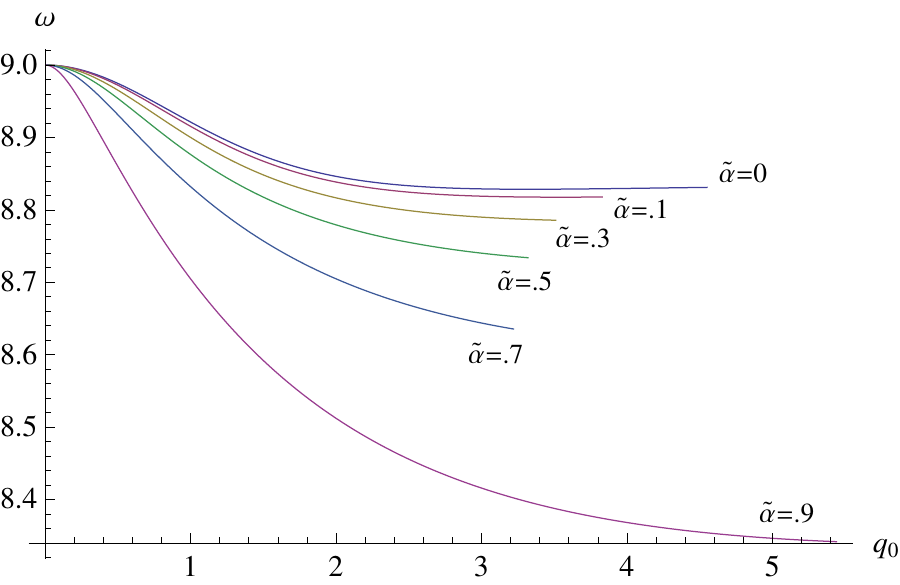}
		}%
		\subfigure[~$\omega$ vs $q_0$ for $D=11$ ($n=9$) for various $\tilde{\alpha}$.]{%
			\label{fig:11Domvq}
			\includegraphics[width=0.48\textwidth]{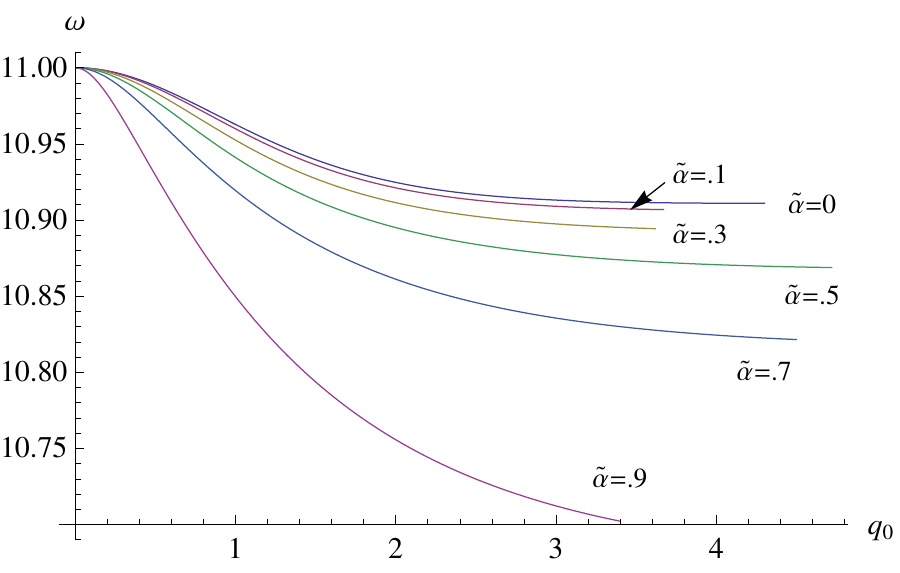}
		}%
	\end{center}
	\caption{%
		The angular velocity $\omega$ plotted against the perturbative parameter $q_0$ in (a) 5, (b) 7, (c) 9, and (d) 11 dimensions for various  values of $\tilde\alpha$.  The termination of the curves in $D=5$ is due to the divergence of the Kretschmann scalar at the end points, except for $\tilde\alpha=0$ which is immune to this divergence.
	}%
	\label{fig:omvq}
\end{figure}

\begin{figure}[ht!]
	\begin{center}
		\subfigure[~Close up of $K$ vs $q_0$ for $D=5$ ($n=3$) for various values of $\tilde\alpha$.]{%
			\label{fig:5DKvqZ}
			\includegraphics[width=0.48\textwidth]{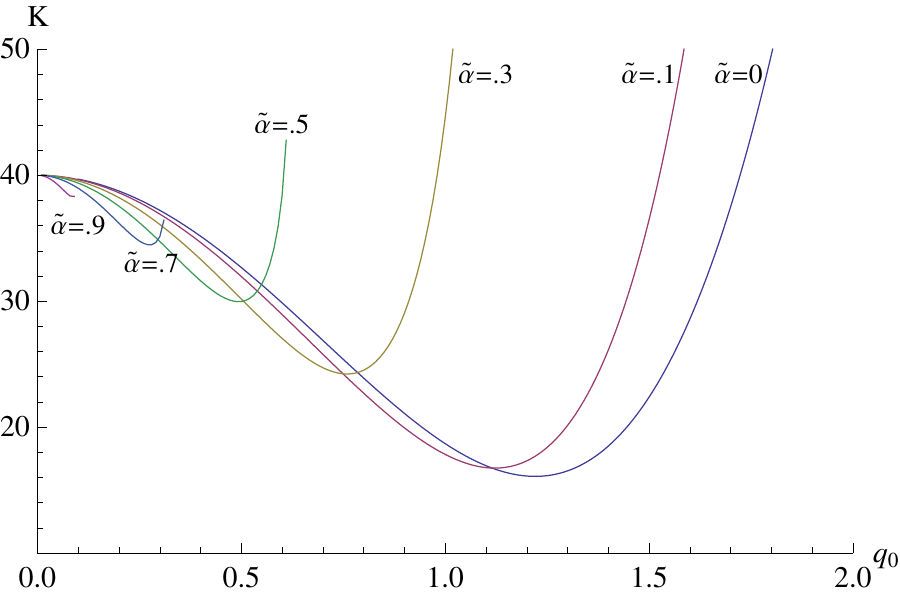}
		}%
		\subfigure[~Close up of $K$ vs $q_0$ for $D=7$ ($n=5$) for various values of $\tilde\alpha$.]{%
			\label{fig:7DKvqZ}
			\includegraphics[width=0.48\textwidth]{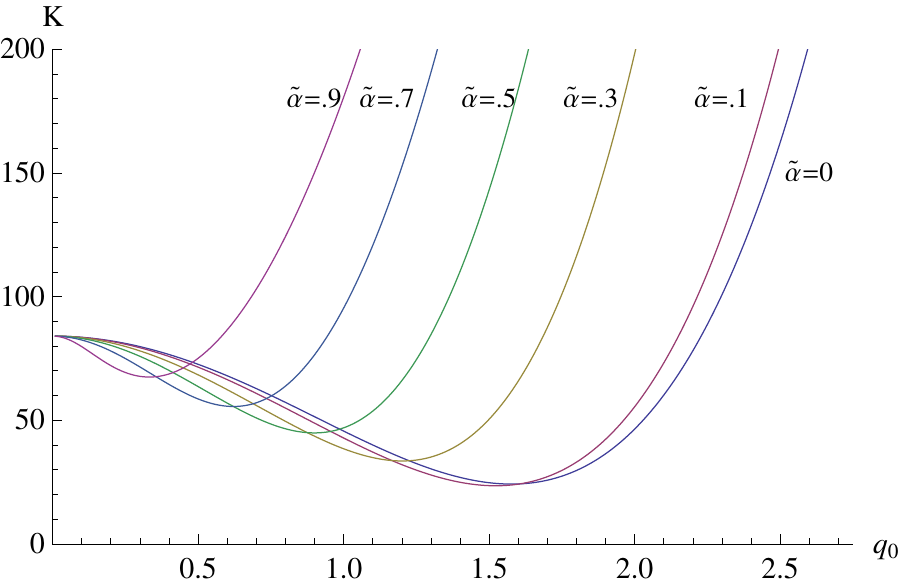}
		}\\ 
		\subfigure[~Close up of $K$ vs $q_0$ for $D=9$ ($n=7$) for various values of $\tilde\alpha$.]{%
			\label{fig:9DKvqZ}
			\includegraphics[width=0.48\textwidth]{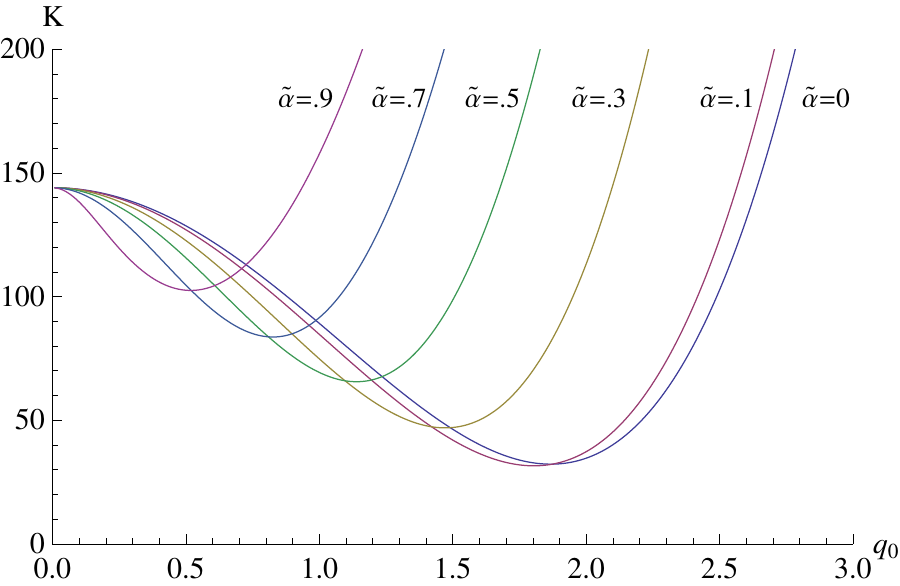}
		}%
		\subfigure[~Close up of ~$K$ vs $q_0$ for $D=11$ ($n=9$) for various values of $\tilde\alpha$.]{%
			\label{fig:11DKvqZ}
			\includegraphics[width=0.48\textwidth]{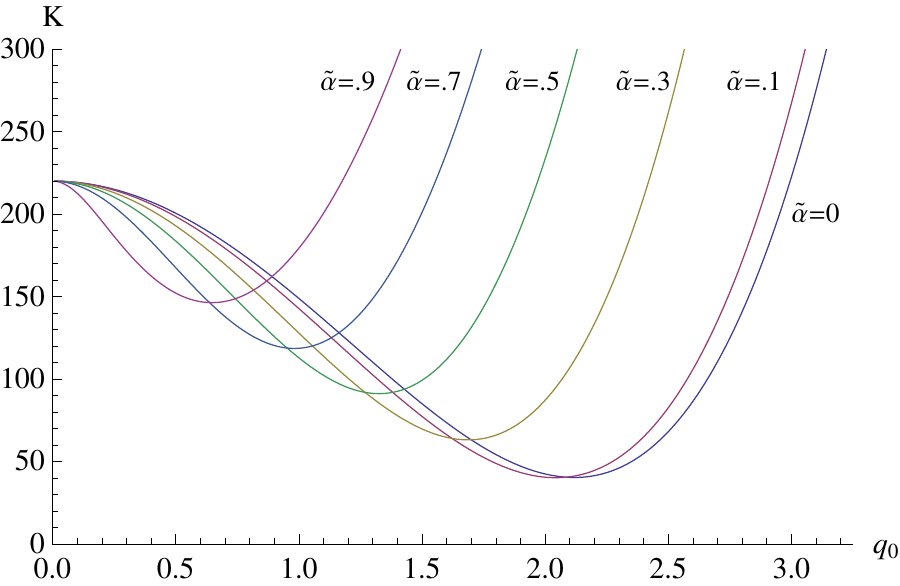}
		}%
	\end{center}
	\caption{%
		Detail of the Kretschmann scalar $K$ plotted against the perturbative parameter $q_0$ in (a) 5, (b) 7, (c) 9, and (d) 11 dimensions for various  values of $\tilde\alpha$.  Except for the scale on the $K$ axis, dimensions $D=7,9,11$ behave qualitatively the same as $\tilde\alpha$ increases, whereas $D=5$ exhibits distinct behaviour.
	}%
	\label{fig:Kvqclose}
\end{figure}

\begin{figure}[ht!]
	\begin{center}
		\subfigure[~$M$ vs $J$ for $D=5$ ($n=3$) for various $\tilde{\alpha}$.]{%
			\label{fig:5DEvJ}
			\includegraphics[width=0.48\textwidth]{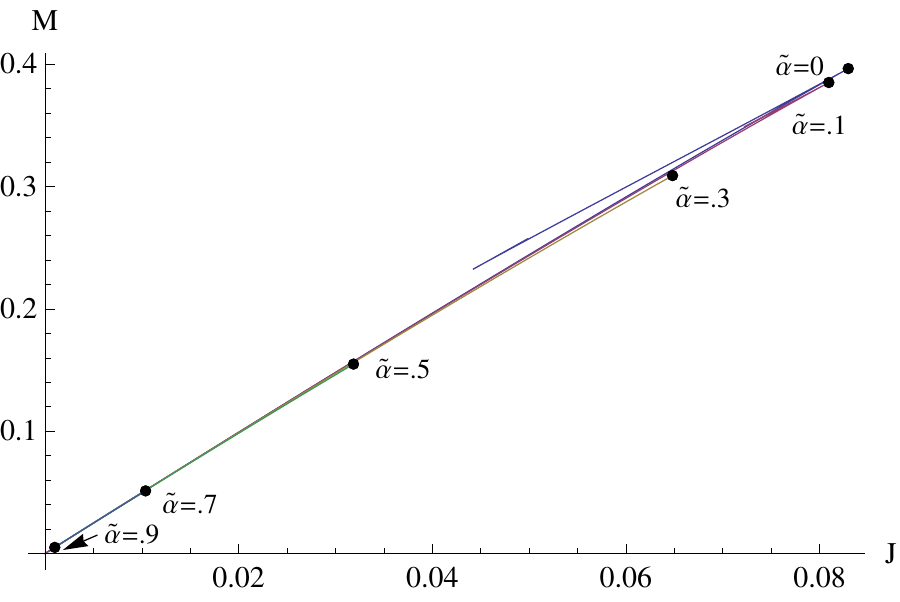}
				\llap{\raisebox{3.0cm}{\includegraphics[width=3.25cm]{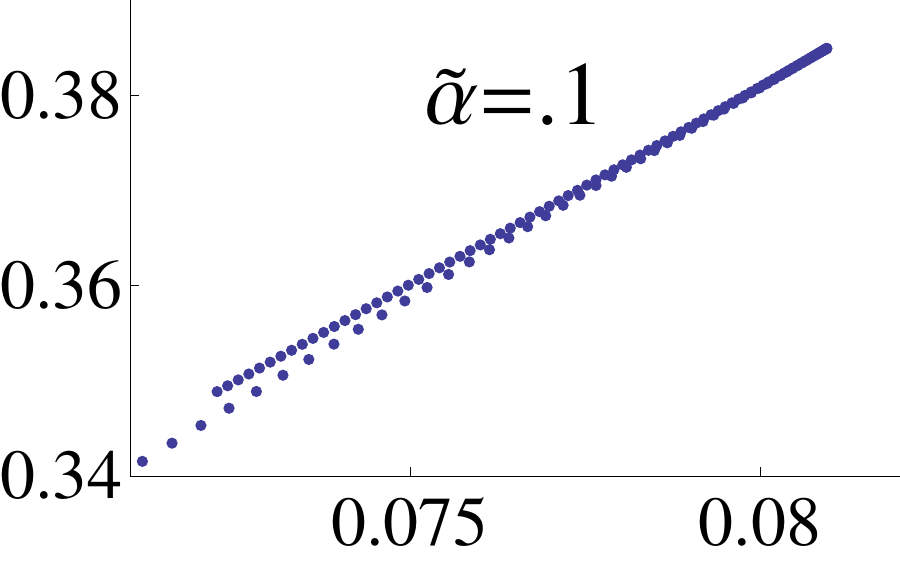}}\hspace*{4.4cm}}
		}%
		\subfigure[~$M$ vs $J$ for $D=7$ ($n=5$) for various $\tilde{\alpha}$.]{%
			\label{fig:7DEvJ}
			\includegraphics[width=0.48\textwidth]{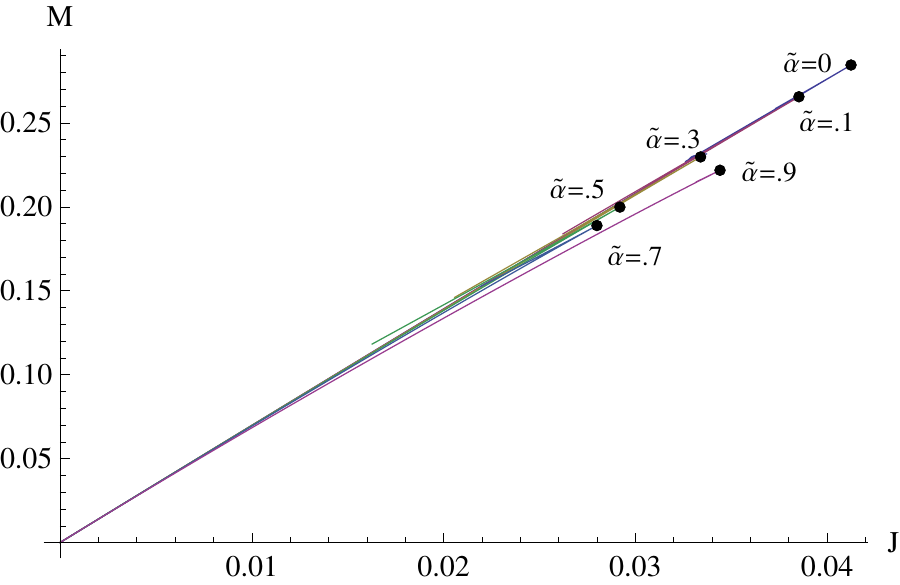}
				\llap{\raisebox{3.0cm}{\includegraphics[width=3.25cm]{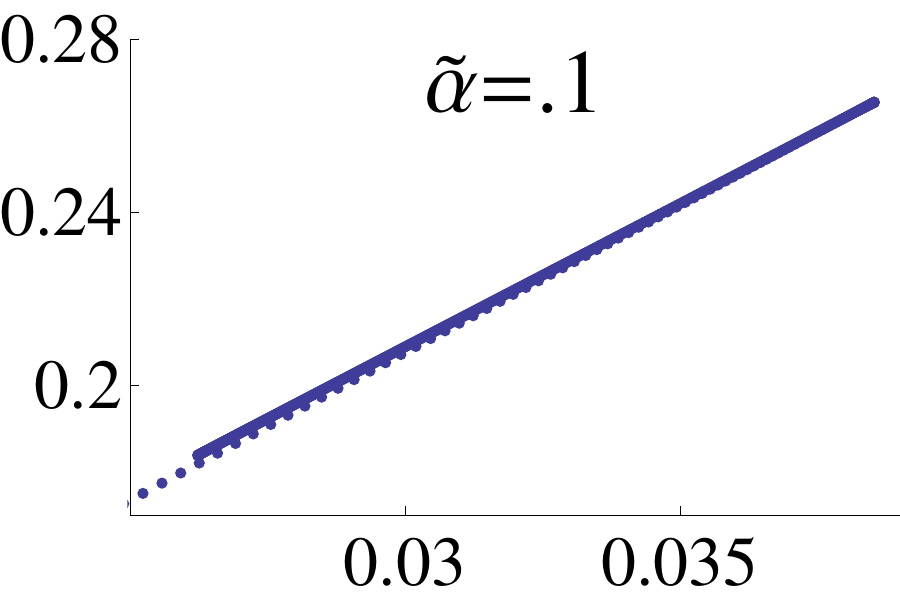}}\hspace*{4.4cm}}
				\llap{\raisebox{0.6cm}{\includegraphics[width=3.25cm]{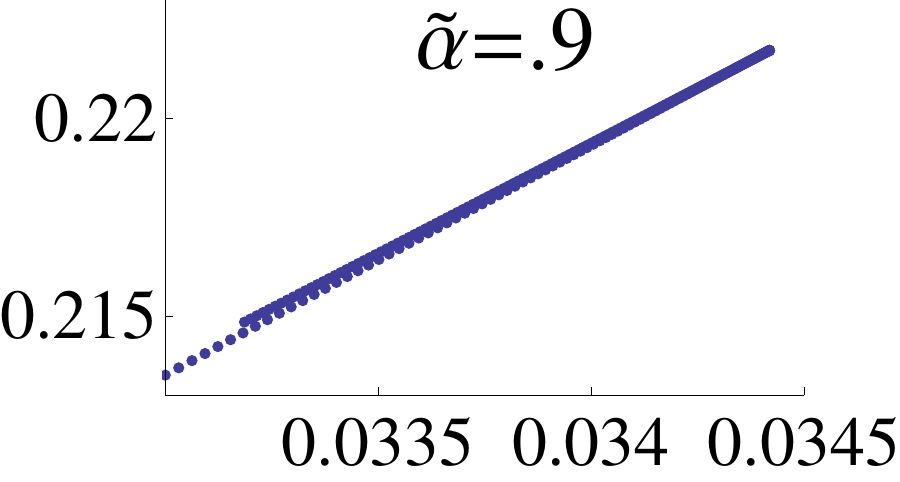}}\hspace*{0.25cm}}
		}\\ 
		\subfigure[~$M$ vs $J$ for $D=9$ ($n=7$) for various $\tilde{\alpha}$.]{%
			\label{fig:9DEvJ}
			\includegraphics[width=0.48\textwidth]{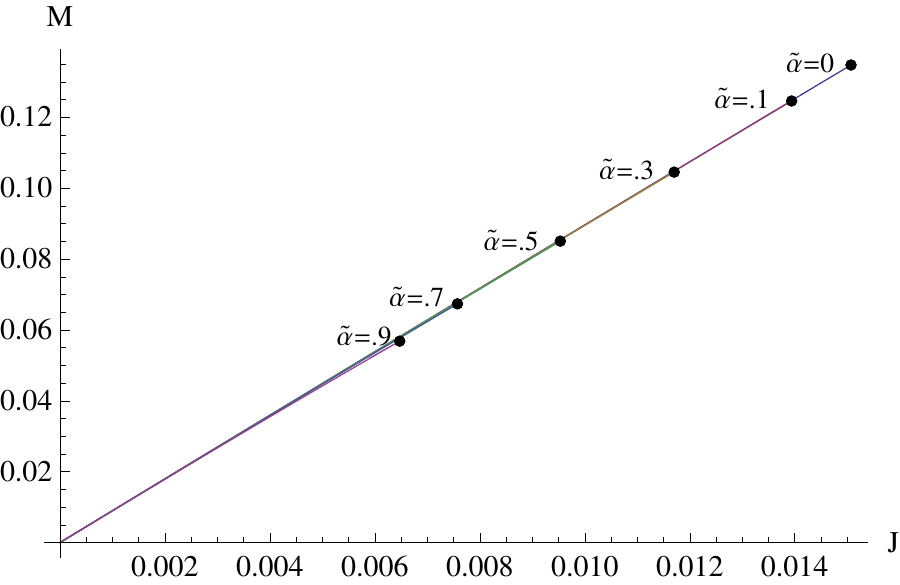}
				\llap{\raisebox{3.0cm}{\includegraphics[width=3.25cm]{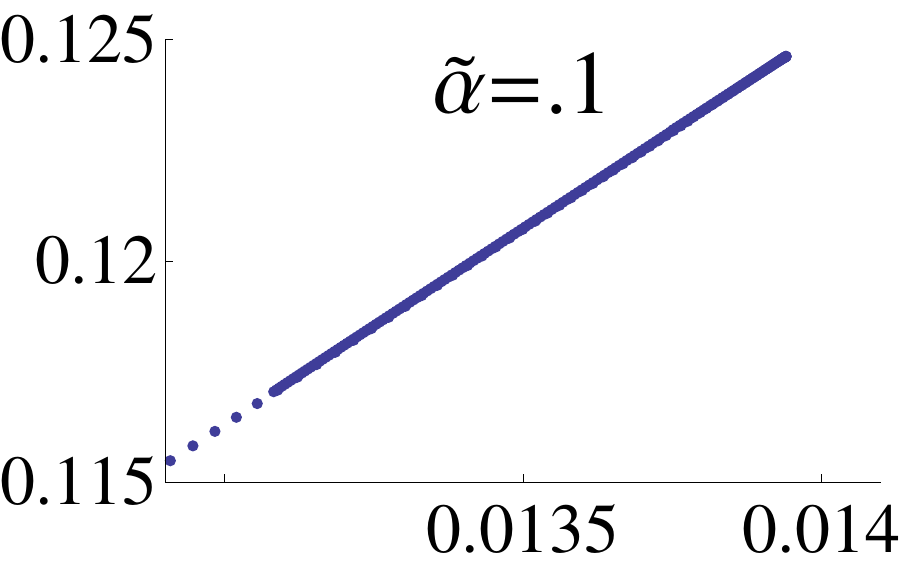}}\hspace*{4.4cm}}
				\llap{\raisebox{0.6cm}{\includegraphics[width=3.25cm]{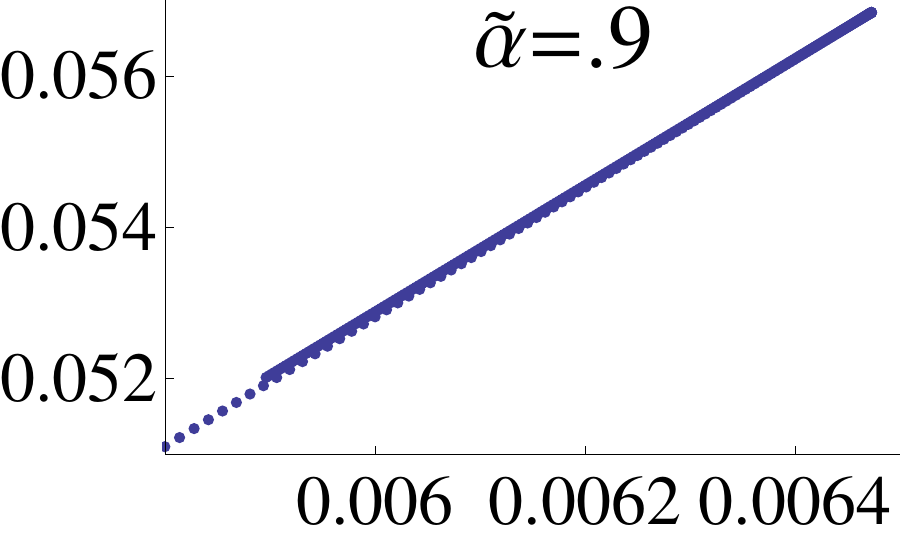}}\hspace*{0.25cm}}
		}%
		\subfigure[~$M$ vs $J$ for $D=11$ ($n=9$) for various $\tilde{\alpha}$.]{%
			\label{fig:11DEvJ}
			\includegraphics[width=0.48\textwidth]{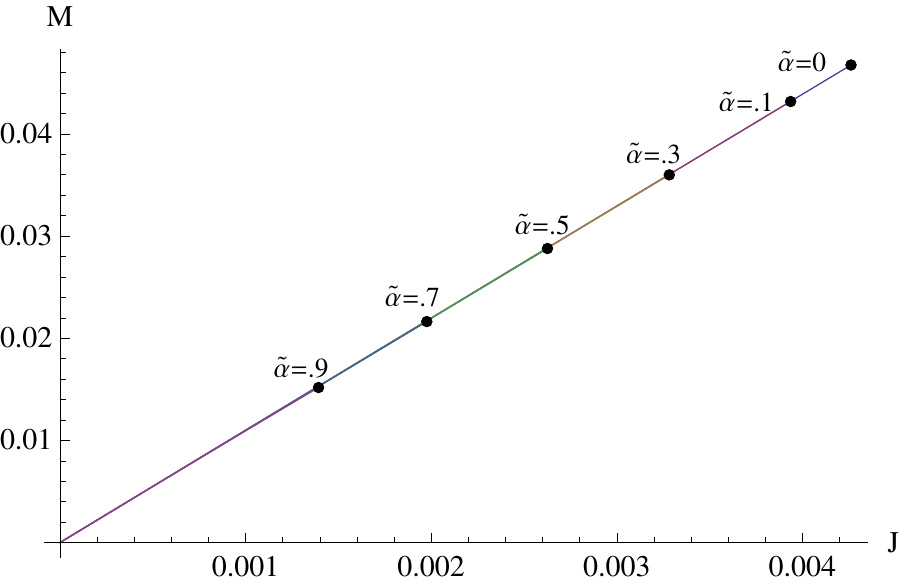}
				\llap{\raisebox{3.0cm}{\includegraphics[width=3.25cm]{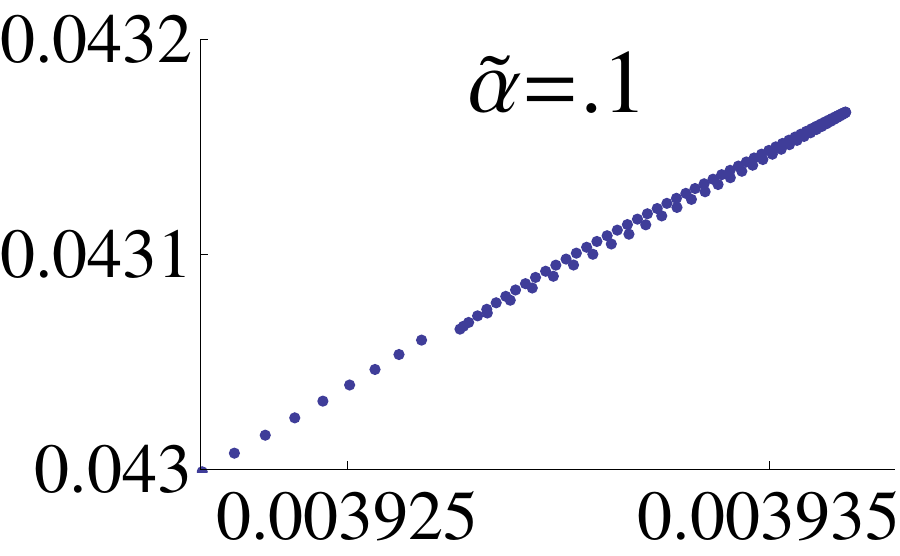}}\hspace*{4.4cm}}
				\llap{\raisebox{0.6cm}{\includegraphics[width=3.25cm]{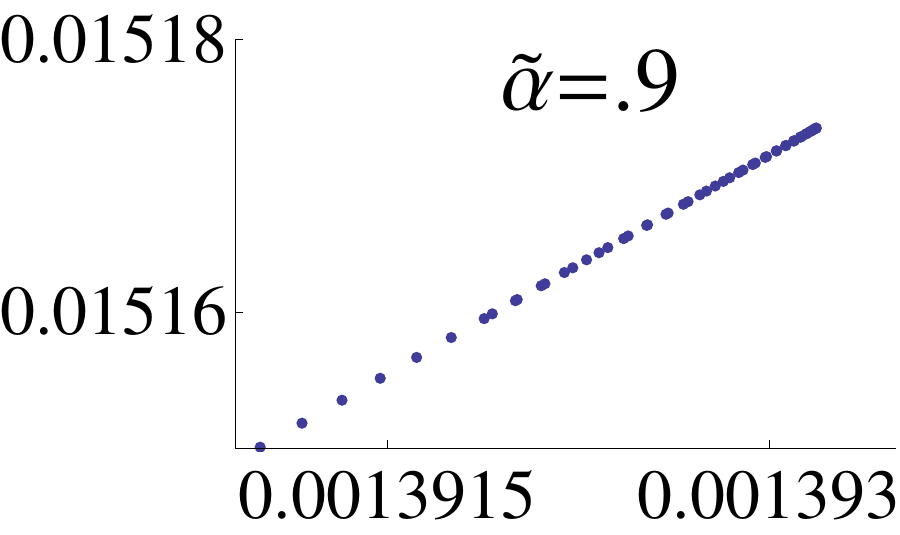}}\hspace*{0.25cm}}
		}%
	\end{center}
	\caption{%
		$M$ vs $J$   in (a) 5, (b) 7, (c) 9, and (d) 11 dimensions for various  values of $\tilde\alpha$.  The black dots correspond to the point at which these curves turn back for a particular value of $\tilde\alpha$. The insets are close ups of the zig-zag patterns for $\tilde{\alpha}=0.1$ (top left corner) and  $\tilde{\alpha}=0.9$ (bottom right corner) for each dimension.
	}%
	\label{fig:EvJ}
\end{figure}

To conclude our analysis, in figure \ref{fig:EvJ} we plot $M$ vs.\ $J$ for each dimension and the values of $\tilde\alpha$ indicated.  The most striking feature is that  for each dimension, all of the solutions lie on almost the same line in the energy versus angular momentum graph regardless of the value of $\tilde\alpha$,at least for sufficiently small values of $(M,J)$.  For a given value of $\tilde\alpha$ these curves ``turn back", making a zig-zag pattern familiar from the Einstein case \cite{Stotyn:2013yka}, albeit much tighter. As $\tilde\alpha$ increases, the turnaround point for the zig-zag pattern is closer to the beginning of the curve (at smaller values of $(M,J)$) for $D\ne7$. As the dimensionality increases the effect of the zig-zag is also suppressed.  It is most  pronounced in $D=5$; for this dimension we found no numerical evidence for a zig-zag pattern for $\alpha\ge0.3\alpha_{\textrm{Max}}$.

\section{Beyond $\alpha_{\textrm{cr}}$}
\label{beyond}

It is possible to numerically analyze solutions for which $\alpha > \alpha_{\textrm{cr}}$. While a full analysis is beyond the scope of this paper, we present some preliminary results.  We find for all boson star solutions that the combinations of
the quantities $C_f$, $C_h$ and $C_\Omega$ in (\ref{eq:EJ}) become negative in this regime, seemingly indicating
they have negative ADM mass.  However the factor of  $(1-\alpha/\alpha_{\textrm{cr}})$ in (\ref{eq:EJ}) also becomes negative, ensuring that the overall signs of $M$ and $J$ do not change in going from $\alpha < \alpha_{\textrm{cr}}$
to $\alpha > \alpha_{\textrm{cr}}$.

\begin{figure}[ht!]
	\begin{center}
		\subfigure[~$M$ vs $\epsilon$ for $\tilde{\alpha}=1.20$ in various dimensions.]{%
			\label{fig:120Mvepsilon}
			\includegraphics[width=0.48\textwidth]{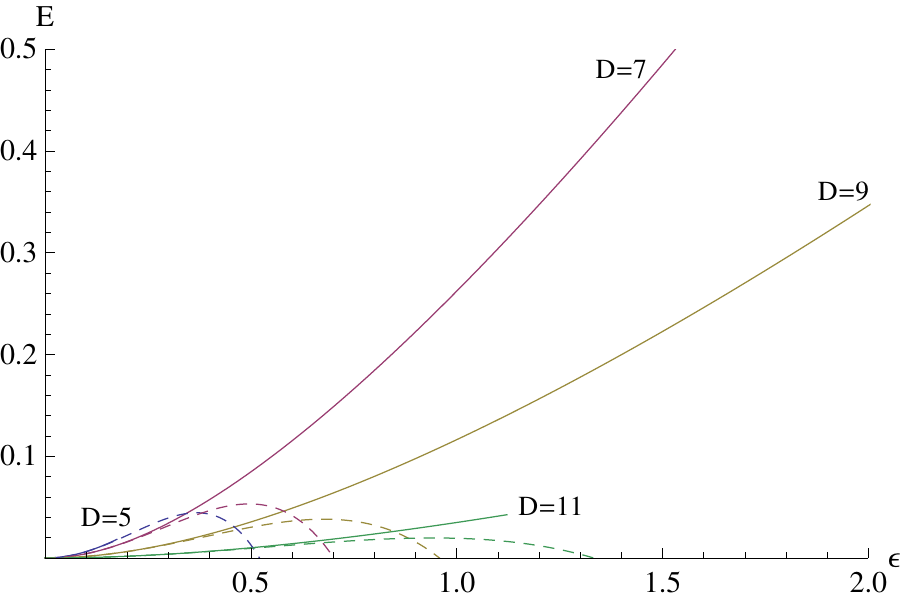}
		}%
		\subfigure[~$J$ vs $\epsilon$ for $\tilde{\alpha}=1.20$ in various dimensions.]{%
			\label{fig:120Jvepsilon}
			\includegraphics[width=0.48\textwidth]{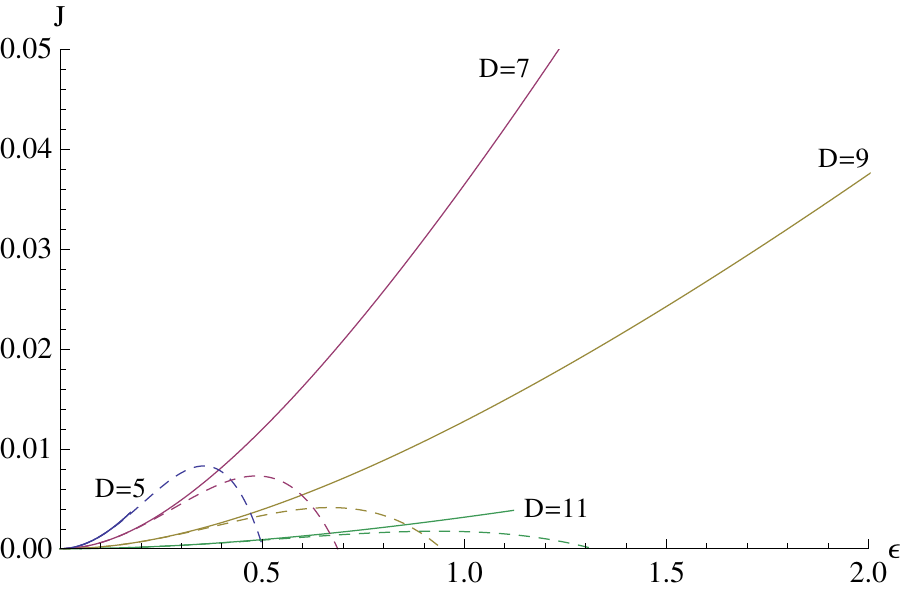}
		}%
	\end{center}
	\caption{%
		(a) The boson star mass $M$ and (b) angular momentum $J$ plotted against the pertubative parameter $\epsilon$ for $\tilde{\alpha}=1.20$ in  $D=5,7,9,11$ ($n=3,5,7,9$).  Perturbative results are  plotted as dashed lines.  Both $M$ and $J$ increase monotonically with increasing $\epsilon$.
	}%
	\label{fig:120MJvepsilon}
\end{figure}

\begin{figure}[ht!]
	\begin{center}
		\subfigure[~$M$ vs $q_0$ for $\tilde{\alpha}=1.20$ for various dimensions.]{%
			\label{fig:120Mvq0}
			\includegraphics[width=0.48\textwidth]{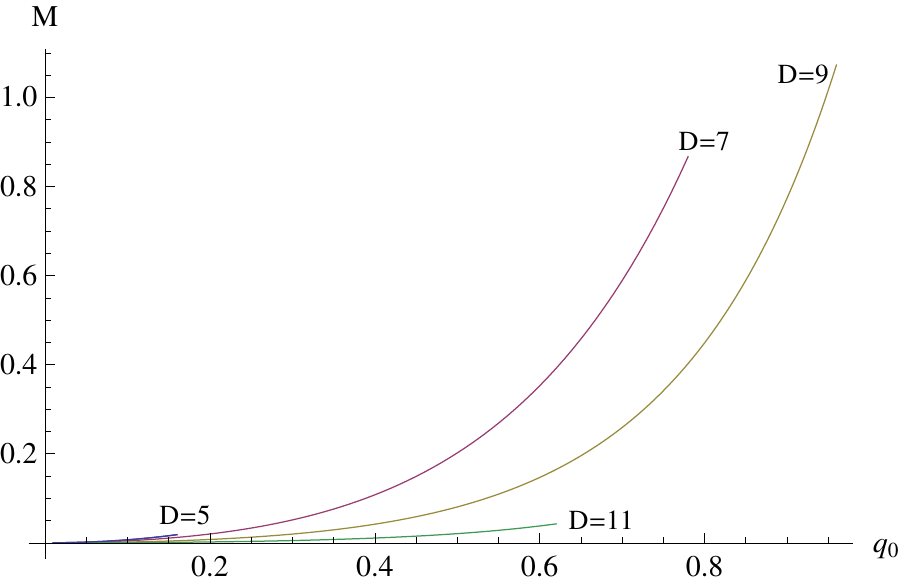}
		}%
		\subfigure[~$J$ vs $q_0$ for $\tilde{\alpha}=1.20$ for various dimensions.]{%
			\label{fig:120Jvq0}
			\includegraphics[width=0.48\textwidth]{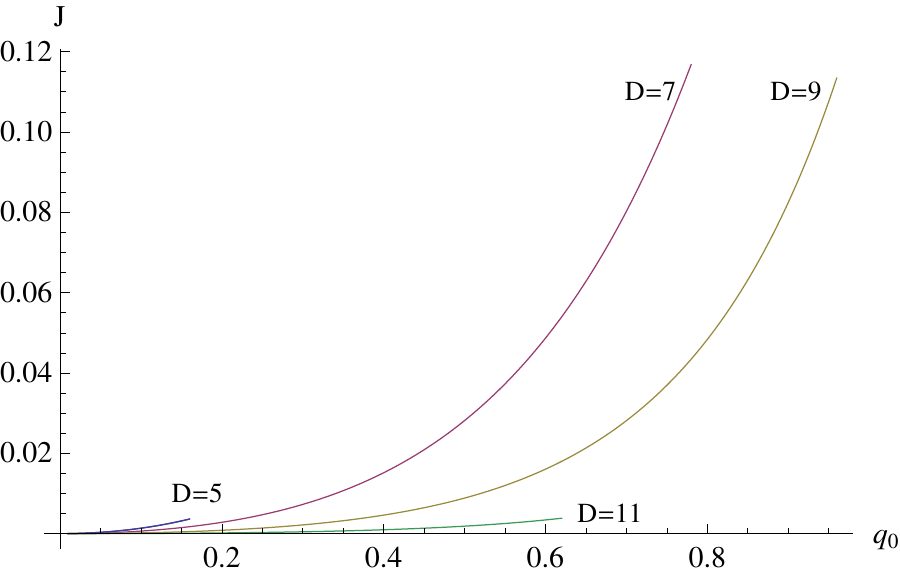}
		}%
	\end{center}
\caption{%
	(a) The boson star mass $M$ and (b) angular momentum $J$ plotted against the parameter $q_0$ for $\tilde{\alpha}=1.20$ in $D=5,7,9,11$ ($n=3,5,7,9$).  Both $M$ and $J$ increase monotonically with increasing $q_0$.
	}%
	\label{fig:120MJvq0}
\end{figure}

\begin{figure}[ht!]
	\begin{center}
		\subfigure[~$\ln(M)$ vs $q_0$ for $\tilde{\alpha}=1.20$ for various dimensions.]{%
			\label{fig:120lnMvq0}
			\includegraphics[width=0.48\textwidth]{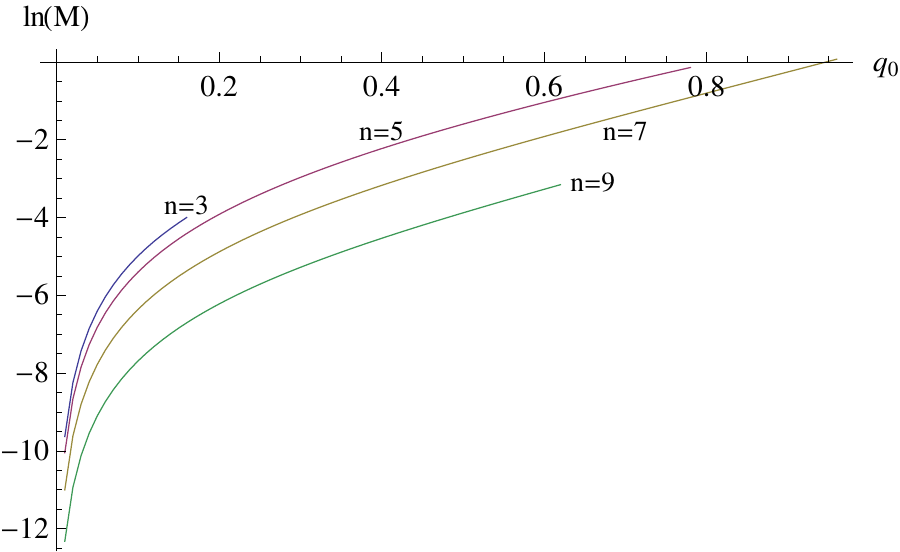}
		}%
		\subfigure[~$\ln(J)$ vs $q_0$ for $\tilde{\alpha}=1.20$ for various dimensions.]{%
			\label{fig:120lnJvq0}
			\includegraphics[width=0.48\textwidth]{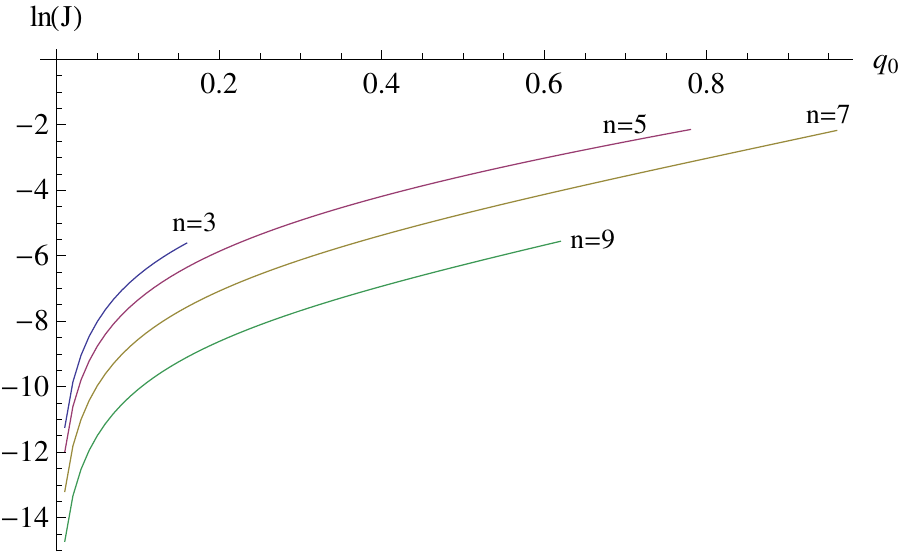}
		}%
	\end{center}
\caption{%
	(a) The natural log of the boson star mass $M$ and (b) angular momentum $J$ plotted against the parameter $q_0$ for $\tilde{\alpha}=1.20$ in $D=5,7,9,11$ ($n=3,5,7,9$).  Both $M$ and $J$ approach exponential growth with increasing $q_0$.
	}%
	\label{fig:120lnMJvq0}
\end{figure}

As before, in order to easily make comparisons between different space-time dimensions, we choose a fixed value of $\tilde{\alpha} = \frac{6}{5}$.  Boson star solutions in this parameter range are qualitatively very different.  Unlike for $\alpha<\alpha_{\text{cr}}$, the spiral behaviour in the $M$ and $J$  curves as functions of $\epsilon$ is not present in any dimension but these quantities instead increase monotonically, as can be seen in figure \ref{fig:120MJvepsilon}.  Similarly, as is seen in figure \ref{fig:120MJvq0}, when the $M$ and $J$ curves are plotted as functions of $q_0$, they also increase monotonically, and plotting the natural log of $M$ and $J$ against $q_0$, s shown in figure \ref{fig:120lnMJvq0}, reveals that the curves approach exponential growth as $q_0$ increases.  At first glance it appears that the growth rate is universal for $D\ge7$; however closer inspection reveals that the growth rate increases slightly with increasing spacetime dimension.  The damped oscillations that appear in the $\alpha<\alpha_{\text{cr}}$ solutions is not present anywhere in these plots.  We omit $\omega$ plotted as a function of $\epsilon$ since, it also increases monotonically with increasing $\epsilon$ and lacks the characteristic spiral behaviour seen for $\alpha<\alpha_{\textrm{cr}}$.

\begin{figure}[ht!]
	\begin{center}
		\subfigure[~$K$ vs $q_0$ for $\tilde{\alpha}=1.20$ in various dimensions.]{%
			\label{fig:120Kvq0}
			\includegraphics[width=0.48\textwidth]{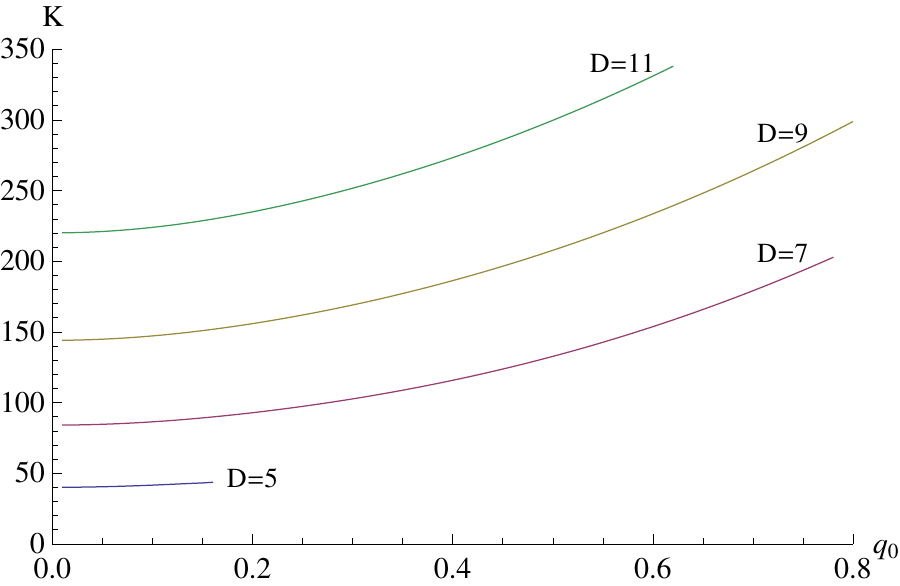}
		}%
		\subfigure[~$q_{h}'^*-q_h'(0)$ vs $q_0$ for $\tilde{\alpha}=1.20$ in various dimensions.]{%
			\label{fig:120qhpstarvq0}
			\includegraphics[width=0.48\textwidth]{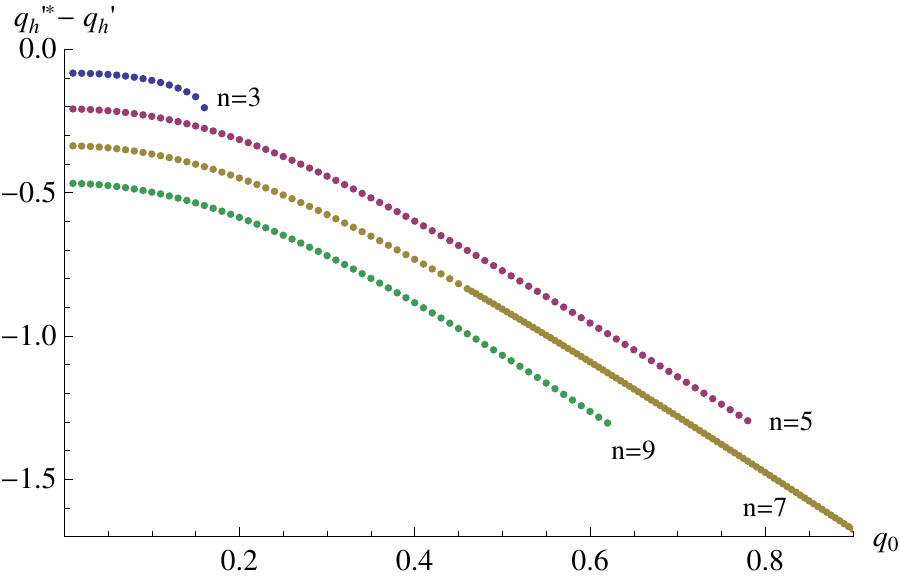}
		}%
	\end{center}
	\caption{%
		(a) The Kretschmann scalar $K$ and (b) $q_{h}'^*-q_h'(0)$ plotted against the parameter $q_0$ for $\tilde{\alpha}=1.20$  in $D=5,7,9,11,$ ($n=3,5,7,9$).
	}%
	\label{fig:120Kqhpstarvq0}
\end{figure}

We again find a difference between boson star behaviour in $D=5$ and $D>5$, namely that there seems to be a maximum value of $q_0$ beyond which the numerics break down.  However, unlike for $\alpha<\alpha_{\mathrm{cr}}$ it can be seen in figure \ref{fig:120Kqhpstarvq0} that in $D=5$ the Kretchmann scalar remains finite for all values of $q_0$, and that $q_{h'}^{*}-q_h'(0)$ is nowhere vanishing and monotonically decreases with increasing $q_0$.  It is unclear whether the breakdown in numerics in $D=5$ is due to a physical effect, but it is clear that these boson stars do not undergo the same drastic change that the $\alpha<\alpha_{\text{cr}}$ solutions do.  A more thorough investigation of this issue is beyond the scope of this paper and is left for future considerations.

\begin{figure}[ht!]
	\begin{center}
		\includegraphics[width=0.48\textwidth]{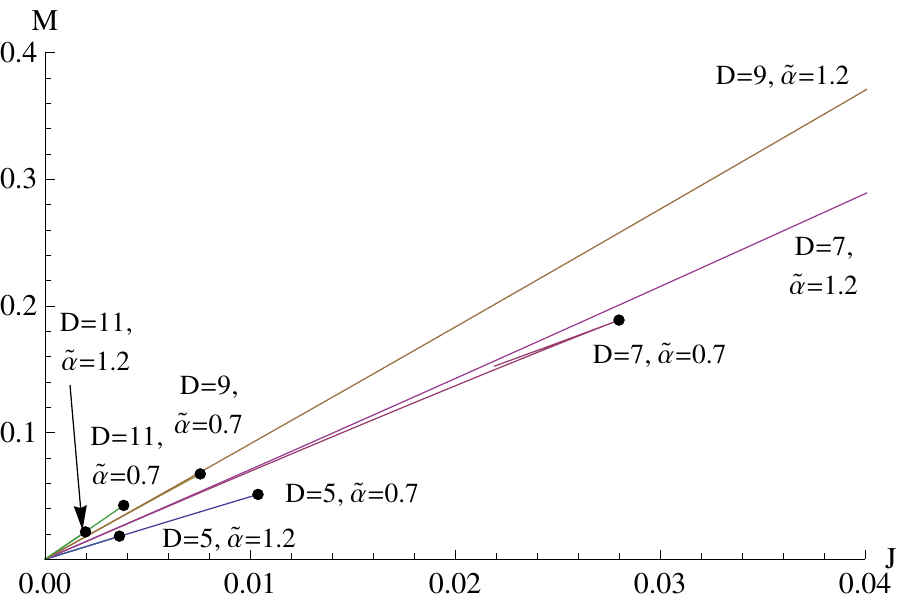}
	\end{center}
	\caption{%
		$M$ vs $J$ for $\tilde\alpha=0.70$ and $\tilde{\alpha}=1.20$ in $D=5,7,9,11$ ($n=3,5,7,9$).  The black dots correspond to the maximum $(M,J)$ value reached on each curve.
	}%
	\label{fig:70120MvJ}
\end{figure}

Finally, in figure \ref{fig:70120MvJ} we plot $M$ versus $J$.  Note that in each dimension the $\alpha>\alpha_{\textrm{cr}}$ solutions are nearly collinear with the $\alpha<\alpha_{\textrm{cr}}$ solutions. However, for $\alpha>\alpha_{\textrm{cr}}$, these lines do not zig-zag and ``turn back'' at any value of $q_0$ due to the lack of damped oscillations in $M$ and $J$.  We choose $\tilde{\alpha}=0.70$ to be representative of the $\alpha<\alpha_{\text{cr}}$ solutions, but any value could have been chosen as they all lie on the same curve (see figure \ref{fig:EvJ}).  It appears that for a specified dimension, the $M$ vs. $J$ curve is universal to all values of $\alpha\neq\alpha_{\text{cr}}$.

Our brief exploration of the $\alpha>\alpha_{\text{cr}}$ parameter space has shown that these boson star solutions have properties that are significantly different from $\alpha<\alpha_{\text{cr}}$, solutions.  This regime is numerically difficult to explore since a large number of Chebyshev points are needed to find solutions for relatively small values of $q_0$. The reason behind this need for increased resolution is not fully understood, as the Kretchmann scalar remains relatively small compared with the $\alpha<\alpha_{\text{crit}}$ case.  Another puzzling feature is that while the $D=5$ case truncates much eariler than the higher dimensional cases, the Kretchmann scalar does not appear the diverge at finite $q_0$ and  $q_h'$ departs from $q_{h}'^{*}$ with increasing $q_0$ in all dimensions.  A full exploration of the reason behind these surprising results will be left for future work.

\medskip
\section{Conclusions}
\label{Conclusion}
 
 We have obtained asymptotically AdS rotating SKV boson star solutions in Einstein-Gauss-Bonnet gravity  coupled to a multiplet of massless scalar fields in all odd spacetime dimensions of interest in string theory, \emph{i.e.} $D=5,7,9,11$, both perturbatively in powers of the scalar field amplitude and numerically.   For the latter, our approach was to employ the same
 relaxation procedure on a Chebyshev grid used in the Einstein case \cite{Stotyn:2013yka,Stotyn:2013spa}.  In all space-time dimensions, the perturbative solutions match the numerical solutions for sufficiently small values of $q_0$, as expected.  Furthermore, each dimension exhibits a critical value of the Gauss-Bonnet coupling constant, $\alpha_{\textrm{cr}}$, for which the solutions no longer have the asymptotic behaviour in (\ref{eq:AsymptoticBC}); numerically we were unable to obtain results for $0.9 \alpha_{\textrm{cr}} < \alpha < \alpha_{\textrm{cr}}$.

The most striking result of the present work is the distinction between the $D=5$ case and its higher dimensional counterparts.  
For $D>5$, the physical quantities ($M$, $J$, $\omega$) undergo damped harmonic oscillations about finite limiting values as $q_0$ increases without bound.  However, for $D=5$ these oscillations are eliminated even for $\alpha = 0.1 \alpha_{\textrm{cr}}$.  Instead, $M$, $J$, and $\omega$ each terminate at a finite critical value $q_{0 \textrm{critical}}$, at which point the Kretschmann scalar diverges.  This situation is redolent of the AdS boson star solutions recently obtained in $D=3$ \cite{Stotyn:2013spa}.  In that case the boson star mass, angular momentum, and angular velocity all monotonically increase with the central energy density up to a critical value, at which point the boson star branch of solutions smoothly connects with the extremal BTZ black hole.  In the present case for $D=5$ there is no evidence that a black hole forms at the critical value of $q_0$, although the diverging Kretschmann scalar indicates the formation of a singularity; this boson star branch is likely dynamically unstable to forming a black hole at some value $q_0<q_{0 \textrm{critical}}$.

Recall that demanding the field equations to have derivatives no higher than 2nd order, the Ricci scalar is the highest-curvature term that can appear in 3 space-time dimensions, while the Gauss-Bonnet term is the highest-curvature term in 5 dimensions. We conjecture that  critical values of $q_0$ will  appear in all odd dimensions in which the gravitational theory includes its highest possible curvature term that maintains 2nd order field equations.  It is conceivably feasible to numerically check this 
for the $D=7$ 3rd-order Lovelock and $D=9$ 4th order Lovelock cases. It would also be interesting to extend our work to that of massive scalar fields, and scalar fields with potential terms to see what effect the scalar field mass has on the critical value of $q_0$.

As with the Einstein case \cite{Dias:2011ss}, one might expect asymptotically AdS space-times in Einstein-Gauss-Bonnet gravity to be non-perturbatively unstable to the formation of black holes.  Heuristically, global AdS is non-perturbatively unstable to  black hole formation because its reflecting boundary conditions imply that finite energy perturbations, given enough time, will come together in sufficient concentration to form a black hole.  Since Gauss-Bonnet gravity tends to increase gravitational attraction, we expect such an effect to be enhanced, though a proper study remains to be carried out. Indeed, since it has been shown that the AdS instability is really due to the high level of symmetry present in global AdS (the normal mode frequencies are all resonant with the AdS frequency \cite{Dias:2012tq}) we expect the boson star solutions constructed in this paper to be non-perturbatively stable, although verifying this explicitly is beyond the scope of the present work.

Further work includes a proper analysis of the $  \alpha > \alpha_{\textrm{cr}}$ and $  \alpha = \alpha_{\textrm{cr}}$ cases.  
We have given a brief discussion of the former case.  It is straightforward to show that transverse traceless 
excitations of the metric about an AdS vacuum obey the equation  
\begin{equation}
\left(1-\frac{\alpha}{\alpha_{\text{cr}}} \right)\left( \nabla^2 h_{a b} +\frac{2}{\ell^2} h_{ab}\right) =  8\pi G T_{ab}
\end{equation}
indicating that for $  \alpha > \alpha_{\textrm{cr}}$ such excitations are ghostlike \cite{Myers:2010ru}.
However the boson star solutions we obtain have positive mass.  While presumably unstable, the details of this instability merit further study.  The $\alpha = \alpha_{\textrm{cr}}$ requires a separate analysis due to the different boundary conditions that must be employed.  It will be interesting to see what features solutions in this class have in common with the $D=3$ case \cite{Stotyn:2013yka}.

 {\bf Note added} As this paper was nearing completion, we became aware of similar work
in $D=5$ \cite{Brihaye:2013zha}, in which rotating  EGB boson stars were studied 
for  scalar fields in a potential. The zero-potential case corresponds to the one we study, though
they evidently discuss only the $\alpha=0$ case when rotation is present and the potential vanishes.  For nonzero potential they also find in EGB gravity that rotating solutions exist up to some maximal value of the central energy density, though they claim the limiting solutions for this case are regular, in contrast to the massless case that we consider here.

\FloatBarrier

\section*{Acknowledgements}

This work was supported in part by the Natural Sciences and Engineering
Research Council of Canada.  We are grateful to R. Myers for helpful discussions.

\appendix

\section{Field Equations}
\label{FEs}
Here we record the field equations we employ in obtaining both our perturbative and numerical solutions.  While it is
possible to rewrite these so that the second-derivative terms are isolated, this produces expressions even more cumbersome
than those presented here, and so we do not carry out this step.  There are 5 coupled second order ODEs for the 5 metric and scalar field functions as follows:
 
\begin{align}
&r^2 \Bigg[2 g \left(f^2 g h (n-2) (n-1)-2 h \Pi ^2 r^2 (\omega -\Omega )^2\right)\nonumber\\
&+f h r \Bigg(g r \left(2 g f''+2 f g''+4 f g \left(\Pi '\right)^2-3 h r^2 \left(\Omega '\right)^2\right)+g f' \left(3 r g'+4 g (n-1)\right)+2 f g (n-1) g'-f r \left(g'\right)^2\Bigg)\nonumber\\
&+2 f g^2 \bigg(h \left(3 h (n-1)-n^2+2 \Lambda  r^2+1\right)+2 \Pi ^2 (h (n-1)-1)\bigg)\Bigg]\nonumber\\
&+2 \alpha  f h (1-n) \Bigg[g^2 (n-3) \Big(f^2 (n-4) (n-2)+2 f (n-4) (3 h-n-1)+3 h (5 h-2 (n+1))+n^2-1\Big)\nonumber\\
&+r \Bigg(g r \Big(2 \left(g f''+f g''\right) (f (n-2)+3 h-n-1)-3 h r^2 \left(\Omega '\right)^2 (f
   (n-2)+5 h-n-1)\Big)\nonumber\\
&+2 g^2 (n-2) r \left(f'\right)^2+g f' \Big(r g' (5 f (n-2)+9 h-3 (n+1))+4 g (n-3) (f (n-2)+3 h-n-1)\Big)\nonumber\\
&+f r \left(g'\right)^2 (-f (n-2)-3 h+n+1)+2 f g (n-3) g' (f (n-2)+3
   h-n-1)\Bigg)\Bigg] = 0
\label{CC}
\end{align}

\begin{align}
&f^2 h r^4 \Omega\left(g'\right)^2+f g^2 r^2 \Bigg(r \Big(2 h r \left(f \Omega ''-\Omega  f''\right)+f \Omega ' \left(3 r h'+2 h (n+2)\right)\Big)-4 h (n-1) r \Omega  f'\nonumber\\
&-2 \Omega  \Big(h (n-1) (f (n-2)+3
   h-n-1)+2 \Pi ^2 (h (n-1)+1)+2 h \Lambda  r^2\Big)-4 f h r^2 \Omega  \left(\Pi '\right)^2+8 \Pi ^2 \omega \Bigg)\nonumber\\
&+g h r^3 \Bigg(r \Omega  \left(-2 f^2 g''+3 f h r^2
   \left(\Omega '\right)^2+4 \Pi ^2 (\omega -\Omega )^2\right)-f g' \left(3 r \Omega  f'+2 f (n-1) \Omega +f r \Omega '\right)\Bigg)\nonumber\\
&+2 \alpha  f (n-1) \Bigg\{g^2 h (n-3) \Omega  \Big(f^2 (n-4) (n-2)+2 f (n-4) (3 h-n-1)+15
   h^2-6 h (n+1)+n^2-1\Big)\nonumber\\
&+r \Bigg[g \bigg(2 h r ((f-1) n-2 f+3 h-1) \left(g \Omega  f''+f \Omega  g''-f g \Omega ''\right)\nonumber\\
&+(f (-n)+2 f-5 h+n+1)
 \left(3 f g r h' \Omega '+3 h^2 r^3 \Omega  \left(\Omega
   '\right)^2\right)\bigg)+2 g^2 h (n-2) r \Omega  \left(f'\right)^2\nonumber\\
&+g h f' \bigg(r \Omega  g' (5 f (n-2)+9 h-3 n-3)
+4 g (n-3) \Omega  ((f-1) n-2 f+3 h-1)-2 f g (n-2) r \Omega '\bigg)\nonumber\\
&+2 f g^2 h n \Omega ' (f
   (-n)+2 f-3 h+n+1)
+f g h g' (f (n-2)+3 h-n-1) \left(2 (n-3) \Omega +r \Omega '\right)\nonumber\\
&+f h r \Omega  \left(g'\right)^2 (-f (n-2)-3 h+n+1)\Bigg]\Bigg\}= 0
\label{CT}
\end{align}

\begin{align}
 &r^2\Bigg\{2 g h \bigg(-2 h^2 \Pi ^2 (\omega -\Omega )^2 \Omega ^2 r^4
+f g h \Big(2 \left(-\omega ^2-2 \Omega  \omega +(h (n-1)+2) \Omega ^2\right) \Pi ^2\nonumber\\
&+h \left(-n^2+3 h (n-1)+2 r^2 \Lambda +1\right) \Omega
   ^2\Big) r^2-f^3 g^2 h (n-1) n\nonumber\\
&+f^2 g \Big(-2 g (h (n-1)+1) \Pi ^2+h^2 (n-2) (n-1) r^2 \Omega ^2+g h \left(n^2-h (n-1)-2 r^2 \Lambda -1\right)\Big)\bigg)\nonumber\\
&-f r \Bigg[f r^3 \Omega ^2 \left(g'\right)^2 h^3-2
   f g r^2 \Omega  g' \left((n-1) \Omega +r \Omega '\right) h^3\nonumber\\
&+g f' \Big(g \left(2 h \left(f g n-2 h (n-1) r^2 \Omega ^2\right)+f g r h'\right)-3 h^2 r^3 \Omega ^2 g'\Big) h\nonumber\\
&+g \Bigg(-f^2 g^2 r\left(h'\right)^2
+2 f g h \left(3 h \Omega  \Omega ' r^3+f g (n+1)\right) h'+h r \bigg(2 f^2 h'' g^2+4 f h \left(f g-h r^2 \Omega ^2\right) \left(\Pi '\right)^2 g\nonumber\\
&+4 f h^2 (n+2) r \Omega  \Omega ' g+h^2 r^2
   \left(3 h r^2 \Omega ^2+f g\right) \left(\Omega '\right)^2-2 h^2 r^2 \Omega  \left(g \Omega  f''+f \Omega  g''-2 f g \Omega ''\right)\bigg)\Bigg)\Bigg]\Bigg\}\nonumber\\
&+2 f (n-1) \alpha  \Bigg\{-2 g^2 h^3 (n-2)
   \Omega ^2 \left(f'\right)^2 r^4+f h^3 (3 h+f (n-2)-n-1) \Omega ^2 \left(g'\right)^2 r^4\nonumber\\
&-2 f g h^3 (3 h+f (n-2)-n-1) \Omega  g' \left((n-3) \Omega +r \Omega '\right) r^3\nonumber\\
&+g h f' \Bigg[h^2 (-9 h-5 f (n-2)+3
   n+3) \Omega ^2 g' r^3+g \bigg(f g (3 h+3 f (n-2)-n-1) r h'\nonumber\\
&+2 h \Big(2 f h (n-2) \Omega  \Omega ' r^3-2 h (n-3) (-2 f+3 h+(f-1) n-1) \Omega ^2 r^2+f g (n-2) (h+(f-1) n-1)\Big)\bigg)\Bigg] r\nonumber\\
&+g \Bigg[g
   (n-3) \bigg(f g \left((f-1)^2 n^2+2 (f-1) (h-f) n+(h-1) (-4 f+3 h+1)\right)\nonumber\\
&-h \Big((n-4) (n-2) f^2+2 (3 h-n-1) (n-4) f+15 h^2+n^2-6 h (n+1)-1\Big) r^2 \Omega ^2\bigg) h^2\nonumber\\
&+r \Bigg(f^2 g^2 (3 h-f
   (n-2)+n+1) r \left(h'\right)^2\nonumber\\
&+2 f g h \Big(3 h (-2 f+5 h+(f-1) n-1) \Omega  \Omega ' r^3+f g (n-1) (-2 f+3 h+(f-1) n-1)\Big) h'\nonumber\\
&+h r \bigg(r^2 \left(3 h (5 h+f (n-2)-n-1) r^2 \Omega ^2+f g (3 h+f
   (n-2)-n-1)\right) \left(\Omega '\right)^2 h^2\nonumber\\
&+4 f g (3 h+f (n-2)-n-1) n r \Omega  \Omega ' h^2\nonumber\\
&+2 (3 h+f (n-2)-n-1) \Big(f \left(f h'' g^2+h^2 r^2 \Omega  \left(2 g \Omega ''-\Omega  g''\right)\right)-g h^2
   r^2 \Omega ^2 f''\Big)\bigg)\Bigg)\Bigg]\Bigg\} = 0
\label{TT}
\end{align}     
  
\begin{align}
&r^2 \Bigg\{2 f h \Big(2 \Pi ^2+h \big(2 \Lambda  r^2+2 (n-3) \Pi ^2+h (n-5)+(3-n) (n+1)\big)\Big) g^2\nonumber\\
&+2 h \Big(f^2 g h (n-2) (n-1)-2 h r^2 \Pi ^2 (\omega -\Omega )^2\Big) g\nonumber\\
&+f r \Bigg[-f h^2 r
   \left(g'\right)^2+f g h \left(2 h (n-1)+r h'\right) g'+g h f' \left(4 g h (n-1)+3 h r g'+2 g r h'\right)\nonumber\\
&+g \bigg(-f g r \left(h'\right)^2+2 f g h n h'+h r \left(h \left(4 f g \left(\Pi '\right)^2-h r^2
   \left(\Omega '\right)^2+2 g f''+2 f g''\right)+2 f g h''\right)\bigg)\Bigg]\Bigg\}\nonumber\\
&-2 f \alpha\Bigg\{g^2 \Big((n-4) (n-2) (n-1) (f-1)^2+2 (h-1) (n-5) (n-4) (f-1)+3 (h-1)^2 (n-9)\Big) (n-3) h^2\nonumber\\
&+r \Bigg[2
   g^2 h (n-2) r \left(h (n-1)+r h'\right) \left(f'\right)^2+g \Bigg(4 g (n-3) \big(h (n-5)+f (n-2) (n-1)+(3-n) (n+1)\big) h^2\nonumber\\
&+r \bigg(-f g (n-2) r \left(h'\right)^2+2 g h \big(3 h (n-3)+(-n-1) (n-3)+f (n-2) (3 n-7)\big) h'+h g'\nonumber\\
&+2 (n-2) \left(h^3 \left(\Omega '\right)^2 r^3+f g h h'' r\right)\bigg)\Bigg) f'\nonumber\\
&-f h r \left(g'\right)^2 \Big(h \big(h (n-5)+f (n-2)
   (n-1)+(3-n) (n+1)\big)+f (n-2) r h'\Big)\nonumber\\
&+f g' \Bigg(2 g \big(h (n-5)+(-n-1) (n-3)+f (n-2) (n-1)\big) (n-3) h^2\nonumber\\
&+r \bigg(-h^3 (n-2) \left(\Omega '\right)^2 r^3
-f g (n-2) \left(h'\right)^2 r+2 f g h (n-2) h'' r+g h \big(3 h
   (n-3)+(3-n) (n+1)\nonumber\\
&+f (n-2) (3 n-5)\big) h'\bigg)\Bigg)\nonumber\\
&+g \Bigg(f g (n-3) (3 h-f (n-2)+n+1) r \left(h'\right)^2\nonumber\\
&+f h (n-2) \left(3 h \left(\Omega '\right)^2 r^4+2 \left(g f''+f g''\right) r^2+2 g (3 h+f
   (n-2)-n-1) (n-3)\right) h'\nonumber\\
&+h r \bigg(4 f h^2 (n-2) \Omega ' \Omega '' r^3+h^2 (-3 h (n-5)+(n-3) (n+1)+f (n-2) (n+3)) \left(\Omega '\right)^2 r^2+\nonumber\\
&2 h \big(h (n-5)+f (n-2) (n-1)+(3-n) (n+1)\big) \left(g f''+f
   g''\right)
+f g (n-3) (6 h+2 f (n-2)\nonumber\\
&-2 (n+1)) h''\bigg)\Bigg)\Bigg]\Bigg\}=0
\label{XX}
\end{align}
    
\begin{equation}
{
\Pi''+\frac{\Pi'}{2fr}\left(2nfgh+2rgh f^\prime +rf(gh)^\prime \right)+\frac{\Pi(\omega-\Omega)^{2}}{f^{2}g}-\frac{\big(1+(n-1)h\big)\Pi}{fhr^{2}}=0}
\label{eq:PiEq}
\end{equation}

where a $'$ denotes differentiation with respect to $r$.  In addition to these second order ODEs, there is a first order ODE in the form of a constraint equation, given by:

\begin{align}
&r^2 \Bigg[4 \left(\frac{1}{2} f^2 g h (n-1) n+f g \left(h \left(\frac{1}{2} (n-1) \left(h-n+2 \Pi ^2-1\right)+\Lambda  r^2\right)+\Pi ^2\right)-h \Pi ^2 r^2 (\omega -\Omega )^2\right)\nonumber\\
&+f r \left(\left(f g\right)' \left(r h'+2 h n\right)+4 f g \left(\frac{1}{2} (n-1) h'-h r \left(\Pi '\right)^2\right)\right)+f h^2 r^4 \left(\Omega '\right)^2\Bigg]\nonumber\\
&-2 \alpha  f (n-1) \Bigg[r \left(f g\right)' \Big(r h'
   (3 f (n-2)+3 h-n-1)+2 h (n-2) (f n+h-n-1)\Big)\nonumber\\
&+g (n-3) \Big(2 f r h' \big(f (n-2)+3 h-n-1\big)+h \big(2 (f-1) (h-1) (n-2)+(f-1)^2 (n-2) n+3 (h-1)^2\big)\Big)\nonumber\\
&+h^2 r^4 \left(\Omega '\right)^2 (3 f (n-2)+3 h-n-1)\Bigg] = 0
\label{RR}
\end{align}

\section{Perturbative Fields}
\label{pertanswers}

In this Appendix we catalogue all of the perturbative gravitational and scalar field solutions to the field equations  in spacetime dimension $D=n+2$ for $n=3,5,7,9$.  The fields are labeled as $F_{n;p}$, where $p$ denotes the order in $\epsilon$.

\subsection*{n=3 results}
\begin{align}
f_{3;2}={}&\frac{6 \ell^4+20 \ell^2 r^2+5 r^4}{9}\nonumber\\
\Omega_{3;2}={}&\frac{6 \ell^4+4 \ell^2 r^2+r^4}{12}\nonumber\\
\Pi_{3;3}={}&\frac{1540 \ell^6+5548 \ell^4 r^2+3935 \ell^2 r^4+900 r^6}{2016}\nonumber\\
 f_{3;4}={}&\Bigg[\frac{\ell^2}{1270080} \bigg(1315860 \ell^{14}+11249595 \ell^{12} r^2+36154839 \ell^{10} r^4+51954798 \ell^8 r^6+40913902 \ell^6 r^8\nonumber\\&
+18512283 \ell^4 r^{10}+4631027 \ell^2 r^{12}+514952 r^{14}\bigg)\nonumber\\&
-\frac{\alpha}{4762800}\bigg(38080980 \ell^{14}+165606315 \ell^{12} r^2+564058887 \ell^{10} r^4+809407494 \ell^8 r^6\nonumber\\&
+616096806 \ell^6 r^8+272183829 \ell^4 r^{10}+67620861 \ell^2 r^{12}+7535096 r^{14}\bigg)\Bigg]\nonumber\\
g_{3;4}={}&\Bigg[\frac{\ell^2}{1270080} \bigg(872290 \ell^{16}+7971520 \ell^{14} r^2+25939155 \ell^{12} r^4+34403186 \ell^{10} r^6+16099475 \ell^8 r^8\nonumber\\&
+3052440 \ell^6 r^{10}+149055 \ell^4 r^{12}+35410 \ell^2 r^{14}+3541 r^{16}\bigg)\nonumber\\&
-\frac{\alpha}{1587600}\bigg(5891650 \ell^{16}+47634040 \ell^{14} r^2+173336925 \ell^{12} r^4+217174806 \ell^{10} r^6\nonumber\\&
+98887425 \ell^8 r^8+21549240 \ell^6 r^{10}+2947095 \ell^4 r^{12}+650010 \ell^2 r^{14}+65001 r^{16}\bigg)\Bigg]\nonumber\\
h_{3;4}={}&\Bigg[\frac{\ell^2 \left(22260 \ell^{12}+183645 \ell^{10} r^2+311661 \ell^8 r^4+260694 \ell^6 r^6+123066 \ell^4 r^8+31869 \ell^2 r^{10}+3541 r^{12}\right)}{1270080}\nonumber\\&
-\frac{\alpha\left(1604820 \ell^{12}+2949765 \ell^{10} r^2+2921877 \ell^8 r^4+1877358 \ell^6 r^6+787362 \ell^4 r^8+195003 \ell^2 r^{10}+21667 r^{12}\right)}{529200}\Bigg]\nonumber\\
\Omega_{3;4}={}&\Bigg[\frac{\ell^2}{2540160} \bigg(1598455 \ell^{14}+8795055 \ell^{12} r^2+19004760 \ell^{10} r^4+20982192 \ell^8 r^6+14139048 \ell^6 r^8\nonumber\\&
+6020712 \ell^4 r^{10}+1505178 \ell^2 r^{12}+167242 r^{14}\bigg)\nonumber\\&
-\frac{\alpha}{635040}\bigg(2196535 \ell^{14}+11355375 \ell^{12} r^2+22425240 \ell^{10} r^4+22910832 \ell^8 r^6+14668248 \ell^6 r^8\nonumber\\&
+6079512 \ell^4 r^{10}+1505178 \ell^2 r^{12}+167242 r^{14}\bigg)\Bigg]\nonumber\\
\Pi_{3;5}={}&\Bigg[\frac{\ell^2}{853493760} \bigg(1128452101 \ell^{18}+10678880150 \ell^{16} r^2+41756607180 \ell^{14} r^4+88872056182 \ell^{12} r^6\nonumber\\&
+115794392980 \ell^{10} r^8+98320298706 \ell^8 r^{10}+55586393870 \ell^6 r^{12}+20395866890 \ell^4 r^{14}\nonumber\\&
+4416801537 \ell^2
   r^{16}+428716940 r^{18}\bigg)\nonumber\\&
-\frac{\alpha}{152562009600}\bigg(928722217423 \ell^{18}+8819401225090 \ell^{16} r^2+33780454205940 \ell^{14} r^4\nonumber\\&
+71249955470946 \ell^{12} r^6+92323311236940 \ell^{10} r^8+78281035435638 \ell^8 r^{10}+44381025035130 \ell^6 r^{12}\nonumber\\&
+16375305364670 \ell^4
   r^{14}+3570276679411 \ell^2 r^{16}+349126822500 r^{18}\bigg)\Bigg]\nonumber
\end{align}

\subsection*{n=5 results}
\begin{align}
f_{5;2}={}&\frac{20 \ell^6+105 \ell^4 r^2+42 \ell^2 r^4+7 r^6}{50}\nonumber\\
\Omega_{5;2}={}&\frac{20 \ell^6+15 \ell^4 r^2+6 \ell^2 r^4+r^6}{60}\nonumber\\
\Pi_{5;3}={}&\frac{152581 \ell^{10}+767826 \ell^8 r^2+1035891 \ell^6 r^4+773598 \ell^4 r^6+313026 \ell^2 r^8+53970 r^{10}}{257400}\nonumber\\
f_{5;4}={}&\Bigg[\frac{\ell^2}{618377760000} \bigg(295557662400 \ell^{20}+3682611572640 \ell^{18} r^2+17103932986140 \ell^{16} r^4\nonumber\\&
+36498131558460 \ell^{14} r^6+47555951613885 \ell^{12} r^8+41236810544215 \ell^{10} r^{10}+24529389790620 \ell^8 r^{12}\nonumber\\&
+10021950443882
   \ell^6 r^{14}+2733953554501 \ell^4 r^{16}+455760809811 \ell^2 r^{18}+35065241462 r^{20}\bigg)\nonumber\\&
-\frac{\alpha}{360720360000}\bigg(5730125681280 \ell^{20}+54710235259680 \ell^{18} r^2+261420130519548 \ell^{16} r^4\nonumber\\&
+542618701892124 \ell^{14} r^6+683370729631161 \ell^{12} r^8+581979609281499 \ell^{10} r^{10}+343653106261036 \ell^8 r^{12}\nonumber\\&
+140092289643826 \ell^6
   r^{14}+38200555296657 \ell^4 r^{16}+6369004169047 \ell^2 r^{18}+490071416494 r^{20}\bigg)\Bigg]\nonumber\\
g_{5;4}={}&\Bigg[\frac{\ell^2}{618377760000} \bigg(198795488864 \ell^{22}+2586519623296 \ell^{20} r^2+11564746522784 \ell^{18} r^4\nonumber\\&
+22932404913236 \ell^{16} r^6+21539953150144 \ell^{14} r^8+12564927910535 \ell^{12} r^{10}+4244040931130 \ell^{10} r^{12}\nonumber\\&
+677658846865
   \ell^8 r^{14}+31598878220 \ell^6 r^{16}+7946956745 \ell^4 r^{18}+1222608730 \ell^2 r^{20}+87329195 r^{22}\bigg)\nonumber\\&
-\frac{\alpha}{360720360000}\bigg(3517759244608 \ell^{22}+44211086834432 \ell^{20} r^2+200305583802688 \ell^{18} r^4\nonumber\\&
+363929580852292 \ell^{16} r^6+319063756957088 \ell^{14} r^8+180755354654875 \ell^{12} r^{10}\nonumber\\&
+61031317677330 \ell^{10} r^{12}+10223405437245
   \ell^8 r^{14}+700305076380 \ell^6 r^{16}+175106329125 \ell^4 r^{18}\nonumber\\&
+26939435250 \ell^2 r^{20}+1924245375 r^{22}\bigg)\Bigg]\nonumber\\
h_{5;4}={}&\Bigg[\frac{\ell^2}{123675552000} \bigg(769728960 \ell^{18}+9390693312 \ell^{16} r^2+20983798836 \ell^{14} r^4+25938182556 \ell^{12} r^6\nonumber\\&
+21295887327 \ell^{10} r^8+12277654675 \ell^8 r^{10}+4978052794 \ell^6 r^{12}+1362335442 \ell^4 r^{14}\nonumber\\&
+227055907 \ell^2
   r^{16}+17465839 r^{18}\bigg)\nonumber\\&
-\frac{\alpha}{369969600}\bigg(1593271680 \ell^{18}+3122255136 \ell^{16} r^2+3848592748 \ell^{14} r^4+3585574564 \ell^{12} r^6\nonumber\\&
+2592129969 \ell^{10} r^8+1419459613 \ell^8 r^{10}+565061926 \ell^6 r^{12}+153939630 \ell^4 r^{14}\nonumber\\&
+25656605 \ell^2 r^{16}+1973585
   r^{18}\bigg)\Bigg]\nonumber\\
\Omega_{5;4}={}&\Bigg[\frac{\ell^2}{46378332000} \bigg(15058082990 \ell^{20}+106176159440 \ell^{18} r^2+311179210446 \ell^{16} r^4\nonumber\\&
+505325794688 \ell^{14} r^6+547149789758 \ell^{12} r^8+420702504651 \ell^{10} r^{10}+234379619045 \ell^8 r^{12}\nonumber\\&
+93593083298 \ell^6
   r^{14}+25525386354 \ell^4 r^{16}+4254231059 \ell^2 r^{18}+327248543 r^{20}\bigg)\nonumber\\&
-\frac{\alpha}{1932430500}\bigg(19521646094 \ell^{20}+131737647392 \ell^{18} r^2+353356087188 \ell^{16} r^4\nonumber\\&
+537690140702 \ell^{14} r^6+562006315442 \ell^{12} r^8+424922932863 \ell^{10} r^{10}+235083023747 \ell^8 r^{12}\nonumber\\&
+93647191352 \ell^6 r^{14}+25525386354 \ell^4
   r^{16}+4254231059 \ell^2 r^{18}+327248543 r^{20}\bigg)\Bigg]\nonumber\\
\Pi_{5;5}={}&\Bigg[\frac{\ell^2}{453635740958400000} \bigg(337633104499816268 \ell^{26}+4271077136958547132 \ell^{24} r^2\nonumber\\&
+23112600885726752792 \ell^{22} r^4+72350175319616674844 \ell^{20} r^6+151159034180233581556 \ell^{18} r^8\nonumber\\&
+228077768805842399885 \ell^{16}
   r^{10}+258620332265962810395 \ell^{14} r^{12}+223871564194330948248 \ell^{12} r^{14}\nonumber\\&
+147920037831416512914 \ell^{10} r^{16}+73557029936994344181 \ell^8 r^{18}+26698941169899723207 \ell^6 r^{20}\nonumber\\&
+6684077146010747418 \ell^4
   r^{22}+1032668691070620996 \ell^2 r^{24}+74253000574956420 r^{26}\bigg)\nonumber\\&
-\frac{\alpha}{4498554431170800000}\bigg(88843628024978738912 \ell^{26}+1122507865343630559388 \ell^{24} r^2\nonumber\\&
+6023372115583107270188 \ell^{22} r^4+18720948444225766469276 \ell^{20} r^6\nonumber\\&
+38914019934303990489424 \ell^{18} r^8+58601438194884845712305 \ell^{16}
   r^{10}\nonumber\\&
+66469446985987754507715 \ell^{14} r^{12}+57615205604166626638152 \ell^{12} r^{14}\nonumber\\&
+38129098662142936918386 \ell^{10} r^{16}+18989139612729591541089 \ell^8 r^{18}\nonumber\\&
+6901386814262466690003 \ell^6
   r^{20}+1729600335134757290322 \ell^4 r^{22}\nonumber\\&
+267442276275622329084 \ell^2 r^{24}+19242295861536333180 r^{26}\bigg)\Bigg]\nonumber
\end{align}

\subsection*{n=7 results}
\begin{align}
f_{7;2}={}&\frac{70 \ell^8+504 \ell^6 r^2+252 \ell^4 r^4+72 \ell^2 r^6+9 r^8}{245}\nonumber\\
\Omega_{7;2}={}&\frac{70 \ell^8+56 \ell^6 r^2+28 \ell^4 r^4+8 \ell^2 r^6+r^8}{280}\nonumber\\
\Pi_{7;3}={}&\frac{1}{62375040}\bigg(28301260 \ell^{14}+181648328 \ell^{12} r^2+342224260 \ell^{10} r^4+415360056 \ell^8 r^6+331552839 \ell^6 r^8\nonumber\\&
+168013320 \ell^4 r^{10}+48974940 \ell^2 r^{12}+6252120 r^{14}\bigg)\nonumber\\
f_{7;4}={}&\Bigg[\frac{\ell^2}{371502619488000} \bigg(96780316472840 \ell^{26}+1589105610832740 \ell^{24} r^2+9521410725967140 \ell^{22} r^4\nonumber\\&
+25965374820716045 \ell^{20} r^6+45556296988624565 \ell^{18} r^8+56539135502871164 \ell^{16} r^{10}\nonumber\\&
+51618427799679364 \ell^{14}
   r^{12}+35302266172890626 \ell^{12} r^{14}+18211924445078266 \ell^{10} r^{16}\nonumber\\&
+7062641146585640 \ell^8 r^{18}+2018090265782824 \ell^6 r^{20}+403651927726529 \ell^4 r^{22}\nonumber\\&
+50460227131621 \ell^2 r^{24}+2968444009876
   r^{26}\bigg)\nonumber\\&
-\frac{\alpha}{835880893848000}\bigg(16866537231824280 \ell^{26}+232049118692571180 \ell^{24} r^2+1414110709187636460 \ell^{22} r^4\nonumber\\&
+3716883650942351175 \ell^{20} r^6+6304111318629442175 \ell^{18} r^8+7700650359344929844 \ell^{16} r^{10}\nonumber\\&
+6986591021140767244
   \ell^{14} r^{12}+4767825442882086806 \ell^{12} r^{14}+2458034340313267006 \ell^{10} r^{16}\nonumber\\&
+953081702075970200 \ell^8 r^{18}+272332684743145624 \ell^6 r^{20}+54472333166578679 \ell^4 r^{22}\nonumber\\&
+6809680934566771 \ell^2
   r^{24}+400602893601676 r^{26}\bigg)\Bigg]\nonumber\\
g_{7;4}={}&\Bigg[\frac{\ell^2}{371502619488000} \bigg(63735166423850 \ell^{28}+1080779423604460 \ell^{26} r^2+6104455301649630 \ell^{24} r^4\nonumber\\&
+15807036948862760 \ell^{22} r^6+21693866896535335 \ell^{20} r^8+20860826548147390 \ell^{18} r^{10}\nonumber\\&
+13769626720091151 \ell^{16}
   r^{12}+5953776409345128 \ell^{14} r^{14}+1535551141626138 \ell^{12} r^{16}\nonumber\\&
+193867029381396 \ell^{10} r^{18}+10157293967310 \ell^8 r^{20}+2709965597136 \ell^6 r^{22}+508118549463 \ell^4 r^{24}\nonumber\\&
+59778652878 \ell^2 r^{26}+3321036271
   r^{28}\bigg)\nonumber\\&
-\frac{\alpha}{835880893848000}\bigg(10750577761041750 \ell^{28}+178690917717167220 \ell^{26} r^2+1006526355431407410 \ell^{24} r^4\nonumber\\&
+2361158122174494840 \ell^{22} r^6+3036925736949961845 \ell^{20} r^8+2850729294425536090 \ell^{18} r^{10}\nonumber\\&
+1868492713518324621
   \ell^{16} r^{12}+807995702469461688 \ell^{14} r^{14}+209605974589864398 \ell^{12} r^{16}\nonumber\\&
+27209163001515516 \ell^{10} r^{18}+1738728305273610 \ell^8 r^{20}+463697435964336 \ell^6 r^{22}+86943269243313 \ell^4 r^{24}\nonumber\\&
+10228619910978
   \ell^2 r^{26}+568256661721 r^{28}\bigg)\Bigg]\nonumber\\
h_{7;4}={}&\Bigg[\frac{\ell^2}{53071802784000} \bigg(153269396280 \ell^{24}+2477855239860 \ell^{22} r^2+6500468085540 \ell^{20} r^4+9976953706005 \ell^{18} r^6\nonumber\\&
+10884488013685 \ell^{16} r^8+9029812070708 \ell^{14} r^{10}+5829937982596 \ell^{12} r^{12}+2930312879858
   \ell^{10} r^{14}\nonumber\\&
+1128813864010 \ell^8 r^{16}+322614952040 \ell^6 r^{18}+64522990408 \ell^4 r^{20}+8065373801 \ell^2 r^{22}+474433753 r^{24}\bigg)\nonumber\\&
-\frac{\alpha}{17058793752000}\bigg(86486350630680 \ell^{24}+182245183220100 \ell^{22} r^2+260482463266740 \ell^{20} r^4\nonumber\\&
+301178906390505 \ell^{18} r^6+289138698210505 \ell^{16} r^8+227450874603524 \ell^{14} r^{10}+143878222471828 \ell^{12} r^{12}\nonumber\\&
+71803208222954
   \ell^{10} r^{14}+27603213721570 \ell^8 r^{16}+7886010815720 \ell^6 r^{18}+1577202163144 \ell^4 r^{20}\nonumber\\&
+197150270393 \ell^2 r^{22}+11597074729 r^{24}\bigg)\Bigg]\nonumber\\
\Omega_{7;4}={}&\Bigg[\frac{\ell^2}{25474465336320} \bigg(4684576834175 \ell^{26}+39868495607983 \ell^{24} r^2+144325351642808 \ell^{22} r^4\nonumber\\&
+301265955958360 \ell^{20} r^6+438282129396128 \ell^{18} r^8+473620370087280 \ell^{16} r^{10}+391916979706512 \ell^{14}
   r^{12}\nonumber\\&
+252029223792528 \ell^{12} r^{14}+126161196878808 \ell^{10} r^{16}+48499288297880 \ell^8 r^{18}+13856939513680 \ell^6 r^{20}\nonumber\\&
+2771387902736 \ell^4 r^{22}+346423487842 \ell^2 r^{24}+20377852226
   r^{26}\bigg)\nonumber\\&
-\frac{\alpha}{2122872111360}\bigg(30060965769595 \ell^{26}+246684097210475 \ell^{24} r^2+809950806744472 \ell^{22} r^4\nonumber\\&
+1586524546896824 \ell^{20} r^6+2238845007906592 \ell^{18} r^8+2387640938645040 \ell^{16} r^{10}\nonumber\\&
+1965266521267920 \ell^{14}
   r^{12}+1261282443509712 \ell^{12} r^{14}+630948024962424 \ell^{10} r^{16}\nonumber\\&
+242504796816952 \ell^8 r^{18}+69284697568400 \ell^6 r^{20}+13856939513680 \ell^4 r^{22}+1732117439210 \ell^2 r^{24}\nonumber\\&
+101889261130 r^{26}\bigg)\Bigg]\nonumber\\
\Pi_{7;5}={}&\Bigg[\frac{\ell^2}{21848348422059134976000} \bigg(8982162145045598581865 \ell^{34}+141066092907373532097690 \ell^{32} r^2\nonumber\\&
+957479816306705256586845 \ell^{30} r^4+3856281625419549220592040 \ell^{28} r^6\nonumber\\&
+10789673943068638679272455 \ell^{26}
   r^8+22792336247876448363452550 \ell^{24} r^{10}\nonumber\\&
+37918308373546561272000663 \ell^{22} r^{12}+50757619733081920144022472 \ell^{20} r^{14}\nonumber\\&
+55227488952051138771625005 \ell^{18} r^{16}+48974717480356234743070466 \ell^{16}
   r^{18}\nonumber\\&
+35284523399336392108933005 \ell^{14} r^{20}+20477493617041202568028488 \ell^{12} r^{22}\nonumber\\&
+9433307528652764598206937 \ell^{10} r^{24}+3370660892553798052350570 \ell^8 r^{26}\nonumber\\&
+900812314937847677993001 \ell^6
   r^{28}+169475537803320197519400 \ell^4 r^{30}\nonumber\\&
+20022251781128501589300 \ell^2 r^{32}+1117614384817280519400 r^{34}\bigg)\nonumber\\&
-\frac{\alpha}{795644021703320165376000}\bigg(21464560511021140995490385 \ell^{34}\nonumber\\&
+336419204700137912471998890 \ell^{32} r^2+2265714121070327795119738725 \ell^{30} r^4\nonumber\\&
+9045794851073754077922660840 \ell^{28} r^6+25157542057967397600077180895 \ell^{26}
   r^8\nonumber\\&
+53007966747953063314567065590 \ell^{24} r^{10}+88148563702087347039590470383 \ell^{22} r^{12}\nonumber\\&
+118053515649034646376315606792 \ell^{20} r^{14}+128548831954201729835945339445 \ell^{18}
   r^{16}\nonumber\\&
+114084293382704825962729579346 \ell^{16} r^{18}+82251607219034803579127543445 \ell^{14} r^{20}\nonumber\\&
+47764004511303199947728571528 \ell^{12} r^{22}+22014741352249962854054900097 \ell^{10}
   r^{24}\nonumber\\&
+7869719713738135878103963770 \ell^8 r^{26}+2104015408041561787435884881 \ell^6 r^{28}\nonumber\\&
+395980508360006822009399400 \ell^4 r^{30}+46797072097258916197449300 \ell^2 r^{32}\nonumber\\&
+2612928661974905984879400
   r^{34}\bigg)\Bigg]\nonumber
\end{align}

\subsection*{n=9 results}
\begin{align}
f_{9;2}={}&\frac{252 \ell^{10}+2310 \ell^8 r^2+1320 \ell^6 r^4+495 \ell^4 r^6+110 \ell^2 r^8+11 r^{10}}{1134}\nonumber\\
\Omega_{9;2}={}&\frac{252 \ell^{10}+210 \ell^8 r^2+120 \ell^6 r^4+45 \ell^4 r^6+10 \ell^2 r^8+r^{10}}{1260}\nonumber\\
\Pi_{9;3}={}&\frac{1}{1025589600}\bigg(356786123 \ell^{18}+2750576510 \ell^{16} r^2+6313029495 \ell^{14} r^4+10221019320 \ell^{12} r^6\nonumber\\&
+11878432160 \ell^{10} r^8+9822117536 \ell^8 r^{10}+5636851220 \ell^6 r^{12}+2135373240 \ell^4 r^{14}+480720240 \ell^2 r^{16}\nonumber\\&
+48768720 r^{18}\bigg)\nonumber\\
f_{9;4}={}\Bigg[&\frac{\ell^2}{859498662576153600}\bigg(133340623834666248 \ell^{32}+2720901281792621928 \ell^{30} r^2\nonumber\\&
+19770363913882326030 \ell^{28} r^4+63513374918908684770 \ell^{26} r^6+135934002405968654888 \ell^{24} r^8\nonumber\\&
+213952704807502670100 \ell^{22}
   r^{10}+257756682509308369842 \ell^{20} r^{12}+242609187466772524790 \ell^{18} r^{14}\nonumber\\&
+180579919814518993740 \ell^{16} r^{16}+107062480455455025288 \ell^{14} r^{18}+50684524900929500343 \ell^{12} r^{20}\nonumber\\&
+19068904767424154265
   \ell^{10} r^{22}+5608737609387866100 \ell^8 r^{24}+1246433118183419250 \ell^6 r^{26}\nonumber\\&
+196811905684882635 \ell^4 r^{28}+19681794632595465 \ell^2 r^{30}+937254465799230 r^{32}\bigg)\nonumber\\&
-\frac{\alpha}{2363621322084422400}\bigg(51858708461340364584 \ell^{32}+919343053698617972424 \ell^{30} r^2\nonumber\\&
+6755425175241524371590 \ell^{28} r^4+20763976427014460249610 \ell^{26} r^6\nonumber\\&
+42888828387704309354344 \ell^{24} r^8+66469172496452595919140 \ell^{22} r^{10}\nonumber\\&
+79612560563631174610986 \ell^{20} r^{12}+74782307435370609291070 \ell^{18} r^{14}\nonumber\\&
+55625640941623976375420 \ell^{16} r^{16}+32972881325499331952424 \ell^{14} r^{18}\nonumber\\&
+15608925102093402787239 \ell^{12} r^{20}+5872467180388490650745 \ell^{10} r^{22}\nonumber\\&
+1727276938801702287300 \ell^8 r^{24}+383856087397493107250 \ell^6 r^{26}\nonumber\\&
+60611238678141544155 \ell^4 r^{28}+6061339452062773545 \ell^2 r^{30}+288644544654444590
   r^{32}\bigg)\Bigg]\nonumber\\
g_{9;4}={}&\Bigg[\frac{\ell^2}{859498662576153600} \bigg(84495843563523392 \ell^{34}+1768012312961670776 \ell^{32} r^2\nonumber\\&
+11936492827687331904 \ell^{30} r^4+37242538451355126890 \ell^{28} r^6+65138644777217991980 \ell^{26} r^8\nonumber\\&
+85422451772928300030 \ell^{24} r^{10}+82904154261107854584 \ell^{22} r^{12}+58562593998521246634 \ell^{20} r^{14}\nonumber\\&
+29286912971461691700 \ell^{18} r^{16}+9866747321987487030 \ell^{16} r^{18}+2041707149885182032 \ell^{14} r^{20}\nonumber\\&
+219306855955192929 \ell^{12} r^{22}+14460723267227622 \ell^{10} r^{24}+4017026312889975 \ell^8 r^{26}\nonumber\\&
+845689750082100 \ell^6 r^{28}+126853462512315 \ell^4 r^{30}+12081282144030 \ell^2 r^{32}+549149188365 r^{34}\bigg)\nonumber\\&
-\frac{\alpha}{787873774028140800}\bigg(10942174841871425152 \ell^{34}+225620421791864115016 \ell^{32} r^2\nonumber\\&
+1510935255902133245184 \ell^{30} r^4+4208312886338787616390 \ell^{28} r^6+6894619427482858451380 \ell^{26} r^8\nonumber\\&
+8848332307639603148370 \ell^{24}
   r^{10}+8534331180859836820104 \ell^{22} r^{12}+6020530420354129493814 \ell^{20} r^{14}\nonumber\\&
+3011536544232528981900 \ell^{18} r^{16}+1015910734943704195530 \ell^{16} r^{18}+211151074883282761392 \ell^{14}
   r^{20}\nonumber\\&
+23184156465480324699 \ell^{12} r^{22}+1720346092586126082 \ell^{10} r^{24}+477878417772403725 \ell^8 r^{26}\nonumber\\&
+100605982688927100 \ell^6 r^{28}+15090897403339065 \ell^4 r^{30}+1437228324127530 \ell^2 r^{32}+65328560187615
   r^{34}\bigg)\Bigg]\nonumber\\
h_{9;4}={}&\Bigg[\frac{\ell^2}{95499851397350400} \bigg(149725307211480 \ell^{30}+3015895473831240 \ell^{28} r^2+8797952237444370 \ell^{26} r^4\nonumber\\&
+15541844723695290 \ell^{24} r^6+20300574058072308 \ell^{22} r^8+21034536810890148 \ell^{20} r^{10}\nonumber\\&
+17783794333160310 \ell^{18}
   r^{12}+12381092092392030 \ell^{16} r^{14}+7089196246488360 \ell^{14} r^{16}\nonumber\\&
+3310396957607208 \ell^{12} r^{18}+1241596238094513 \ell^{10} r^{20}+365184210262725 \ell^8 r^{22}+81152046725050 \ell^6 r^{24}\nonumber\\&
+12813481061850 \ell^4
   r^{26}+1281348106185 \ell^2 r^{28}+61016576485 r^{30}\bigg)\nonumber\\&
-\frac{\alpha}{29180510149190400}\bigg(162655860396589560 \ell^{30}+362008206104224680 \ell^{28} r^2\nonumber\\&
+572645003750910090 \ell^{26} r^4+763224818178352530 \ell^{24} r^6+877801432409857956 \ell^{22} r^8\nonumber\\&
+862035819332396436 \ell^{20} r^{10}+713514589774183470 \ell^{18}
   r^{12}+492832261815045110 \ell^{16} r^{14}\nonumber\\&
+281416001757949320 \ell^{14} r^{16}+131301969045434376 \ell^{12} r^{18}+49236215518455161 \ell^{10} r^{20}\nonumber\\&
+14481164174921325 \ell^8 r^{22}+3218036483315850 \ell^6
   r^{24}+508111023681450 \ell^4 r^{26}\nonumber\\&
+50811102368145 \ell^2 r^{28}+2419576303245 r^{30}\bigg)\Bigg]\nonumber\\
\Omega_{9;4}={}&\Bigg[\frac{\ell^2}{59687407123344000} \bigg(6598585567818972 \ell^{32}+64940197030107672 \ell^{30} r^2+273777806813998500 \ell^{28} r^4\nonumber\\&
+681865813132481250 \ell^{26} r^6+1221884704722504550 \ell^{24} r^8+1674173405090497584 \ell^{22} r^{10}\nonumber\\&
+1807245829726785084 \ell^{20} r^{12}+1565834578187766250 \ell^{18} r^{14}+1101647489722639650 \ell^{16} r^{16}\nonumber\\&
+632417867298394200 \ell^{14} r^{18}+295268817053306040 \ell^{12} r^{20}+110708213943786165 \ell^{10} r^{22}\nonumber\\&
+32561239395231225 \ell^8 r^{24}+7235830976718050 \ell^6 r^{26}+1142499627902850 \ell^4 r^{28}\nonumber\\&
+114249962790285 \ell^2 r^{30}+5440474418585 r^{32}\bigg)\nonumber\\&
-\frac{\alpha}{532923277887000}\bigg(8546242063804172 \ell^{32}+81523891654804872 \ell^{30} r^2+307318194150454500 \ell^{28} r^4\nonumber\\&
+715769933670769250 \ell^{26} r^6+1245320391503010550 \ell^{24} r^8+1686136326474110784 \ell^{22} r^{10}\nonumber\\&
+1811885990514247284 \ell^{20}
   r^{12}+1567209651830709250 \ell^{18} r^{14}+1101953061643293650 \ell^{16} r^{16}\nonumber\\&
+632466115496392200 \ell^{14} r^{18}+295273641873105840 \ell^{12} r^{20}+110708443697109965 \ell^{10} r^{22}\nonumber\\&
+32561239395231225 \ell^8
   r^{24}+7235830976718050 \ell^6 r^{26}+1142499627902850 \ell^4 r^{28}\nonumber\\&
+114249962790285 \ell^2 r^{30}+5440474418585 r^{32}\bigg)\Bigg]\nonumber\\
\Pi_{9;5}={}&\Bigg[\frac{\ell^2}{12928610477426464084962297600000} \bigg(2963660262582405195600812804780 \ell^{42}\nonumber\\&
+54960718928326817581432660068800 \ell^{40} r^2+438680016278348277072828543787740 \ell^{38} r^4\nonumber\\&
+2095459575089784573411052833767630 \ell^{36}
   r^6+7126896792786249923435515161591030 \ell^{34} r^8\nonumber\\&
+18828704309439914264745324330571272 \ell^{32} r^{10}+40318687255770901091487447290055224 \ell^{30} r^{12}\nonumber\\&
+71564603098088232189439094493589402 \ell^{28}
   r^{14}+106606113471306396324075091246911420 \ell^{26} r^{16}\nonumber\\&
+134129087251912145628690383503063720 \ell^{24} r^{18}+142873004760249240821838450887004074 \ell^{22} r^{20}\nonumber\\&
+128775176906725704442407617007965163 \ell^{20}x
   r^{22}+97921202768932011707091446069275799 \ell^{18} r^{24}\nonumber\\&
+62477592200815789501550742105430020 \ell^{16} r^{26}+33171546764185616749586054262623220 \ell^{14} r^{28}\nonumber\\&
+14480110437447756571401543030317327 \ell^{12}
   r^{30}+5107957999464578930096857586903749 \ell^{10} r^{32}\nonumber\\&
+1419986713892478010760754706555122 \ell^8 r^{34}+299476749663332693852942194632980 \ell^6 r^{36}\nonumber\\&
+45035990011802132234460421933080 \ell^4
   r^{38}+4302621620873952400266415533360 \ell^2 r^{40}\nonumber\\&
+196278755076119819922240261840 r^{42}\bigg)\nonumber\\&
-\frac{\alpha}{6691827369268356151636800000}\bigg(188034260736405759920667016480 \ell^{42}\nonumber\\&
+3478119834944534785677233867800 \ell^{40} r^2+27521476972944344576431164066840 \ell^{38} r^4\nonumber\\&
+129974087795678824227161887469830 \ell^{36} r^6+438663601013588577076579436614230
   \ell^{34} r^8\nonumber\\&
+1154908376153092917794909859416952 \ell^{32} r^{10}+2470488821740365277752497363025784 \ell^{30} r^{12}\nonumber\\&
+4385163888580609561595948244197282 \ell^{28} r^{14}+6535069763202647770491146331262470 \ell^{26}
   r^{16}\nonumber\\&
+8226462646319309998709010636408020 \ell^{24} r^{18}+8767061720853343920255745069666384 \ell^{22} r^{20}\nonumber\\&
+7905447916801781766415703529158883 \ell^{20} r^{22}+6013636958064400713013562652664609 \ell^{18}
   r^{24}\nonumber\\&
+3838221370012324936903467456571320 \ell^{16} r^{26}+2038449297489327713347987874728770 \ell^{14} r^{28}\nonumber\\&
+890063930280989733044566047178207 \ell^{12} r^{30}+314051732713817341652961541727309 \ell^{10}
   r^{32}\nonumber\\&
+87324111529482417321012534860802 \ell^8 r^{34}+18420525186923716655392197784180 \ell^6 r^{36}\nonumber\\&
+2770650184390283183984243528280 \ell^4 r^{38}+264748247119254298796127491760 \ell^2
   r^{40}\nonumber\\&
+12079435074425924874820591440 r^{42}\bigg)\Bigg]\nonumber
\end{align}

\section{Conserved Charges and Thermodynamic Quantities}
\label{conscharges}

Here we catalogue the thermodynamic charges and potentials entering the first law of thermodynamics 
 in spacetime dimension $D=n+2$, for $n=3,5,7,9$.  As the boson star temperature is zero, the first law reads $dM = \omega dJ$. 

\subsection*{n=3}
\begin{align}
M_3={}&\frac{5 \pi \ell^2 \epsilon
   ^2}{24}+\frac{77951 \pi  \ell^4 \epsilon ^4}{508032 \left(\ell^2-4 \alpha \right)}+{\cal O}(\epsilon^6)\nonumber\\
J_3={}&\frac{\pi  \ell^3 \epsilon
   ^2}{24}+\frac{83621 \pi  \ell^5 \epsilon ^4}{2540160 \left(\ell^2-4 \alpha \right)}+{\cal O}(\epsilon^6)\nonumber
\end{align}

\begin{align}
\ell\omega_3={}&5-\frac{15 \ell^2 \epsilon ^2}{28 \left(\ell^2-4 \alpha \right)}-\frac{\ell^4 \left(3211271921
   \ell^2-14548856100 \alpha \right) \epsilon ^4}{5085400320 \left(\ell^2-4 \alpha \right)^3}+{\cal O}(\epsilon^6)\nonumber
\end{align}

\subsection*{n=5}
\begin{align}
M_5={}&\frac{7 \pi ^2 \ell^4 \epsilon ^2}{160}+\frac{314018183 \pi ^2 \ell^6 \epsilon ^4}{17667936000 \left(\ell^2-24 \alpha \right)}+{\cal O}(\epsilon^6)\nonumber\\
J_5={}&\frac{\pi ^2 \ell^5 \epsilon ^2}{160}+\frac{327248543 \pi ^2 \ell^7 \epsilon ^4}{123675552000 \left(\ell^2-24 \alpha \right)}+{\cal O}(\epsilon^6)\nonumber
\end{align}

\begin{align}
\ell\omega_5={}&7-\frac{514 \ell^2 \epsilon ^2}{2145 \left(\ell^2-24 \alpha \right)}-\frac{\ell^4  \left(1023791506512739 \ell^2-26707161768137988 \alpha \right)\epsilon ^4}{5355421941870000 \left(\ell^2-24 \alpha \right)^3}+{\cal O}(\epsilon^6)\nonumber
\end{align}

\subsection*{n=7}
\begin{align}
M_7={}&\frac{3 \pi ^3 \ell^6 \epsilon ^2}{560}+\frac{1100829437 \pi ^3 \ell^8 \epsilon ^4}{943498716160 \left(\ell^2-60 \alpha \right)}+{\cal O}(\epsilon^6)\nonumber\\
J_7={}&\frac{\pi ^3 \ell^7 \epsilon ^2}{1680}+\frac{10188926113 \pi ^3 \ell^9 \epsilon ^4}{76423396008960 \left(l^2-60 \alpha \right)}+{\cal O}(\epsilon^6)\nonumber
\end{align}

\begin{align}
\ell\omega_7={}&9-\frac{4135 \ell^2 \epsilon ^2}{37128 \left(\ell^2-60 \alpha \right)}-\frac{\ell^4 \left(1188837899283033191 \ell^2-76263642821315268516 \alpha \right) \epsilon ^4}{20666078485800523776 \left(\ell^2-60 \alpha \right)^3}+{\cal O}(\epsilon^6)\nonumber
\end{align}

\subsection*{n=9}
\begin{align}
M_9={}&\frac{11 \pi ^4 \ell^8 \epsilon ^2}{24192}+\frac{1065755263141 \pi ^4 \ell^{10} \epsilon ^4}{20836331213967360 \left(\ell^2-112 \alpha \right)}+{\cal O}(\epsilon^6)\nonumber\\
J_9={}&\frac{\pi ^4 \ell^9 \epsilon ^2}{24192} + \frac{1088094883717 \pi ^4 \ell^{11} \epsilon ^4}{229199643353640960 \left(\ell^2-112 \alpha \right)}+{\cal O}(\epsilon^6)\nonumber
\end{align}

\begin{align}
\l\omega_9={}&11-\frac{754 \ell^2 \epsilon ^2}{14535 \left(\ell^2-112 \alpha \right)}-\frac{\ell^4  \left(971718029741243000453 \ell^2-115487747450683234769436 \alpha \right)\epsilon ^4}{58241769402253879260000 \left(\ell^2-112 \alpha
   \right)^3}+{\cal O}(\epsilon^6)\nonumber
\end{align}

\section{Kretschmann Scalar and Critcal $q_{h}'$ Values}
\label{Kscalar}

\subsection*{n=3}
\begin{align}
K_3={}&\frac{1}{\alpha ^2 \ell^4 \left(144 \alpha ^2-576 \alpha ^2 q_h'(0)^2+9 \ell^4-72 \alpha  \ell^2+48 \alpha  \ell^2
   q_0^2\right)}\nonumber\\&
\times\Bigg[\ell^2 \left(-\left(144 \alpha ^2-576 \alpha ^2 q_h'(0)^2+9 \ell^4-72 \alpha  \ell^2+48 \alpha  \ell^2 q_0^2\right)^{3/2}\right)\nonumber\\&
-2 \ell^2 \left(48 \alpha ^2+3 \ell^4-24 \alpha  \ell^2+8 \alpha  \ell^2 q_0^2\right) \sqrt{144
   \alpha ^2-576 \alpha ^2 q_h'(0)^2+9 \ell^4-72 \alpha  \ell^2+48 \alpha  \ell^2 q_0^2}\nonumber\\&
+\left(24 \alpha ^2+4 \ell^4-12 \alpha  \ell^2+8 \alpha  \ell^2 q_0^2\right) \left(144 \alpha ^2-576 \alpha ^2 q_h'(0)^2+9 \ell^4-72 \alpha
    \ell^2+48 \alpha  \ell^2 q_0^2\right)\nonumber\\&
+\left(48 \alpha ^2+3 \ell^4-24 \alpha  \ell^2+8 \alpha  \ell^2 q_0^2\right)^2\Bigg]
\end{align}

\begin{equation}
q_{h}'^{*}={}-\frac{\sqrt{144 \alpha ^2+9 \ell^4-72 \alpha  \ell^2+48 \alpha  \ell^2 q_0^2}}{24 \alpha }\nonumber
\end{equation}

\subsection*{n=5}
\begin{align}
K_5={}&\frac{1}{48 \alpha ^2 \ell^4 \left(14400 \alpha ^2-9216 \alpha ^2 q_h'(0)^2+25 \ell^4-1200 \alpha  \ell^2+480 \alpha  \ell^2 q_0^2\right)}\nonumber\\&
\times\Bigg[-2 \ell^2 \left(14400 \alpha ^2-9216 \alpha ^2 q_h'(0)^2+25 \ell^4-1200 \alpha  \ell^2+480 \alpha  \ell^2 q_0^2\right)^{3/2}\nonumber\\&
-4 \ell^2 \left(2880 \alpha ^2+5 \ell^4-240 \alpha  \ell^2+48 \alpha  \ell^2 q_0^2\right)
   \sqrt{14400 \alpha ^2-9216 \alpha ^2 q_h'(0)^2+25 \ell^4-1200 \alpha  \ell^2+480 \alpha  \ell^2 q_0^2}\nonumber\\&
-\left(14400 \alpha ^2-9216 \alpha ^2 q_h'(0)^2+25 \ell^4-1200 \alpha  \ell^2+480 \alpha  \ell^2
   q_0^2\right)^2\nonumber\\&
+\left(17280 \alpha ^2+37 \ell^4-1440 \alpha  \ell^2+576 \alpha  \ell^2 q_0^2\right) \left(14400 \alpha ^2-9216 \alpha ^2 q_h'(0)^2+25 \ell^4-1200 \alpha  \ell^2+480 \alpha  \ell^2 q_0^2\right)\nonumber\\&
+2 \left(2880
   \alpha ^2+5 \ell^4-240 \alpha  \ell^2+48 \alpha  \ell^2 q_0^2\right)^2\Bigg]
\end{align}

\begin{equation}
q_{h}'^{*}={}-\frac{\sqrt{14400 \alpha ^2+25 \ell^4-1200 \alpha  \ell^2+480 \alpha  \ell^2 q_0^2}}{96 \alpha }\nonumber
\end{equation}

\subsection*{n=7}
\begin{align}
K_7={}&\frac{1}{225 \alpha ^2 \ell^4 \left(176400 \alpha ^2-43200 \alpha ^2 q_h'(0)^2+49 \ell^4-5880 \alpha  \ell^2+1680 \alpha  \ell^2 q_0^2\right)}\nonumber\\&
\times\Bigg[-2 \ell^2 \left(176400 \alpha ^2-43200 \alpha ^2 q_h'(0)^2+49 \ell^4-5880 \alpha  \ell^2+1680 \alpha  \ell^2 q_0^2\right)^{3/2}\nonumber\\&
-4 \ell^2 \left(25200 \alpha ^2+7 \ell^4-840 \alpha  \ell^2+120 \alpha  \ell^2 q_0^2\right)
   \sqrt{176400 \alpha ^2-43200 \alpha ^2 q_h'(0)^2+49 \ell^4-5880 \alpha  \ell^2+1680 \alpha  \ell^2 q_0^2}\nonumber\\&
-2 \left(176400 \alpha ^2-43200 \alpha ^2 q_h'(0)^2+49 \ell^4-5880 \alpha  \ell^2+1680 \alpha  \ell^2
   q_0^2\right)^2\nonumber\\&
+\left(378000 \alpha ^2+114 \ell^4-12600 \alpha  \ell^2+3600 \alpha  \ell^2 q_0^2\right) \left(176400 \alpha ^2-43200 \alpha ^2 q_h'(0)^2+49 \ell^4-5880 \alpha  \ell^2+1680 \alpha  \ell^2 q_0^2\right)\nonumber\\&
+2\left(25200 \alpha ^2+7 \ell^4-840 \alpha  \ell^2+120 \alpha  \ell^2 q_0^2\right)^2\Bigg]
\end{align}

\begin{equation}
q_{h}'^{*}={}-\frac{\sqrt{529200 \alpha ^2+147 \ell^4-17640 \alpha  \ell^2+5040 \alpha  \ell^2 q_0^2}}{360 \alpha }
\end{equation}

\subsection*{n=9}
\begin{align}
K_9={}&\frac{1}{28224 \alpha ^2 \ell^4 \left(112896 \alpha ^2-14336 \alpha ^2 q_h'(0)^2+9 \ell^4-2016 \alpha  \ell^2+448 \alpha  \ell^2
   q_0^2\right)}\nonumber\\&
\times\Bigg[-270 \ell^2 \left(112896 \alpha ^2-14336 \alpha ^2 q_h'(0)^2+9 \ell^4-2016 \alpha  \ell^2+448 \alpha  \ell^2 q_0^2\right)^{3/2}\nonumber\\&
-60 \ell^2 \left(112896 \alpha ^2+9 \ell^4-2016 \alpha  \ell^2+224 \alpha  \ell^2 q_0^2\right)
   \sqrt{112896 \alpha ^2-14336 \alpha ^2 q_h'(0)^2+9 \ell^4-2016 \alpha  \ell^2+448 \alpha  \ell^2 q_0^2}\nonumber\\&
-1215 \left(112896 \alpha ^2-14336 \alpha ^2 q_h'(0)^2+9 \ell^4-2016 \alpha  \ell^2+448 \alpha  \ell^2
   q_0^2\right)^2\nonumber\\&
+\left(142248960 \alpha ^2+11835 \ell^4-2540160 \alpha  \ell^2+564480 \alpha  \ell^2 q_0^2\right)\nonumber\\&
\times\left(112896 \alpha ^2-14336 \alpha ^2 q_h'(0)^2+9 \ell^4-2016 \alpha  \ell^2+448 \alpha  \ell^2
   q_0^2\right)\nonumber\\&
+10 \left(112896 \alpha ^2+9 \ell^4-2016 \alpha  \ell^2+224 \alpha  \ell^2 q_0^2\right)^2\Bigg]
\end{align}

\begin{equation}
q_{h}'^{*}={}-\frac{\sqrt{1580544 \alpha ^2+126 \ell^4-28224 \alpha  \ell^2+6272 \alpha  \ell^2 q_0^2}}{448 \alpha }\nonumber
\end{equation}

\bibliographystyle{unsrt}
\bibliography{bosonstarrefs}

\end{document}